\documentclass[a4paper,11pt]{article}
\pdfoutput=1

\usepackage{jcappub}% for details on the use of the package, please
\renewcommand{\afterAuthorSpace}{\vskip 8pt}
\renewcommand{\afterAffiliationSpace}{\vskip 0pt}
\usepackage[T1]{fontenc} % if needed
\usepackage{graphicx}
\usepackage{float}
\usepackage{placeins}
\usepackage{tablefootnote}
\usepackage{pbox}
\usepackage{booktabs}
\usepackage{makecell}
\usepackage{multirow}
\usepackage{array}
\usepackage{mleftright}
\usepackage[ISO]{diffcoeff}
\usepackage{siunitx}
\usepackage{hyperref}
\usepackage[capitalise]{cleveref}

\makeatletter
\input{aas_macros.sty}
\let\jnl@style=\relax
\makeatother

\DeclareSIUnit{\pc}{pc}
\DeclareSIUnit{\kpc}{\kilo\pc}
\DeclareSIUnit{\Mpc}{\mega\pc}
\DeclareSIUnit{\Gpc}{\giga\pc}
\DeclareSIUnit{\hHubble}{\text{\ensuremath{h}}}
\crefname{section}{\S}{\S}
\crefformat{section}{\S#2#1#3}
\crefmultiformat{section}{\S#2#1#3}{ and~\S#2#1#3}{, \S#2#1#3}{ and~\S#2#1#3}
\AddToHook{cmd/appendix/before}{%
    \crefalias{chapter}{appendix}%
    \crefalias{section}{appendix}%
    \crefalias{subsection}{appendix}%
}

\title{A Finely-Binned Measurement of the Connected Even-Parity Galaxy 4-Point Correlation Function of DESI Year 1 Luminous Red Galaxies}
\author[1]{William Ortolá Leonard,}
\author[2,3,4,*]{Alex Krolewski,}
\author[1,6]{Zachary Slepian,}
\author[1]{Alessandro Greco,}
\author[5]{Simon May,}
\author[6]{J.~Aguilar,}
\author[7]{S.~Ahlen,}
\author[8]{O.~Alves,}
\author[9,10]{A. ~Aviles,}
\author[11]{F. ~Beutler,}
\author[12,13]{D.~Bianchi,}
\author[14]{D.~Brooks,}
\author[15,16]{A. Carnero Rosell}
\author[6]{T.~Claybaugh,}
\author[17]{A.~de la Macorra,}
\author[18,19]{Biprateep~Dey,}
\author[6,20]{S. ~Ferraro,}
\author[21,22]{J.~E.~Forero-Romero,}
\author[23,24,25]{E.~Gazta\~{n}aga,}
\author[26]{Satya~{Gontcho A Gontcho},}
\author[27]{G.~Gutierrez,}
\author[28,29,30]{K. Honscheid,}
\author[31,8]{D.~Huterer,}
\author[32]{M.~Ishak,}
\author[33]{S. ~Juneau,}
\author[18]{T.~Karim,}
\author[34]{R. Kehoe}
\author[35]{D.~Kirkby,}
\author[14]{O.~Lahav,}
\author[6]{M.~Landriau,}
\author[36]{L.~Le~Guillou,}
\author[37,38]{M. ~Manera}
\author[33]{A. ~Meisner,}
\author[39,38]{R.~Miquel,}
\author[19]{J.~A.~Newman,}
\author[40,41]{E.~Paillas,}
\author[3,42,2]{W.~J.~Percival,}
\author[43]{F.~Prada,}
\author[44]{I.~P\'erez-R\`afols,}
\author[45]{C.~Ravoux,}
\author[46]{R. ~Ruggeri,}
\author[47,48]{L. Samushia,}
\author[49]{E.~Sanchez,}
\author[50]{C.~Saulder,}
\author[6]{D.~Schlegel,}
\author[6]{J. H. ~Silber,}
\author[25,16]{M.~Siudek,}
\author[8]{G.~Tarl\'{e},}
\author[33]{B.~A.~Weaver}

\emailAdd{wortola@ufl.edu}
\emailAdd{akrolews@caltech.edu}
\emailAdd{zslepian@ufl.edu}
\emailAdd{alessandro.greco@ufl.edu}
\emailAdd{simon.may@uni-bielefeld.de}

\abstract{We present the connected even-parity galaxy 4-Point Correlation Function (4PCF) of DESI Year 1 (Y1) Luminous Red Galaxies (LRGs). This is one of the first measurements of the connected 4PCF on data, and shows a clear detection at $\sim$12 to 17$\sigma$ in the full-sky analysis. We test the robustness of the signal by varying three factors: hemisphere (north vs. south), redshift range (either the full range $0.4 < z < 1.1$ or the higher-number-density interval $0.4 < z < 0.8$), and sample completeness (the full sample vs.\ only the most complete regions within the number density-motivated cut). Finally, we cross-correlate different patches that are spatially well-separated, and thus should have largely independent noise. This approach, at leading order, is able to remove mismatch between the covariance matrix of the data and that of the mocks; however, in certain cases at the cost of sensitivity. We find $\sim$15$\sigma$ evidence for an even-parity 4PCF in this approach. Overall, the studies of the 4PCF presented here probe sensitively any mismatch between mock catalogs and data, and also open up the prospect of fitting a model,  constraining cosmological parameters and galaxy biases, and searching for Baryon Acoustic Oscillation (BAO) features.}

\newcommand{\vecx}{\mathbf{x}}
\newcommand{\vecr}{\mathbf{r}}
\newcommand{\vecs}{\mathbf{s}}
\newcommand{\hatr}{\widehat{\mathbf{r}}}
\newcommand{\PP}{\mathcal{P}}
\newcommand{\tj}[6]{\begin{pmatrix} {#1} & {#2} & {#3}\\ {#4} & {#5} & {#6}\end{pmatrix}}

\begin{document}

\maketitle
%\flushbottom

\section{Introduction}

Inflation (see \textit{e.\,g.}\ \cite{guth_2007} for an accessible review) generates a nearly Gaussian Random Field (GRF), which can be characterized completely by its 2-Point Correlation Function (2PCF) or the Fourier-space analog, the power spectrum. Very small, and as yet unmeasured, corrections beyond the GRF are expected, and this is termed primordial non-Gaussianity (PNG) (see \cite{bartolo}, for a review). However, the late-time galaxy field we observe is far from Gaussian: galaxies are biased tracers of the underlying matter \cite{Desjacques_2018}, and furthermore, the underlying matter has undergone non-linear evolution due to gravity \cite{Bernardeau_2002, goroff, bert_jain}, which generates higher-order correlations such as the 3-Point (3PCF) and 4-Point Correlation Functions (4PCF) or their Fourier-space analogs, the bispectrum and trispectrum \cite{Encyclopedia_Slepian}.

Much work has focused on measuring the 2PCF or power spectrum, including on the dataset we study in this paper, the Dark Energy Spectroscopic Instrument (DESI) \cite{DESI_Science_PartI, DESI_Science_PartII, Instrument_Overview_DESI, Optical_corrector_DESI, Survey_Operations_DESI, Fiber_system_DESI, Data_Processing_DESI} Year~1 (Y1) Luminous Red Galaxies (LRGs) \cite{desi_y1_2ps, Zhou_LRG_Selection, DESI_DR1_paper, adame2024desi}. The 3PCF and bispectrum have also fairly recently entered regular use as a tool for cosmology, with a series of papers presenting them for DESI's predecessor (and the largest previous galaxy redshift survey), the Sloan Digital Sky Survey (SDSS), Baryon Oscillation Spectroscopic Survey (BOSS) and extended BOSS (eBOSS), \textit{e.\,g.}\ \cite{se_3pcf_bao, se_boss_3pcf, gil2015power, gualdi_2022_data_trispec}. A particularly exciting BOSS result was the first detection of Baryon Acoustic Oscillation (BAO) features \cite{SE_BAO_2016, esw_07, Eisenstein_05, moresco_3pcf_bao} in the 3PCF \cite{se_3pcf_bao} and the bispectrum \cite{pearson_bao_bispec}, with earlier evidence (a few $\sigma$) having been seen in the 3PCF in 2009 \cite{gaztanaga_2009_3pcf}. Both the 3PCF and the bispectrum were used to measure cosmological parameters \cite{se_boss_3pcf, se_3pcf_bao, pearson_bao_bispec, gil2015power, Gil-Marin_2017_RSDMeasurement, ivanov2023cosmology, Ivanov_2022, philcox2022boss} as well as constrain baryon–dark matter relative velocity biasing (3PCF) \cite{se_rv_boss, se_rsd_3pcf, se_rv} (also studied in the power spectrum \cite{Beutler_rv}). Newer results have also measured BAO with DESI \cite{DESI_ResultsII_BAO, DESI_III_BAO_from_Galaxies, Kamalinejad_3PCF_BAO} and constraint cosmological parameters \cite{DESI_VI_Cosmological_Constraints_BAO}.

The 4PCF is gaining attention in cosmology due to its ability to probe new aspects of the large-scale structure (LSS) of the Universe. In particular, the 4PCF enables the detection of parity-odd modes in the data, a method first proposed by \cite{cahn_parity} (for related works, see \cite{jeong_fossil, jamieson_pops}; see \cite{shiraishii} for a similar idea in the context of the Cosmic Microwave Background (CMB), and \cite{inomata} for discussion of detecting parity violation with the CMB lensing 4-point function). These ``odd'' modes are inaccessible to lower-order statistics as parity requires a 3D observable to probe \cite{cahn_parity}. This approach was applied to BOSS data by \cite{hou_parity}, yielding a $7.1\sigma$ detection, and by \cite{phil_parity}, who reported 2.9$\sigma$ evidence. However, a subsequent study of the same data set by \cite{Krowleski_No_PV_det}, using a different method that is more robust to mismatch between the mock and the data covariances, found evidence of these modes to be at a much lower level, ranging from none to $2.5\sigma$ depending on the specific sky patch combinations used. Theoretically, parity-violating models for the trispectrum have been developed, such as \cite{reinhard_axion, niu_axion_trispectrum}. These theoretical models, combined with the increasing volume of available data, will help determine whether the previous measurements of parity-odd modes were genuine or the result of systematic effects.

\cite{Hou_Even_4PCF} recently measured the even-parity 4PCF signal of the DESI Y1 LRG sample---see \cite{DESI_Catalog_Construction} for the details of the DESI catalog construction---using ten radial bins. They measured the NGC and SGC signals over the full LRG redshift range, $0.4<z<1.1$, while their reduced-region analyses used the narrower redshift range $0.4<z<0.8$ for the NGC and SGC subregions. Previously, the even-parity 4PCF had been measured in BOSS \cite{phil_4pcf}. Theoretical modeling of the even-parity 4PCF has also recently been developed \cite{Ortola_4PCF}, building on earlier work \cite{gualdi_2021_trispec, gualdi_joint_bi_tri, bertolini, fry_peebles_4pcf, Zwicky_1961_CatalogueGalaxies, fry_bbgky, fry_n_point_theory}. In parallel, both the Gaussian \cite{hou_covar} and next-to-leading order \cite{ortola_cov_I, ortola_cov_II} 4PCF covariance have been modeled. On another front, \cite{Williamson_4PCF_MHD} presented the first measurement of the 4PCF in a suite of magnetohydrodynamic turbulence simulations. The Fourier-space analog of the 4PCF, the Trispectrum, has also been studied and measured \cite{gualdi_2021_trispec, gualdi_2022_data_trispec, gualdi_joint_bi_tri}. A fuller history of the 4PCF and Trispectrum work is given in \cite{Encyclopedia_Slepian}. 

In this work, we measure the connected (\textit{i.e.}, beyond-Gaussian) 4PCF of the DESI Y1 LRGs. We perform the measurement over the full redshift range, $0.4<z<1.1$, as well as over the narrower interval, $0.4<z<0.8$, for both the NGC and SGC. We use an 18-bin approach for the measurement: first, a galaxy is selected as the center of the coordinate system (the primary galaxy). Eighteen radial bins are then constructed around the primary, and tetrahedra are formed only with secondary galaxies located at separations between $r_{\rm min}=20$ [Mpc$/h$] and $r_{\rm max}=160$ [Mpc$/h$]. The resulting radial bin width is $\Delta=7.8$ [Mpc$/h$]. Additionally, our work emphasizes assessing the statistical agreement between the data and the mocks. Obtaining agreement is crucial because, when measuring parity-odd modes, the $\chi^2$ statistic effectively involves the product of two 4PCFs (\textit{i.\,e.}, an 8PCF). As explained in \cite{Krowleski_No_PV_det}, if the parity-even 8PCF differs significantly between the mocks and the data, it could lead to a false parity violation signal. We currently cannot measure the 8PCF, but we can look at the agreement between mocks and data at the 4PCF level. 

The structure of this work is as follows. In \cref{Sec:DESI_data_Y1}, we describe the DESI Y1 data and its properties, followed by our 4PCF detection method in \cref{Sec:4PCF_Method}. In \cref{sec:msd_4}, we present the measurements of the 4PCF and comparison with the mocks, and in \cref{sec:Results} we show our results. Finally, we end with a discussion in \cref{Sec:Discussion}. 

\section{DESI Year 1 Luminous Red Galaxy Sample}
\label{Sec:DESI_data_Y1}

We perform our analysis using the DESI Y1 LRG dataset, as it provides a crucial tracer sample for mapping large-scale structure at intermediate redshifts, specifically within the range $0.4 < z < 1.1$, covering observations from May 2021 to June 2022. The celestial footprint of DESI LRGs covers a broad range in right ascension (RA) and declination (DEC), enabling robust measurements of galaxy clustering and cosmological parameters across multiple sky regions with diverse observing conditions.

\begin{figure}
    \centering
    \includegraphics[width=0.5\linewidth]{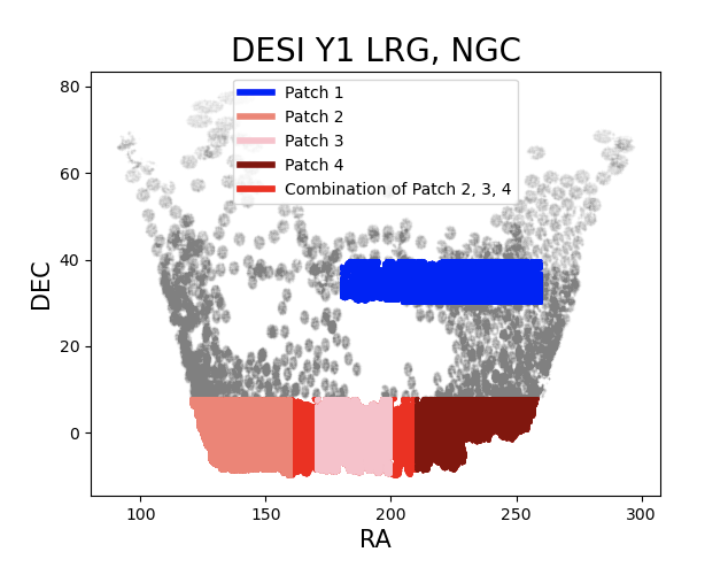}%
    \hfill%
    \includegraphics[width=0.5\linewidth]{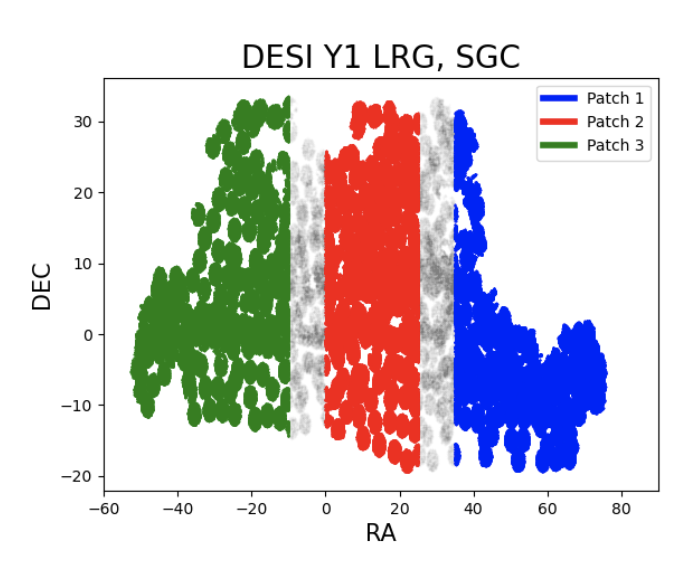}

    \caption{DESI Y1 footprints in the North Galactic Cap (NGC, left) and South Galactic Cap (SGC, right) showing the full sample (gray), regions, and patches (reproduced from \cite{Slepian_Parity_Meas_DESI}). Patches~2, 3, and 4 in the NGC are separated by the two vertical scarlet stripes, so that we have four patches in the NGC and three in the SGC. In addition to these \textit{patches}, we define two larger \textit{regions} in the NGC (upper, Region~1, in blue corresponding to Patch~1, and lower, Region~2, being Patches~2, 3, and 4 plus the scarlet stripes) and one in the SGC (its entirety). All \textit{regions} are cut to $0.4 < z < 0.8$, as motivated by the flatness in redshift distribution up to $z=0.8$ (\cref{fig:number_density}), while the \textit{patches} contain the entire redshift range, $0.4 < z < 1.1$. The data and mocks used for this work have been made available online \cite{Zenodo_repo_with_Data}.}
    \label{fig:footprint}
\end{figure}

\subsection{Survey Footprint, Redshift Cuts, and Subsamples}

The DESI Y1 LRG sample used in this work spans two spatially-separated regions of the sky: the North Galactic Cap (NGC) and the South Galactic Cap (SGC), as illustrated in \cref{fig:footprint}. This division arises from obscuration by the Milky Way disk, which prevents continuous sky coverage. Because DESI operates from Kitt Peak National Observatory in the Northern hemisphere, observational conditions favor the northern sky, both in terms of airmass and horizon accessibility. As a result, the NGC covers a larger angular footprint than the SGC.

For the Y1 data analyzed here, the total sky coverage reaches about \SI{3557}{deg\squared} in the NGC and \SI{1916}{deg\squared} in the SGC. These correspond to effective volumes of \SI{7.2}{\per\hHubble\cubed\Gpc\cubed} and \SI{3.9}{\per\hHubble\cubed\Gpc\cubed}, respectively. A summary of these values is provided in \cref{tab:region-summary}. Completeness maps for the Year~1 footprint are presented in \cite{desi_y1_2ps}.

\begin{figure}
    \centering
    \includegraphics[width=0.55\textwidth]{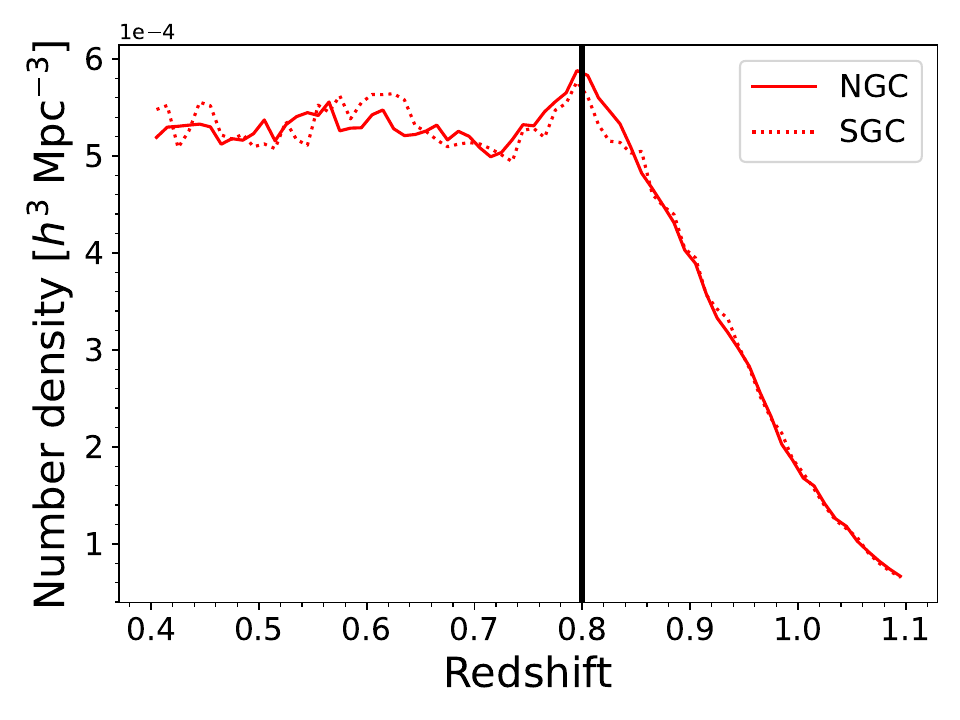}

    \caption{Number density for the full LRG samples (NGC solid, SGC dotted). The black vertical line marks $z=0.8$, above which the number density is no longer approximately constant and begins to decrease rapidly. It also delineates the two redshift ranges analyzed in this work: $0.4<z<0.8$ and $0.4<z<1.1$. The figure has been taken from \cite{Slepian_Parity_Meas_DESI}.}
    \label{fig:number_density}
\end{figure}

\paragraph{Redshift selections.}
We consider two redshift ranges throughout our analysis. The primary sample spans $0.4 < z < 1.1$, while a restricted subsample applies an upper cut at $z = 0.8$. This latter range is chosen to probe sensitivity to the redshift distribution, $n(z)$. As shown in \cref{fig:number_density}, $n(z)$ remains relatively flat up to $z \simeq 0.8$ before declining rapidly at higher redshifts. Comparing measurements between these two ranges therefore allows us to assess the impact of the high-redshift tail on our results.

\paragraph{Angular regions and patches.}
In addition to the full NGC and SGC footprints, we analyze a set of smaller subsamples designed to isolate the effects of survey geometry, angular completeness, and redshift depth. We distinguish between \emph{regions}, which are defined by a combination of angular and redshift cuts, and \emph{patches}, which are purely angular subdivisions.
These subdivisions are illustrated in \cref{fig:footprint}. Within the NGC, we identify a high-completeness, contiguous region (``Region~2'') of approximately \SI{1600}{deg\squared}, which we analyze both over the full redshift range and with the $z < 0.8$ cut applied. In the SGC, we consider the full angular footprint for both redshift ranges. That is, the SGC redshift-cut sample ($0.4 < z < 0.8$) also uses the full SGC footprint. Finally, for analyses that benefit from internal cross-correlations, we subdivide the footprint into seven approximately equal-area patches. The NGC is divided into four patches drawn from two high-completeness contiguous regions, while the SGC is divided into three patches selected to match the NGC patch areas as closely as possible. All seven patches span the full redshift range $0.4 < z < 1.1$.

\begin{table}
    \centering
    \begin{tabular}{l ccc S[table-format=4.0] S[table-format=1.1]}
        \toprule
        \textbf{Sector} & \textbf{RA} & \textbf{DEC} & \textbf{\boldmath$z$} & {\textbf{Angular Size}} & {\textbf{\boldmath $V_{\mathrm{eff}}$}} \\
               & (deg) & (deg) & & {(deg$^2$)} & {(\si{\Gpc\cubed})} \\
        \midrule
        Full NGC      &   &   & $[0.4, 1.1]$     &  3557 & 7.2 \\
        NGC Patch 1   & $[\phantom{-}180, \phantom{-}260]$  & $[\phantom{-}30, 40]$    & $[0.4, 1.1]$     &  488  & 1.0 \\
        NGC Patch 2   & $[\phantom{-}110, \phantom{-}160]$  & $[-10, \phantom{0}8]$    & $[0.4, 1.1]$     &  506  & 1.0 \\
        NGC Patch 3   & $[\phantom{-}170, \phantom{-}200]$  & $[-10, \phantom{0}8]$    & $[0.4, 1.1]$     &  426  & 0.9 \\
        NGC Patch 4   & $[\phantom{-}210, \phantom{-}260]$  & $[-10, \phantom{0}8]$    & $[0.4, 1.1]$     &  458  & 0.9 \\
        NGC Region 1  & $[\phantom{-}180, \phantom{-}260]$  & $[\phantom{-}30, 40]$    & $[0.4, 0.8]$   & 488   &  1.0   \\
        NGC Region 2  & $[\phantom{-}110, \phantom{-}260]$  & $[-10, \phantom{0}8]$    & $[0.4, 0.8]$      &   1666  & 3.4 \\[\smallskipamount]
        Full SGC &   &   & $[0.4, 1.1]$     & 1916  & 3.9 \\
        SGC Patch 1 & $[\phantom{0}{-}55, \phantom{0}{-}10]$  & $[-17,35]$   & $[0.4, 1.1]$  & 480  &  1.0 \\
        SGC Patch 2 & $[\phantom{-00}0, \phantom{-0}25]$  & $[-17, 35]$   & $[0.4, 1.1]$  &  537 &  1.1 \\
        SGC Patch 3 & $[\phantom{-0}35, \phantom{-0}75]$  & $[-17, 35]$   & $[0.4, 1.1]$  & 479 & 1.0  \\
        SGC Region & $[\phantom{0}{-}55, \phantom{-0}75]$  & $[-17, 35]$   & $[0.4, 0.8]$   &  480  &  1.0 \\
        \bottomrule
    \end{tabular}

    \caption{Summary of the parts of the sky used in this analysis, including their corresponding intervals in Right Ascension (RA), Declination (dec), and redshift ($z$), angular size, and effective comoving volume. These parts are drawn from the DESI Y1 footprint and include both full and redshift-limited ($z$-cut) patches in the NGC and SGC. The effective volume is computed for each part over its specified redshift range, assuming a flat $\Lambda$CDM cosmology with \textit{Planck} 2018 parameters. The data and mocks used for this work have been made available online \cite{Zenodo_repo_with_Data}.}
    \label{tab:region-summary}
\end{table}

\section{Methodology}
\label{Sec:4PCF_Method}

Here we briefly outline the method to measure the connected 4PCF, which includes only non-Gaussian contributions. For details of the algorithm, see \cite{encore} (though we used a GPU version of the code, called \textsc{cadenza}), and for a fuller discussion of the method for the parity-even 4PCF, including subtraction of the disconnected piece, \cite{phil_4pcf}. Here we simply recapitulate sufficient detail for our results to be understandable.

\subsection{The 4PCF and the Isotropic Basis Functions}

We measure the isotropic 4PCF, \textit{i.\,e.}\ the clustering of tetrahedra over and above what a spatially random distribution of points would have. We average over translations and rotations in 3D space and have
\begin{equation}
    \zeta(\vecr_1,\vecr_2,\vecr_3) = \sum_{\ell_i} \zeta_{\ell_1 \ell_2 \ell_3}(r_1, r_2, r_3) \mathcal{P}_{\ell_1 \ell_2 \ell_3}(\hatr_1, \hatr_2, \hatr_3),
\end{equation}
where we have split the 4PCF $\zeta$ into its radial and angular parts, $\zeta_{\ell_1 \ell_2 \ell_3}$ and $\mathcal{P}_{\ell_1 \ell_2 \ell_3}$, on the right-hand side. The angular part is described in terms of the isotropic basis functions of \cite{cahn_iso, iso_gen}, given, for three arguments, by 
\begin{equation}
    \label{eq:iso_function_def}
    \mathcal{P}_{\ell_1 \ell_2 \ell_3}(\hatr_1, \hatr_2, \hatr_3) \equiv (-1)^{\ell_1 +\ell_2 + \ell_3}\sum_{m_i} \begin{pmatrix}
    \ell_1 & \ell_2 & \ell_3\\
    m_1 & m_2 & m_3
    \end{pmatrix} Y_{\ell_1 m_1}(\hatr_1)
        Y_{\ell_2 m_2}(\hatr_2)
        Y_{\ell_3 m_3}(\hatr_3),
\end{equation}
with $Y_{\ell m}$ being the spherical harmonics. There are only three arguments because we always consider one galaxy to be at the origin of coordinates, and the three $\hatr_i$ are the direction (unit) vectors pointing to the other three galaxies in the tetrahedron, relative to this origin. The $2 \times 3$ ``matrix'' is a Wigner 3-$j$ symbol, trivially related to the Clebsch–Gordan coefficients by a phase and a rescaling\footnote{\url{https://mathworld.wolfram.com/Wigner3j-Symbol.html}} 
\begin{equation}
\begin{pmatrix}
j_1 & j_2 & j_3 \\
m_1 & m_2 & m_3
\end{pmatrix}
=
\frac{(-1)^{j_1-j_2-m_3}}{\sqrt{2j_3+1}}
\,
\langle j_1 m_1,\; j_2 m_2 \mid j_3,\,-m_3\rangle.
\end{equation}
To obtain the radial part we exploit the basis' orthonormality and integrate both sides against $\mathcal{P}^*$, where the star means complex conjugation. We then obtain
\begin{equation}
    \label{eq:4PCF_Coeff}
    \zeta_{\ell_1 \ell_2 \ell_3}(r_1, r_2, r_3) = \int \dl\hatr_1 \dl\hatr_2 \dl\hatr_3 \;
    \mathcal{P}^*_{\ell_1 \ell_2 \ell_3}(\hatr_1, \hatr_2, \hatr_3) \,
    \zeta(\vecr_1, \vecr_2, \vecr_3),
\end{equation}
with the left-hand side being the 4PCF coefficient we measure from the data. 

We note that the parity of these functions may be deduced from the behavior of their constituent spherical harmonics under parity, $Y_{\ell m}(-\hatr) \to (-1)^{\ell} Y_{\ell m}(\hatr)$. Hence the parity is
\begin{equation}
    \mathcal{P}_{\ell_1 \ell_2 \ell_3}(-\hatr_1, -\hatr_2, -\hatr_3) = (-1)^{\ell_1 + \ell_2 + \ell_3}
    \mathcal{P}_{\ell_1 \ell_2 \ell_3}(\hatr_1, \hatr_2, \hatr_3).
\end{equation}
Thus, an even sum of the $\ell_i$ implies even parity, and vice versa, while an odd sum implies odd parity, and vice versa.

Now, we notice that for an odd sum of the $\ell_i$, the 3-$j$ symbol in \cref{eq:iso_function_def} receives a phase of $(-1)$ under non-cyclic permutations of its columns.\footnote{\url{https://dlmf.nist.gov/34.3}, Eq.~34.3.9} Thus, it matters what $\ell_1, \ell_2, \ell_3$ are: there is not a symmetry under interchange of the indices. Hence, if the sum of the $\ell_i$ is odd, one needs a way to assign observable meaning to the $\ell_i$. This point is detailed more fully in \cite{cahn_iso}. Here, since our focus is on the even-parity modes, we will not discuss this further, since it is not a necessary consideration for the even-parity $\mathcal{P}_{\ell_1 \ell_2 \ell_3}$ we will require. The isotropic basis functions are further explored, and a generating function given, in \cite{iso_gen}.

\subsection{Estimator in the Isotropic Basis Functions and Algorithm for Evaluating It}

In practice, our algorithm proceeds by forming the quantity 
\begin{equation}
   \zeta(\vecr_1,\vecr_2,\vecr_3) =
   \frac{1}{V} \int \dl\vecx \,
   \delta(\vecx)
   \delta(\vecx + \vecr_1)
   \delta(\vecx + \vecr_2)
   \delta(\vecx + \vecr_3),
\end{equation}
where $\delta$ is the galaxy density contrast, the galaxy at point $\vecx$ is the ``primary'' and the other three are the ``secondaries'' (as first suggested in \cite{se_3pt_alg}), and $V$ is the survey volume. Then, we insert it into \cref{eq:4PCF_Coeff} to find an estimator---in terms of radial binning $(a,b,c)$ of the distances $(r_1, r_2, r_3)$---for the 4PCF as \cite{phil_4pcf}:
\begin{align}
    \label{eq:4PCF_radial_def}
    \hat{\zeta}^{abc}_{\ell_1\ell_2\ell_3} &=
    \frac{1}{V} \int \dl\vecx \dl\vecr_1 \dl\vecr_2 \dl\vecr_3 \,
    \delta(\vecx) \delta(\vecx+\vecr_1) \delta(\vecx+\vecr_2) \delta(\vecx+\vecr_3)
    \nonumber \\
    &\qquad
    \times \PP^*_{\ell_1\ell_2\ell_3}(\hatr_1, \hatr_2, \hatr_3) \frac{\Theta_a(r_1)}{V_a} \frac{\Theta_b(r_2)}{V_b} \frac{\Theta_c(r_3)}{V_c},
\end{align}
where $\Theta_i$ equals unity if the magnitude of the vector is in the bin $i$ and zero otherwise; $V_i$ are the bin volumes. Then, we need to subtract the disconnected piece of the 4PCF, as shown in Eq.~(21) of \cite{phil_4pcf}:
\begin{align}
    \hat{\zeta}_{\ell_1\ell_2\ell_3}^{(\mathrm{disconn}),\; abc} &=
    (-1)^{\ell_1+\ell_2+\ell_3} \sum_{m_1m_2m_3} \tj{\ell_1}{\ell_2}{\ell_3}{m_1}{m_2}{m_3} \hat{\xi}_{\ell_1m_1}^{a}(r_1) \, \hat{\xi}_{\ell_2 m_2\ell_3m_3}^{bc}(r_2,r_3)
    \nonumber
    \\
    &\qquad + \text{2 perms.},
\end{align}
where we have defined
\begin{align}
    \hat{\xi}_{\ell m}^{a}
    &\equiv
    \frac{1}{V}
    \int \dl\vecs \dl\vecr \,
    \delta(\vecs) \delta(\vecs+\vecr) \,
    Y^{*}_{\ell m}(\hatr)\frac{\Theta_a(r)}{V_a}
    \,
    ,
    \\
    \hat{\xi}_{\ell m \ell' m'}^{bc}
    &\equiv
    \frac{1}{V}
    \int \dl\vecs \dl\vecr \dl\vecr' \,
    \delta(\vecs+\vecr) \delta(\vecs+\vecr')\,
    Y_{\ell m}(\hatr)Y^{*}_{\ell' m'}(\hatr')\frac{\Theta_b(r)}{V_b}\frac{\Theta_c(r')}{V_c}
    \,
    ,
\end{align}
to obtain the connected 4PCF,
\begin{equation}
    \hat{\zeta}^{(\mathrm{conn}),\; abc}_{\ell_1\ell_2\ell_3} =
    \hat{\zeta}_{\ell_1\ell_2\ell_3}^{abc} - \hat{\zeta}_{\ell_1\ell_2\ell_3}^{(\mathrm{disconn}),\; abc}
    \,
    . 
\end{equation}

To evaluate these estimators for a survey with a nonuniform geometry and selection function, we represent the density fluctuation field using a generalization of the Landy--Szalay estimator \cite{landy_szalay},
\begin{equation}
\delta \rightarrow \frac{D-R}{R},
\end{equation}
where $D$ and $R$ denote the appropriately weighted data and random density fields, respectively. The random field traces the expected galaxy density in the absence of clustering, including the angular and radial selection functions of the survey. As described in \S3 of \cite{se_3pt_alg} for the 3PCF, with the analogous construction applying here, each factor of $\delta$ is weighted by the corresponding random density. In the shot-noise limit, this gives the optimally weighted field
\begin{equation}
R\delta = D-R.
\end{equation}
The numerator of the estimator can therefore be evaluated using products of the difference field $D-R$, while the corresponding products of the random field provide the normalization and account for the survey window. Further details of this estimator and its normalization are given in \cite{encore}. The GPU implementation used in this work, \textsc{cadenza}, employs the same estimator and weighting prescription.

%The density fluctuation field, in practice, is evaluated with a generalization of the Landy–\allowbreak Szalay estimator \cite{landy_szalay}, so that
%\begin{equation}
%    \delta \to \frac{D - R}{R},
%\end{equation}
%where $D$ represents a data particle and $R$ a random particle. As further described in \cite{se_3pt_alg}, \S3, in the context of the 3PCF (but analogous here), each $\delta$ is then optimally weighted (in the shot-noise limit) by $R$, and the weights are normalized. Fuller details are given in \cite{encore} and the GPU version used for this work, \textsc{cadenza}, is the same in this regard.

\subsection{Computational Scaling of the 4PCF Algorithm}

Now, for efficiency, \cref{eq:4PCF_radial_def} is evaluated by binning radially first, and then obtaining the harmonic coefficients using the factorized representation of $\mathcal{P}_{\ell_1\ell_2\ell_3}$, and finally multiplying these and summing over the $m_i$, around each primary. This results in a formally order $N^2$ algorithm, with $N$ the number of objects; this is because the number of neighbors around a primary scales as $N$, and the cost of obtaining the harmonic coefficients on the set of all spherical shells about the primary is linear in $N$. This must then be done about each primary. 

In more detail, the cost of constructing the harmonic coefficients for all primaries therefore scales as
\begin{equation}
T_{\mathrm{harm}}\propto N\bar{n}V_{\mathrm{max}} \sim N^2,
\end{equation}
where $\bar{n}$ is the number density and $V_{\mathrm{max}}$ is the volume enclosed by $r_{\mathrm{max}}$. After the harmonic coefficients have been constructed, they must be combined and summed over the angular-momentum indices for every primary. The cost of this contraction can be written schematically as
\begin{equation}
T_{\mathrm{contr}}^{(p)}\propto NC_p,
\end{equation}
where $p$ denotes the order of the correlation function and $C_p$ depends on the adopted radial and angular truncations, but not directly on the number of neighbors. The total runtime therefore has the approximate form
\begin{equation}
T_p \sim N\left(\bar{n}V_{\mathrm{max}}+C_p\right).
\end{equation}
For the 3PCF, the contraction is relatively inexpensive and the construction of the harmonic coefficients generally dominates, giving a runtime controlled primarily by $N\bar{n}V_{\mathrm{max}}$. For the 4PCF, however, the products over the three sets of harmonic coefficients and the associated sums over the $m_i$ make $C_4$ considerably larger. For the number densities and angular truncations used in this work, this contraction dominates the runtime, resulting in an approximately linear dependence on $N$ and only a weak dependence on $\bar{n}$. Note, however, that at sufficiently high number density and fixed volume, the harmonic-accumulation term would eventually dominate.

This distinction also affects how the random catalog may be divided into subsets. For the 3PCF, whose runtime is dominated by the $N\bar{n}V_{\mathrm{max}}$ term, the density of each random subset directly affects the computational cost. An optimal density of approximately $1.76$ times the data density was found in \cite{se_3pt_alg}. For the 4PCF, the dominant contraction cost is instead proportional to the total number of primaries processed and is largely independent of how a fixed total random sample is divided. There is therefore no analogous optimal random-subset density: the random catalog may be split in any convenient manner, provided that the desired total random oversampling is obtained.

%the scaling is $N \bar{n} V_{\mathrm{max}}$, where $\bar{n}$ is the number density and $V_{\mathrm{max}}$ is the volume of a sphere out to the maximum radius to which we compute correlations. However, for the 4PCF, the in-practice scaling is as $N$, \textit{i.\,e.}\ linear, as the harmonic expansion around each primary is not what dominates, but rather, forming the products over the $\ell_i$ and $m_i$ and summing over the latter. As this must be done around each primary, the cost is linear in $N$.

%This linear in-practice scaling is very advantageous from a simplicity perspective. For the 3PCF, the harmonic expansion dominates over forming combinations around each primary, and so the scaling in practice is $N \bar{n}V_{\mathrm{max}} \sim N^2$; this means that there is an optimal number density of random particles to use, found to be 1.76 times that of the data \cite{se_3pt_alg}; hence, to achieve the typically desired $50\times$ over-sampling of randoms relative to data, one uses 32 random catalogs (because $32 \times 1.76 \approx 50$). In contrast, for the 4PCF, due to the linearity, there is no optimal $\bar{n}$ for the randoms; any splitting can be used, as long as one achieves the desired $50\times$ over-sampling in the end. 

\newpage
\section{Measured 4PCF: Consistency of Mocks and Data}
\label{sec:msd_4}

We present the measurements of the connected galaxy 4PCF (\textit{i.\,e.}\ with the purely Gaussian contribution subtracted) from the DESI Y1 NGC and SGC samples and compare these with expectations derived from mock catalogs for the LRG sample. The measurements are performed in twelve different areas of the sky, as summarized in \cref{tab:region-summary}; we describe the main characteristics of each area in \cref{Sec:DESI_data_Y1}. To interpret the measurements, we use two complementary sets of mock catalogs: \textsc{EZmocks}, which are approximate realizations designed to efficiently reproduce large-scale clustering statistics \cite{EZMock_Catalogue}, and high-fidelity \textsc{Abacus} simulations, which are based on a full $N$-body evolution and provide accurate nonlinear predictions \cite{Abacus_Mocks}. For the \textsc{Abacus} catalogs, we consider two treatments of DESI fiber assignment: fast fiber assignment (\textsc{FFA}) and alternative merged target ledger (\textsc{altMTL}). The \textsc{altMTL} catalogs are processed using the full DESI fiber-assignment pipeline, reproducing the ordering and cadence of the observed tiles and propagating the resulting changes to the target ledger as the survey progresses. They therefore provide the more detailed realization of the observational selection process. In contrast, \textsc{FFA} is a computationally efficient emulator that assigns targets using observation probabilities inferred from quantities such as the local angular target density and the number of overlapping tiles. 

We show a selection of figures in this section and make the remainder available \href{https://doi.org/10.5281/zenodo.20185242}{online} \cite{zenodo_complete_results}.\footnote{%
    URL: \url{https://doi.org/10.5281/zenodo.20185242} \cite{zenodo_complete_results}%
}
All figures show six different combinations of modes, labeled $(\ell_1, \ell_2, \ell_3)$, with each panel corresponding to a unique combination. DESI measurements, shown as blue dashed lines, are plotted as a function of radial bins on side lengths ranging from \SIrange{20}{160}{\per\hHubble\Mpc} (\cref{fig:NGC_full_Meas,fig:SGC_full_Meas}). These are compared against predictions from \textsc{EZmocks}, represented by solid red lines with shaded pink $1\sigma$ bands, and \textsc{Abacus} mocks, with light blue lines and shaded regions.

The DESI measurements show visually strong consistency with the ensemble mean of the mocks, particularly within the $1\sigma$ bounds, though non-negligible fluctuations are also apparent. These fluctuations persist across bins and suggest the presence of a highly correlated measurement. The fact that the DESI results lie well within the mock variance across most bins adds confidence to the robustness of the measurement. 

To assess the level of agreement between the DESI data and the mock catalogs, we compute a $T^2$ statistic (defined in \cref{eq:T2_statistics_def}) for each $\ell_i$ combination (\cref{fig:NGC_full_comp,fig:SGC_full_comp}). This statistic quantifies the squared deviation of the DESI measurement from the mean of the mocks, weighted by inverse covariance. We have also included the probability-to-exceed (PTE) values (calculated based on the \textsc{EZmocks}), which indicate how likely it is to observe a $T^2$ at least as large as that of DESI, under the assumption that the mocks accurately represent the true underlying signal.

We emphasize that \cref{fig:NGC_full_Meas,fig:SGC_full_Meas} should not be interpreted as a statistical assessment of the level of agreement (or tension) between the DESI measurements and the mocks. The shaded regions show the scatter of the mock realizations in each individual radial bin, while the different bins are generally correlated. This means that the value of the DESI measurements relative to the mocks at the same radial bins cannot be interpreted by treating the bins as independent. Statistical comparisons at the individual multipole level are instead presented in \Cref{fig:NGC_full_comp,fig:SGC_full_comp}, where the correlations between radial bins are accounted for through the covariance entering the $T^2$ statistic. Nevertheless, \Cref{fig:NGC_full_Meas,fig:SGC_full_Meas} provide a useful visual guide to the expected appearance of the 4PCF signal: although we do not include a perturbation-theory prediction here, the mean signals measured from the \textsc{EZmocks} and \textsc{Abacus} catalogs provide an empirical indication of the expected shape and approximate amplitude of the connected 4PCF within these mock models.

It is also important to note, however, that this is not a detection test against a null hypothesis. The \textsc{EZmocks} contain connected 4PCF contributions that arise naturally from nonlinear gravitational evolution and galaxy bias modeling.
The PTE values should thus be interpreted as diagnostics of model consistency. Low PTE values may signal mismatches between data and mocks due to limitations in modeling, systematics, or new physical effects, while high PTE values suggest that the observed 4PCF is well captured by the simulations.

\subsection{North Galactic Cap}

\begin{figure}
    \centering
    \includegraphics[width=\textwidth]{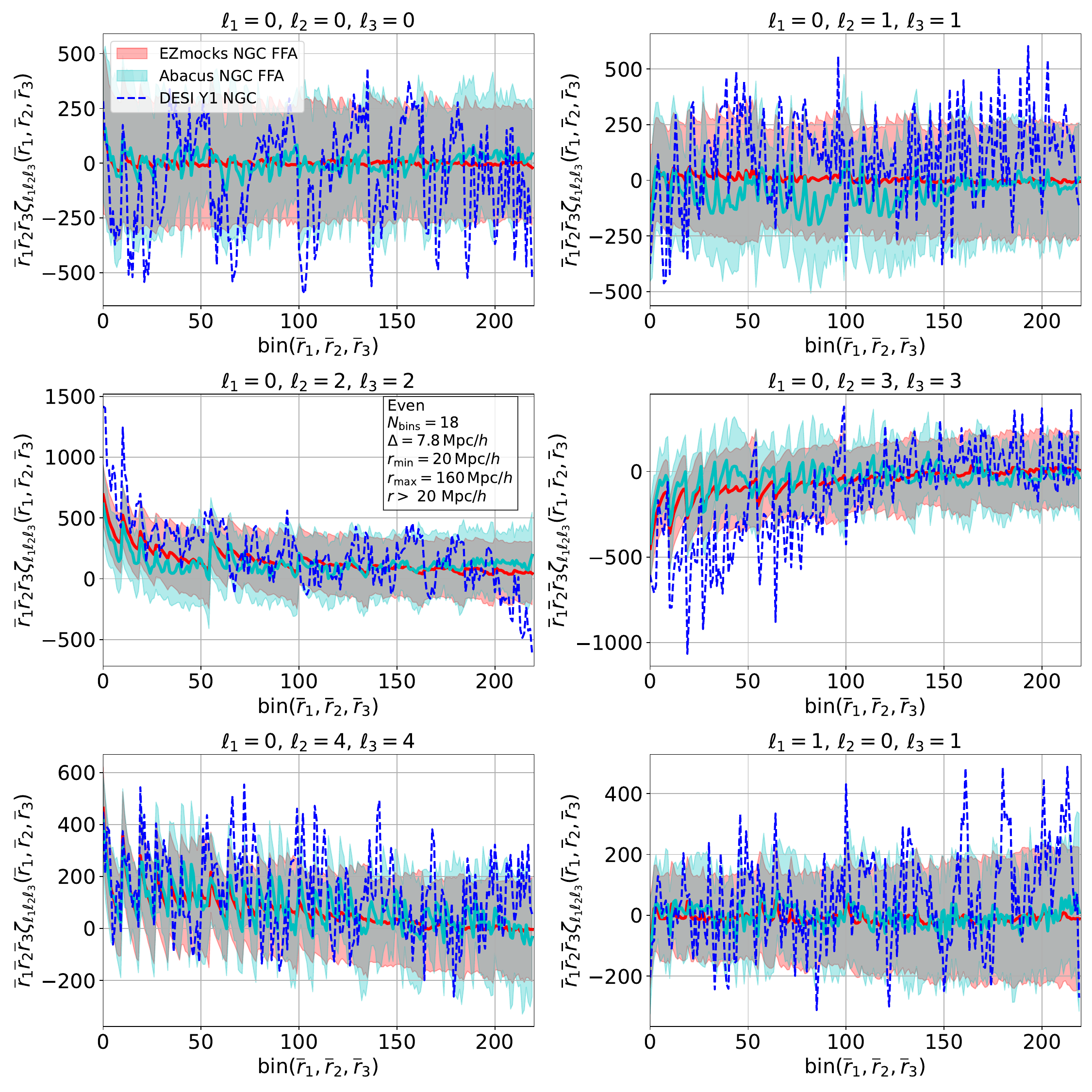}

    \caption{Measurement of the connected galaxy 4PCF isotropic-basis-projected modes from the DESI Y1 NGC (blue dashed line) compared to mock catalogs. The red solid line and pink shaded region represent the mean and $1\sigma$ deviations from the \textsc{EZMocks}, while the light blue line and shaded region show results from \textsc{Abacus} mocks; the overlap between \textsc{EZMocks} and \textsc{Abacus} appears gray. Each panel corresponds to a different $(\ell_1, \ell_2, \ell_3)$ combination as specified in the title. The notation $\bar{r}_i$ denotes the representative separation associated with the $i$-th radial bin, which we take to be the bin center. Thus, while $r_i$ denotes a continuous pair separation, $\bar{r}_i$ labels the interval of separations contained in a single bin as used in the measurement. The horizontal axis, $\mathrm{bin}(\bar{r}_1,\bar{r}_2, \bar{r}_3)$, is therefore a one-dimensional indexing of all allowed sets of tetrahedron side-length combinations $(\bar{r}_1, \bar{r}_2, \bar{r}_3)$ included in the analysis. Specifically, each point along the horizontal axis corresponds to one unique configuration $(\bar{r}_1, \bar{r}_2, \bar{r}_3)$ comprising all sets of separations matching the given radial bins. The inset box indicates the binning configuration used: $N_{\mathrm{bins}} = 18$ radial bins of width $\Delta = \SI{7.8}{\per\hHubble\Mpc}$, spanning $r_{\min} = \SI{20}{\per\hHubble\Mpc}$ to $r_{\max} = \SI{160}{\per\hHubble\Mpc}$, with all triangle sides restricted to $r > \SI{20}{\per\hHubble\Mpc}$. The data generally exhibit consistency with the mocks within the $1\sigma$ regions, arguing for the robustness of the measurement and providing evidence for a statistically significant non-zero connected 4PCF.
    The $\ell_1 = 0, \ell_2 = \ell_3 = 3$ panel (right-hand column, middle row) shows a mild deviation from the mock mean over the first $\sim$80 bin configurations. However, as shown in \cref{fig:NGC_full_comp}, this feature is consistent with the expected statistical fluctuations of the mocks, and no significant discrepancy is observed overall.}
    \label{fig:NGC_full_Meas}
\end{figure}

\begin{figure}
    \centering
    \includegraphics[width=\textwidth]{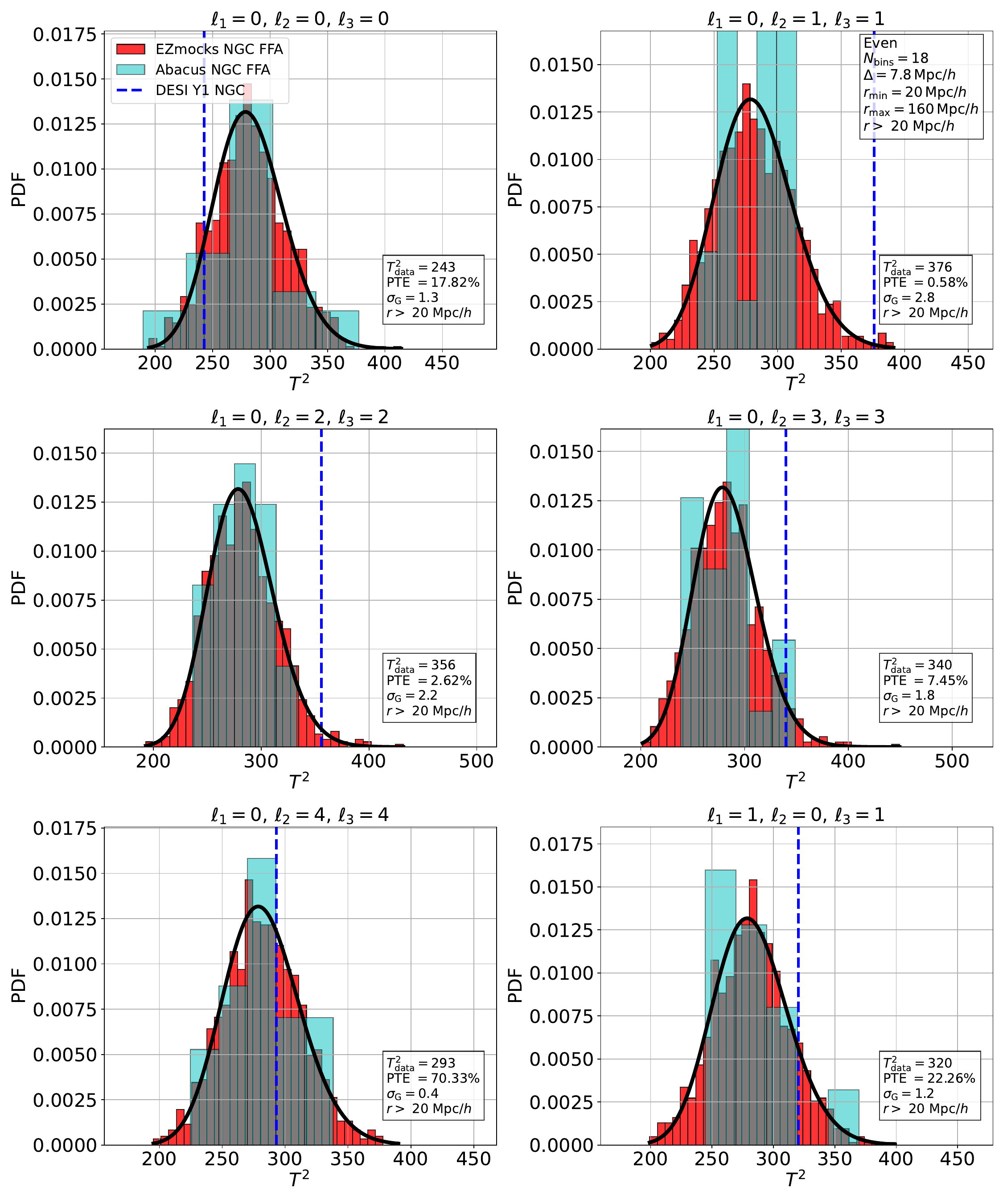}

    \caption{Probability Distribution Functions (PDFs) of the $T^2$ statistics for the 4PCF modes measured from \textsc{EZMocks} (red) and \textsc{Abacus} mocks (light blue) in the DESI Y1 NGC. The black curve shows a $T^2$ distribution. The vertical blue dashed line indicates the value of $T^2$ measured from DESI Y1 data. Each panel corresponds to a different $(\ell_1, \ell_2, \ell_3)$ combination. The Probability-To-Exceed (PTE) and the corresponding number of standard deviations, $\sigma_{\mathrm{G}}$, on a Gaussian are reported in each panel. The binning scheme and scale cuts applied are indicated in the inset at the upper right panel.}
    \label{fig:NGC_full_comp}
\end{figure}

We now present the results of measuring the 4PCF in the full NGC sample, as well as in its individual patches and regions, and assess the consistency between the data and the mock catalogs. The results for the full NGC are shown in \cref{fig:NGC_full_Meas,fig:NGC_full_comp}. Corresponding results for NGC \textit{patches} are presented in \cref{sec:Additional_Mst} in \cref{fig:NGC_P1_meas,fig:NGC_P1_comp}, while those for NGC \textit{regions} are shown in \cref{fig:NGC_R1_meas,fig:NGC_R1_comp}. For brevity, only a representative subset of figures is included in this work.\footnote{%
    The complete set of results is made available online, see \cite{zenodo_complete_results}.%
}

%\subsubsection{Full NGC}\label{sec:NGC_full}

\newpage
\subsection{South Galactic Cap}

\begin{figure}
    \centering
    \includegraphics[width=\textwidth]{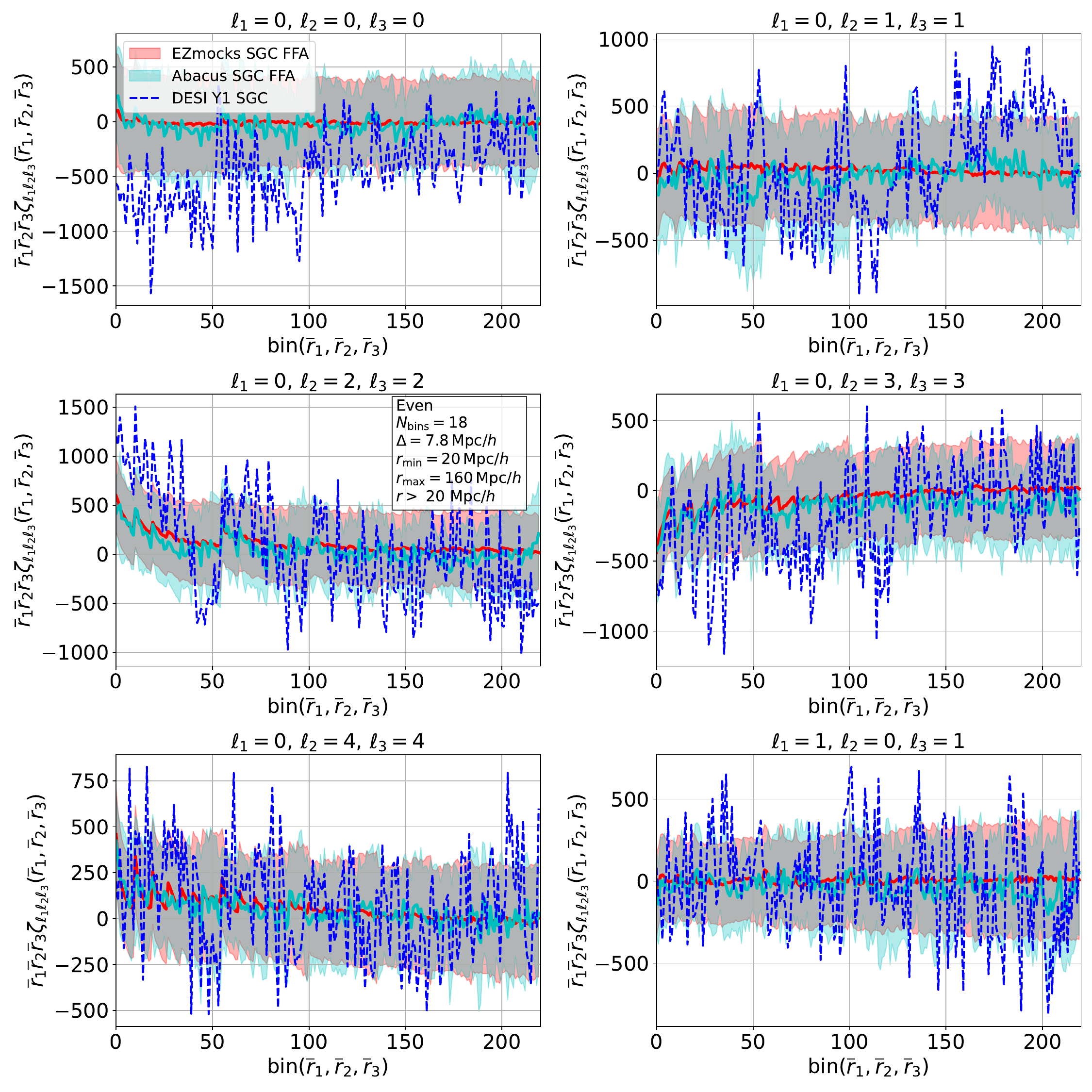}

    \caption{Same as \cref{fig:NGC_full_Meas}, but for the SGC. The $\ell_1=\ell_2=\ell_3=0$ panel (left-hand column, first row) shows a mild deviation from the mock mean over the first $\sim$100 bin configurations; however, as shown in \cref{fig:SGC_full_comp}, this feature is consistent with the expected statistical fluctuations of the mocks, and no significant discrepancy is observed overall.}
    \label{fig:SGC_full_Meas}
\end{figure}

\begin{figure}
    \centering
    \includegraphics[width=\textwidth]{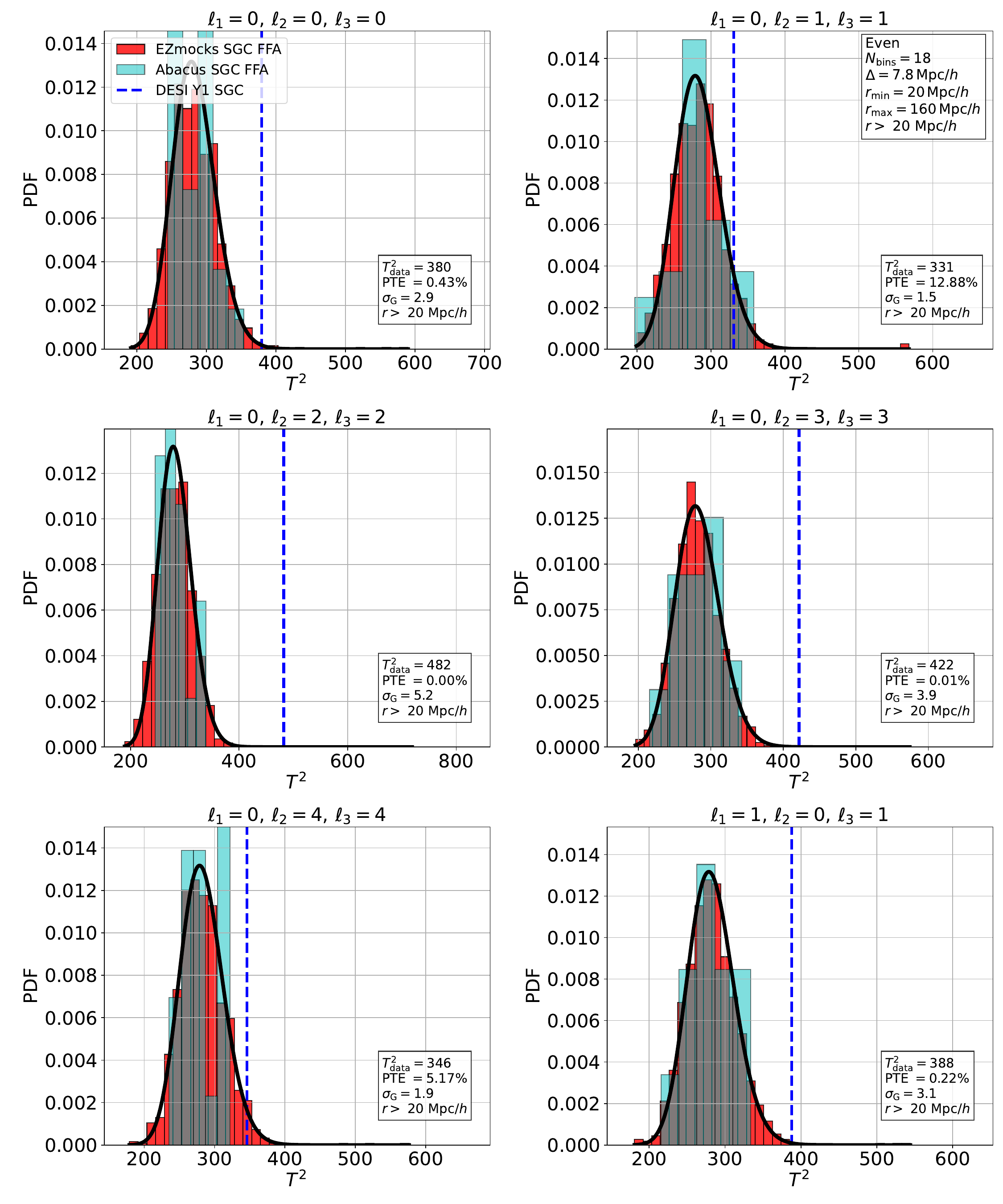}

    \caption{Same as \cref{fig:NGC_full_comp}, but for the SGC. While the $\ell_1=0,\ell_2=2,\ell_3=2$ configuration (left-hand column, middle panel) visually appears consistent with the mocks in \cref{fig:SGC_full_Meas}, the corresponding $T^2$ statistic shows a deviation at $>5\sigma$. This likely reflects the accumulation of small, correlated offsets that are not visually significant in the 4PCF measurement but become statistically significant when combined into an inverse-covariance-weighted global statistic.}
    \label{fig:SGC_full_comp}
\end{figure}

We now present the results of measuring the 4PCF in the full SGC sample, as well as in its individual patches and regions, and assess the consistency between the data and the mock catalogs. The results for the full SGC are shown in \cref{fig:SGC_full_Meas,fig:SGC_full_comp}. We note that the corresponding results for SGC \textit{patches} are presented in \cref{sec:Additional_Mst} in \cref{fig:SGC_P2_meas,fig:SGC_P2_comp}, while those for the SGC \textit{region} are shown in \cref{fig:SGC_R1_meas,fig:SGC_R1_comp}. For brevity, only a representative subset of figures is included in this work.\footnote{%
    The complete set of results is made available online, see \cite{zenodo_complete_results}.
}

%\subsubsection{Full SGC}
%\label{sec:SGC_full}

\clearpage

\section{Results}
\label{sec:Results}

In this section, we present the detection significance of the connected 4PCF measured from the DESI LRG sample using two complementary statistical approaches: An empirical and an analytical covariance. Each analysis is further divided into two parts, an auto correlation and a cross correlation. In the first subsection, \cref{Sec:Empirical_Cov_Results}, we show the results obtained from the $T^2$ analysis:
\begin{equation}
    \label{eq:T2_statistics_def}
    T^2 =\left(\boldsymbol{\zeta} - \boldsymbol{\zeta}_{\mathrm{null}}\right)^{\mathrm{T}} \widehat{\mathbf{C}}^{-1}_{\mathrm{Emp.}} \left(\boldsymbol{\zeta}-\boldsymbol{\zeta}_{\mathrm{null}}\right),
\end{equation}
quantifying how strongly the connected 4PCF departs from the null hypothesis, $\boldsymbol{\zeta}_{\mathrm{null}} = 0$ (which corresponds to a purely Gaussian random field), with an empirical covariance $\widehat{\mathbf{C}}_{\mathrm{Emp.}}$\rule{0pt}{2.5ex} (see \cref{sec:emp_cov} for more details), and $\boldsymbol{\zeta}$ representing the 4PCF data vector, where $\mathrm{T}$ denotes transpose. The analysis is performed with two configurations---the auto-correlation, where the 4PCF vectors stem from only one specific region:
\begin{equation}
    T^2 =\boldsymbol{\zeta}^{\mathrm{T}} \widehat{\mathbf{C}}^{-1}_{\mathrm{Emp.}}\boldsymbol{\zeta},
\end{equation}
and the cross-correlation, where the 4PCF vectors stem from two different regions:
\begin{equation}
    T^2_{\times} = \frac{1}{N_{\times}}\sum_{\mu<\nu}\boldsymbol{\zeta}^{\mathrm{T}}_\mu \, \widehat{\mathbf{C}}^{-1}_{\mathrm{Emp.}} \, \boldsymbol{\zeta}_\nu
    ,
\end{equation}
with $\mu$ and $\nu$ representing the galactic caps, regions or patches described in \cref{tab:region-summary}, and
\begin{equation}
    N_{\times} = \frac{N_{\mathrm{p}}(N_{\mathrm{p}}-1)}{2},
\end{equation}
where $N_{\mathrm{p}}$ is the number of regions or patches. 
In the second subsection, \cref{sec:Analytical_Cov_Results}, we perform a similar study using the $\chi^2$ analysis:
\begin{equation}
    \chi^2
    =
    \boldsymbol{\zeta}^{\mathrm{T}}
    \widehat{\mathbf{C}}^{-1}_{\mathrm{Ana.}}
    \boldsymbol{\zeta}
    ,
    \qquad
    \chi^2_{\times}
    =
    \frac{1}{N_{\times}}
    \sum_{\mu<\nu}
    \boldsymbol{\zeta}^{\mathrm{T}}_\mu
    \widehat{\mathbf{C}}^{-1}_{\mathrm{Ana.}}
    \boldsymbol{\zeta}_\nu
    ,
\end{equation}
with an analytical covariance, $\widehat{\mathbf{C}}_{\mathrm{Ana.}}$ (we used the Gaussian analytical covariance derived in \cite{hou_covar}; see \cref{Sec:Cov-analytic} for more details). Again, we separate the results into auto- and cross-correlation configurations. 

These two methods provide consistent yet distinct ways of quantifying the significance of the 4PCF signal, offering complementary insights into the robustness of the measurement across different analysis choices. In both analyses, the auto-correlation refers to measurements obtained within a single region of the DESI survey, while the cross-correlation refers to measurements that combine spatially-separated regions to test the coherence of the signal across the survey volume. The cross-correlation analysis is performed only with the \textsc{EZmock} catalogs, since the \textsc{Abacus} mocks include replicated volumes; such replication can induce artificial correlations between nominally independent regions, potentially biasing the cross-correlation signal and leading to spurious detections.

The principal result of these analyses is a statistically significant detection of the connected 4PCF in the DESI LRG sample. The significance increases substantially when small tetrahedra are included, indicating that configurations extending to smaller separations contain a large fraction of the detected signal. For example, in the full-sample auto-correlation analysis, the combined NGC$+$SGC measurement reaches $17.1\sigma$ using the empirical-covariance $T^2$ statistic and $8.6\sigma$ using the analytical-covariance $\chi^2$ statistic when small tetrahedra are included. The corresponding cross-correlation measurements also show significant correlations between spatially separated regions, demonstrating that the detected 4PCF structure is not restricted to a single portion of the survey. In contrast, several analyses excluding small tetrahedra yield substantially weaker detections, particularly in the cross-correlation measurements.

 In the $T^2$ analyses, one can observe a discrepancy between the mocks and data in some cases when the small tetrahedra are included; \textit{i.e.}, with sides extending down to $7.8,h^{-1}{\rm Mpc}$. In particular, three cases show the clearest tension: The full NGC, the full SGC for 300 and 500 retained eigenvalues, and the NGC Region 2. There are also other cases, including some in which the small tetrahedra are excluded, where the data lie near the extreme tail of the \textsc{EZmock} distribution, although the discrepancy is not as pronounced as in the three cases highlighted above. Overall, the tension is strongest when the analysis retains the maximum amount of small-scale information. This behavior may indicate that the mocks reproduce the data less accurately as we reach down to the $7.8,h^{-1}{\rm Mpc}$ scales. In this regime, the connected 4PCF might be more sensitive to the Halo Occupation Distribution (HOD) and galaxy biasing, which may not be fully constrained by constructing the mocks to reproduce the 2PCF \cite{Galaxy_Halo_DESI_DR2}. We discuss this further in \Cref{Sec:Discussion}. 

When interpreting these results, it is also important to account for the finite number of available mock realizations. A $\sim3\sigma$ fluctuation corresponds to a tail probability of order $10^{-3}$, so with only $\sim1000$ \textsc{EZmocks} one expects only of order one realization to reach such an extreme value. Consequently, in cases with a $\sim3\sigma$ discrepancy, it is possible for the data to lie to the right of all of the mocks in the corresponding $T^2$ or $\chi^2$ distribution without this implying a substantially larger tension than that associated with the $\sim3\sigma$ result itself.

For the $\chi^2$ analysis, no significant discrepancies between the data and the mocks are found. In some cases the data lie near the tail of the \textsc{Abacus} altMTL mock distribution, but these differences are not as significant as those found in the $T^2$ analysis. The $\chi^2$ results do, however, show a discrepancy between the different mock catalogs: the \textsc{Abacus} FFA and EZmock distributions lie close together, while the \textsc{Abacus} altMTL distribution is shifted toward larger $\chi^2$ values. Since the analytic covariance allows the analysis to use all of the available degrees of freedom, the resulting $\chi^2$ statistic can produce stronger detections and can therefore also be more sensitive to tensions between different types of mocks.

%The DESI measurements also generally show a stronger departure from the null hypothesis than the mock catalogs. Although the mock distributions are themselves displaced from the null because the mocks contain a nonzero connected 4PCF, the DESI statistic frequently lies beyond the bulk of these distributions, particularly for the NGC and for measurements including small tetrahedra. This behavior is present in both the $T^2$ and $\chi^2$ analyses, although its magnitude depends on the covariance treatment and the number of retained eigenmodes. It therefore cannot be attributed exclusively to a single covariance construction.

\subsection{4PCF Detection with Empirical Covariance}
\label{Sec:Empirical_Cov_Results}

We present the $T^2$ analysis, performed by selecting the \numlist{50; 300; 500} eigenmodes (we use the eigenvector and eigenmode interchangeably throughout this paper) with the highest signal-to-noise ratio (SNR) from the analytic covariance matrix (see \cref{sec:emp_cov} for more details). We then construct an empirical covariance by projecting the mock-based covariance onto these eigenmodes, which allows us to estimate the detection significance from the corresponding test statistic. \Cref{tab:T2_summary_Table} compiles the most important detection significances from all configurations; further detection significances are in \cref{tab:T2_det_table} (\cref{sec:Additional_Mst_significances}). 

\begin{figure}
    \centering
    \includegraphics[width=\textwidth]{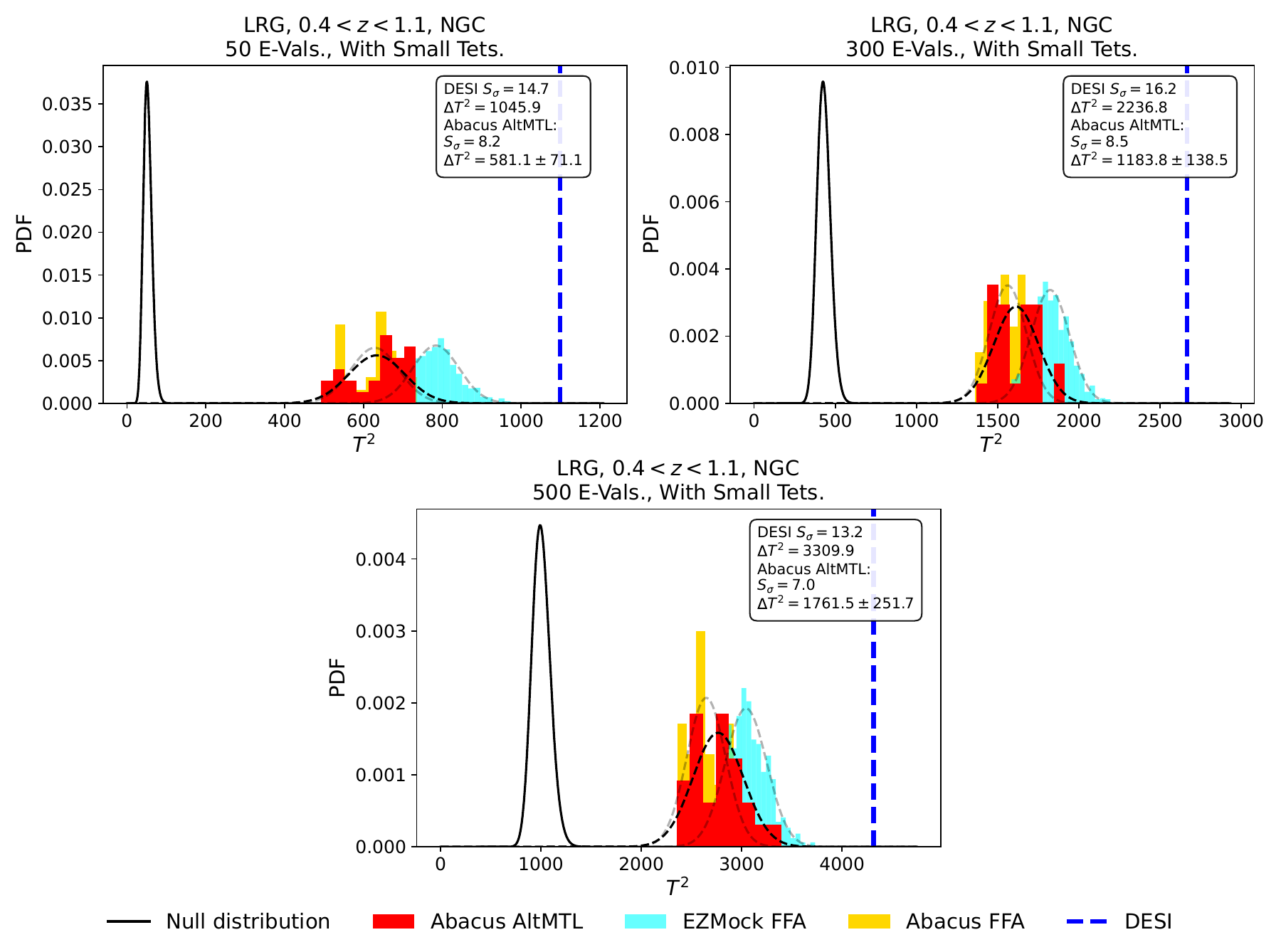}

    \caption{Detection significance of the connected 4PCF in the NGC LRG sample ($0.4 < z < 1.1$). Each panel shows the distribution of the test statistic $T^2$ computed using \numlist{50; 300; 500} eigenvalues (from left to right) of the covariance, as explained in \cref{sec:emp_cov}. The solid black curve shows the theoretical null distribution \cref{eq:Null_dist_def}, while the vertical dashed blue line indicates the measured DESI value and the colored histograms show  values measured from the mock catalogs. The dashed black line is a Gaussian fit to the \textsc{Abacus} altMTL mock distribution. We use \textsc{Abacus} altMTL as the main mock comparison because these mocks include a more realistic treatment of fiber assignment \cite{DESI_Altmtl_Mock_Production}. The main purpose of this figure is to visualize the detection significance of the DESI and mock 4PCF measurements relative to the null distribution. The large separation between the DESI value and the null distribution shows that the measured 4PCF is detected with high significance. We note that, although the DESI and mock 4PCF measurements show broad mode-by-mode consistency in \cref{fig:NGC_full_comp}, the DESI value of the compressed statistic $T^2$ can appear offset from the mock distributions shown here. This comparison should therefore be interpreted with care, since $T^2$ is an inverse-covariance-weighted quadratic statistic and is sensitive to differences along the retained covariance eigenmodes, as well as to the empirical covariance estimate itself. The origin of this offset is not fully determined here, and may be related to the covariance estimation and eigenmode truncation rather than to a simple mode-by-mode disagreement between DESI and the mocks. The inset in each panel reports the corresponding detection significance in $\sigma$, $S_{\sigma}$, for both DESI and the \textsc{Abacus} altMTL mock distribution. It also gives the difference in $T^2$, $\Delta T^2$, between the null and both DESI and the \textsc{Abacus} altMTL mock distribution. } %as a function of the number of evaluated eigenvalues.}
    \label{fig:ST_NGC_Full_T2_Auto}
\end{figure}

\begin{figure}
    \centering
    \includegraphics[width=\textwidth]{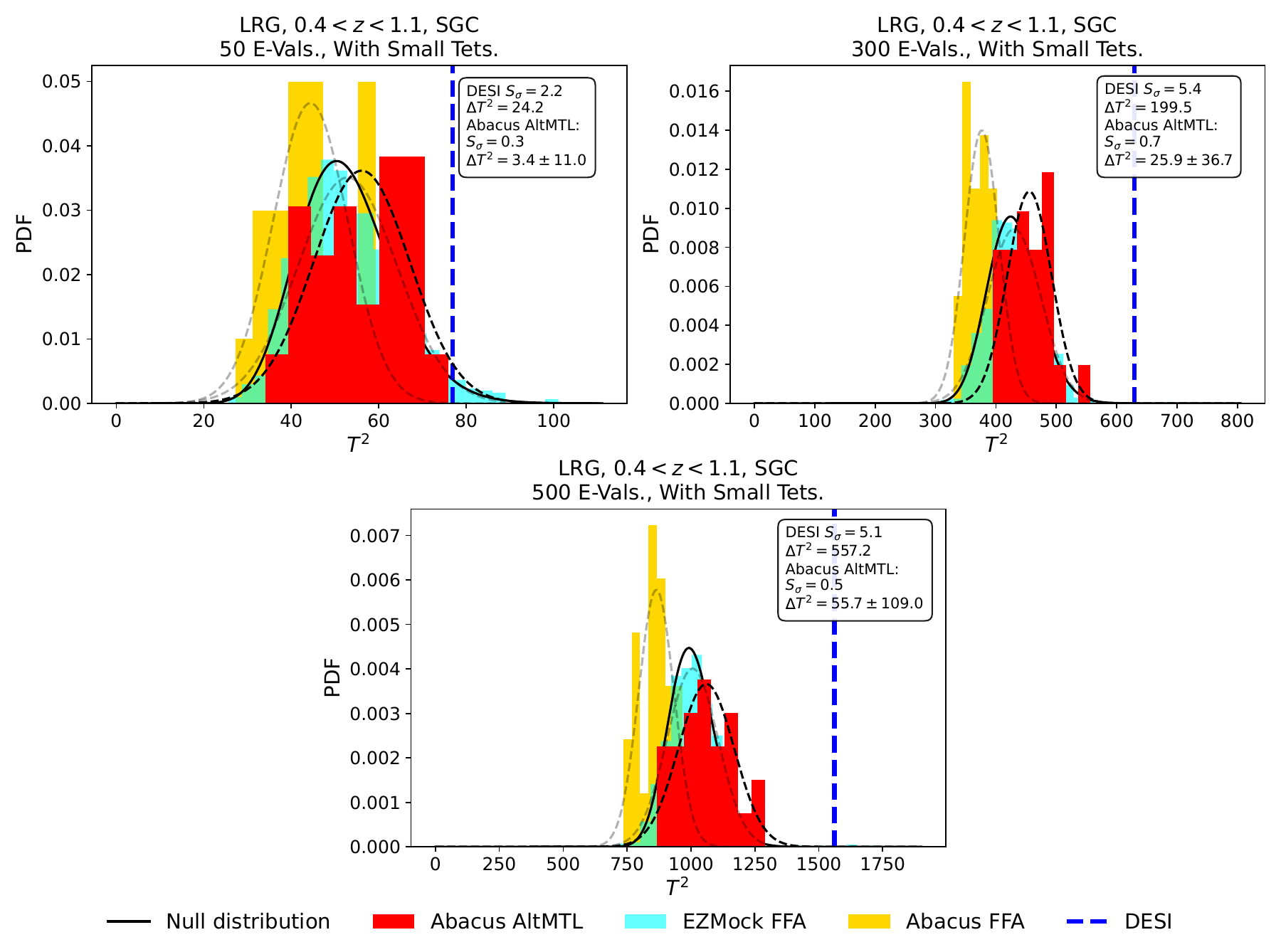}

%    \vspace*{-2ex}
    \caption{Detection significance of the 4PCF in the SGC LRG sample ($0.4 < z < 1.1$).
    Same as \cref{fig:ST_NGC_Full_T2_Auto}, but for the SGC.}
    \label{fig:ST_SGC_Full_T2_Auto}
\end{figure}

Since the analyses of this section use an empirical covariance constructed from a finite number of mock realizations, the null distribution of $T^2$ is not taken to be a simple $\chi^2$ distribution---as we use in \cref{sec:Analytical_Cov_Results}. Instead, the black curve shown in \crefrange{fig:ST_NGC_Full_T2_Auto}{fig:ST_NGC_and_SGC_Cross_Regions_T2_Auto} corresponds to the finite-sample-corrected null distribution of $T^2$. This is the probability distribution expected for the statistic shown in \cref{eq:T2_statistics_def} under the null hypothesis, after accounting for the fact that $\widehat{\mathbf{C}}_{\mathrm{Emp.}}$ is estimated from a finite number of mocks. Therefore, the black curve represents the distribution of $T^2$ values that would be expected if there were no true connected 4PCF signal, but the covariance were estimated in the same way as in the data analysis.

\begin{table}
    \centering
    {\textbf{\boldmath Detection Significance of the $T^2$ Analyses}}

    \smallskip

    % ---- COLUMN GROUP HEADERS ----
    \begin{tabular}{m{7em} @{} c @{\;} *{8}{S[table-format=2.1]}}
        \toprule
          \textbf{Sector} & \textbf{Analysis}
        & \multicolumn{2}{c}{\textbf{NGC}} 
        & \multicolumn{2}{c}{\textbf{SGC}} 
        & \multicolumn{2}{c}{\textbf{NGC $\times$ SGC}} &
         \multicolumn{2}{c}{\textbf{NGC $+$ SGC}}
        \\
        %& & & & & & \multicolumn{2}{c}{$N(N-1)/2$} & \\[-4pt]
        &  & {no-ST} & {ST} & {no-ST} & {ST} & {no-ST} & {ST}  & {no-ST} & {ST}  \\
        \midrule
        % ---- ROWS ----
        Full\newline($0.4 < z < 1.1$) & Auto & 3.9 & 16.2 & 3.9 & 5.4 & {---} & {---} & 5.5 & 17.1 \\
        Full\newline($0.4 < z < 1.1$) & Cross & {---} & {---} & {---} & {---}  & 3.4 & 8.2 & {---} & {---} \\
        Patches\newline($0.4 < z < 1.1$) & Cross & 1.0 & 10.7 & 0.4 & 7.2  & 1.3 & 14.8 & {---} & {---} \\
        Regions\newline($0.4 < z < 0.8$) & Cross &  0.4 & 2.2  & {---} & {---} & 2.2 & 10.2 & {---} & {---} \\
        \bottomrule
    \end{tabular}

    \caption{Detection significance of the DESI dataset using the $T^2$ analysis for the different sectors of the Galactic Caps. Columns labeled ``ST'' include small tetrahedra, while those labeled ``no-ST'' exclude them. Here, ``small tetrahedra'' refers to configurations with minimum side lengths extending down to the smallest radial bin of \SI{7.8}{\per\hHubble\Mpc}, whereas ``no-ST'' restricts all side lengths to be at least \SI{20}{\per\hHubble\Mpc}. We list the detection-significance values corresponding to an empirical covariance (\cref{sec:emp_cov}) constructed with 300 eigenvalues. The column labeled NGC~$\times$~SGC corresponds to the detection significance obtained from the cross-correlation between the NGC and SGC measurements for the sector specified in each row. In other words, it quantifies the significance of the correlation between the 4PCF data vectors measured in the two Galactic Caps. For the patch and region analyses, this cross-correlation is performed using the all the patches or regions from the NGC and SGC. The column labeled NGC~$+$~SGC corresponds to the detection significance obtained from the auto-correlation analysis after combining the NGC and SGC measurements. Across all sectors, including small tetrahedra increases the detection significance relative to the corresponding no-ST measurement. This is expected, since the ST measurements include smaller-scale configurations, where more nonlinear information is present.} %The increase is especially pronounced for the patch and NGC~$\times$~SGC cross-correlation analyses, indicating that the small-scale configurations contribute significantly to the measured detection significance.}%Therefore, NGC~$\times$~SGC gives the detection significance of the signal between the two galactic caps, while NGC~+~SGC gives the significance of the combined NGC~+~SGC.}
    \label{tab:T2_summary_Table}
\end{table}

Following \cite{phil_4pcf, sellentin}, the finite-sample corrected null distribution is:
\begin{equation}
    \label{eq:Null_dist_def}
    P(T^2) = \frac{N-p+1}{N p} \; f_{T}\mleft(\frac{N-p+1}{N p}T^2;\,n, p\mright),
\end{equation}
with $N=n-1$. Here $f_{T}$ is the probability density function defined in Eq.~(46) of \cite{phil_4pcf} (a $t$-distribution), $n$ is the number of mock realizations used to estimate the covariance, and $p$ is the number of retained data-vector components/eigenmodes. In this analysis we use $n=1000$, while $p$ is set by the number of retained eigenvalues after the covariance regularization described in \cref{sec:emp_cov}. The black curve in the figures is precisely the distribution in \cref{eq:Null_dist_def}.

\Cref{tab:T2_summary_Table} summarizes the detection significance obtained from the $T^2$ analysis using 300 eigenvalues. A clear trend is that the inclusion of small tetrahedra increases the detection significance in every sector considered. This behavior is consistent with expectations: By extending the analysis to smaller radial bins, the ST measurements include configurations that are more sensitive to nonlinear clustering and therefore contain an additional 4PCF signal. The effect is modest in some full-cap measurements, such as SGC, but becomes much more pronounced in the patch and cross-correlation analyses. In particular, the NGC$\times$SGC patch cross-correlation increases from $1.3\sigma$ in the no-ST case to $14.8\sigma$ when small tetrahedra are included, showing that small-scale configurations contribute significantly to the 4PCF signal.

\subsubsection{Detection in Auto-Correlation}

In this subsection, we present the detection significance for the auto-correlation analysis of the connected 4PCF in the DESI LRG sample using the $T^2$ statistic. We remind the reader that the auto-correlation refers to measurements obtained within a single region of the DESI survey. \Cref{fig:ST_NGC_Full_T2_Auto,fig:ST_SGC_Full_T2_Auto} show the distributions of $T^2$ for the mock catalogs alongside the value measured from the DESI data. For brevity, only a representative subset of figures is included in this work.\footnote{%
The complete set of results is made available online; see \cite{zenodo_complete_results}.}

\subsubsection{Detection in Cross-Correlation}

\begin{figure}
    \centering
    \includegraphics[width=0.5\textwidth]{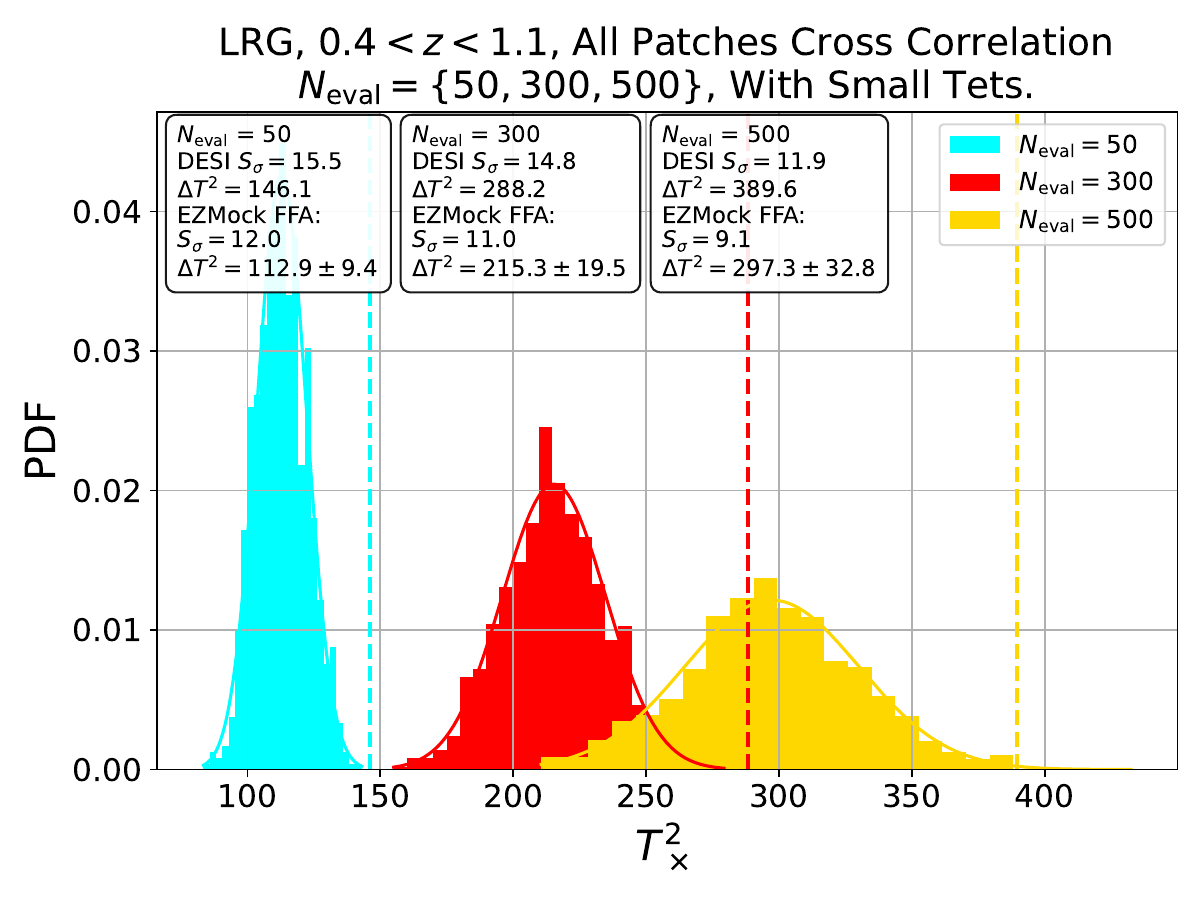}%
    \includegraphics[width=0.5\textwidth]{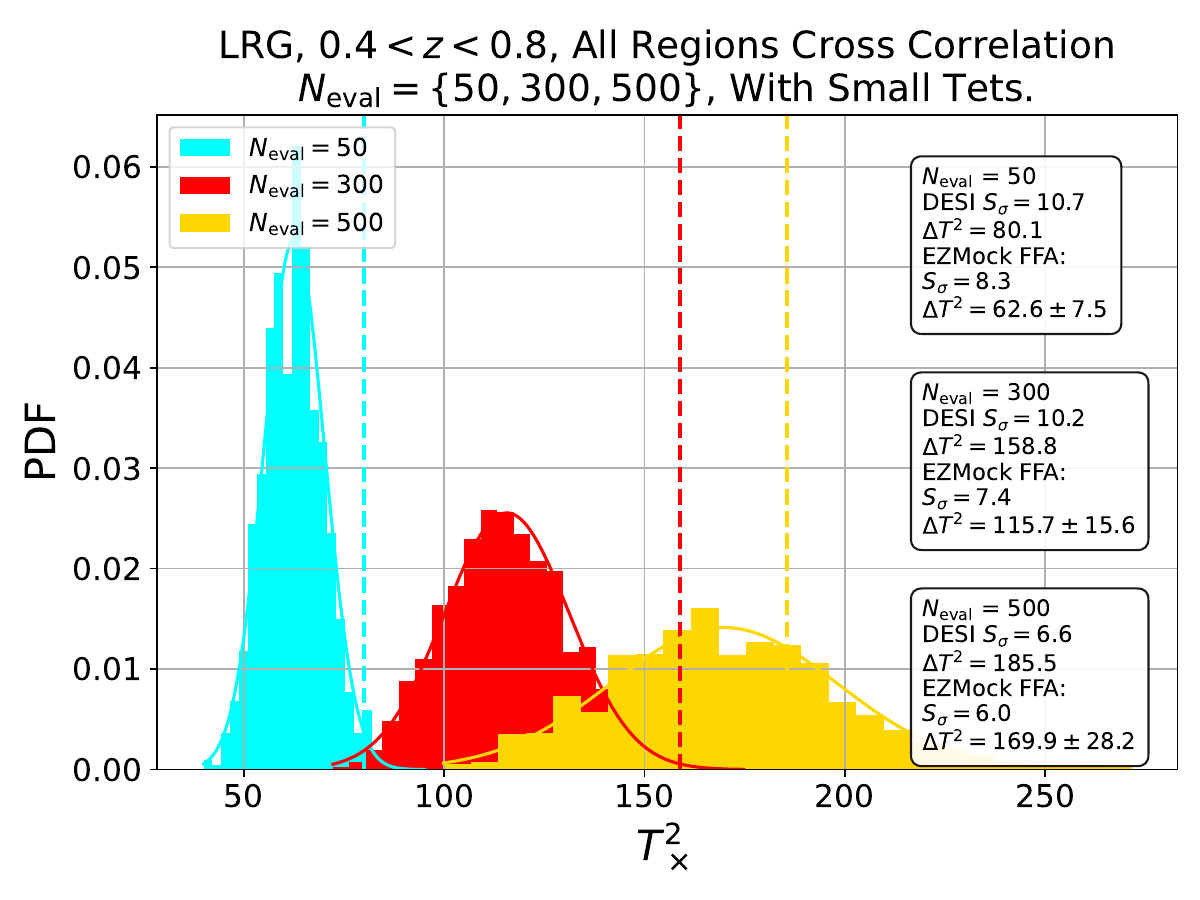}

    \caption{Detection significance of the 4PCF cross-correlation between the NGC and SGC patches (left) and regions (right) of the LRG sample. We show the probability density of the test statistic $T_{\times}^2$ computed using $N_{\mathrm{eval}} = \numlist{50; 300; 500}$ eigenvalues of the covariance matrix, including all tetrahedra. Colored histograms represent results from the \textsc{EZMocks} FFA simulations for each $N_{\mathrm{eval}}$, as indicated in the legend. Solid colored lines show Gaussian fits to the mock distributions, while the vertical dashed lines mark the corresponding DESI measurements. The inset boxes report the detection significances in units of $\sigma$, denoted $S_\sigma$, and the corresponding $\Delta T^2$, defined relative to zero---since the expectation for the null distribution is to be centered around zero.}
    \vspace*{2ex}
    \label{fig:ST_and_NoST_Cross_Patches_T2_Auto}
\end{figure}

\begin{figure}
    \centering
    \includegraphics[width=0.5\textwidth, height=6cm]{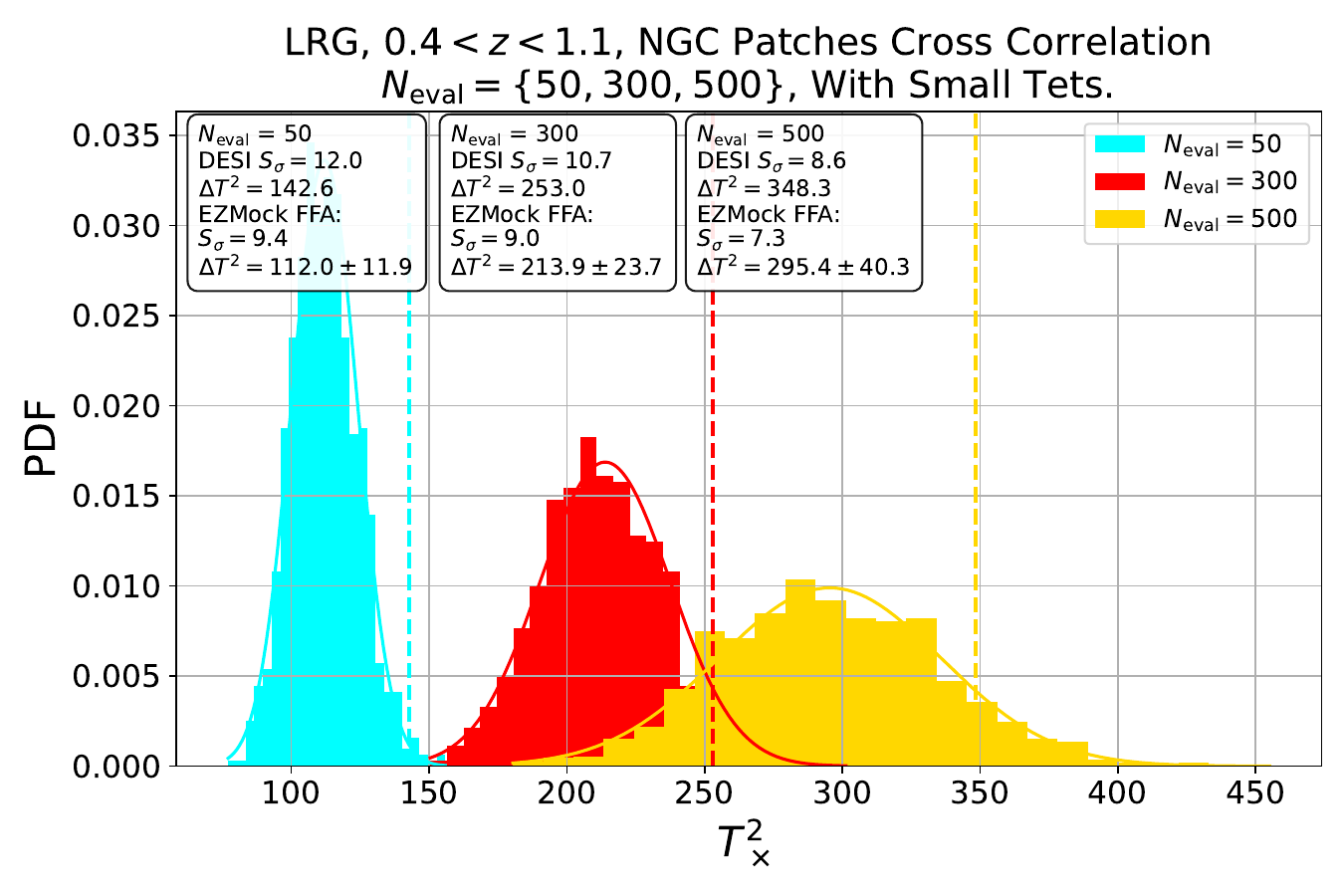}%
    \includegraphics[width=0.5\textwidth, height=6cm]{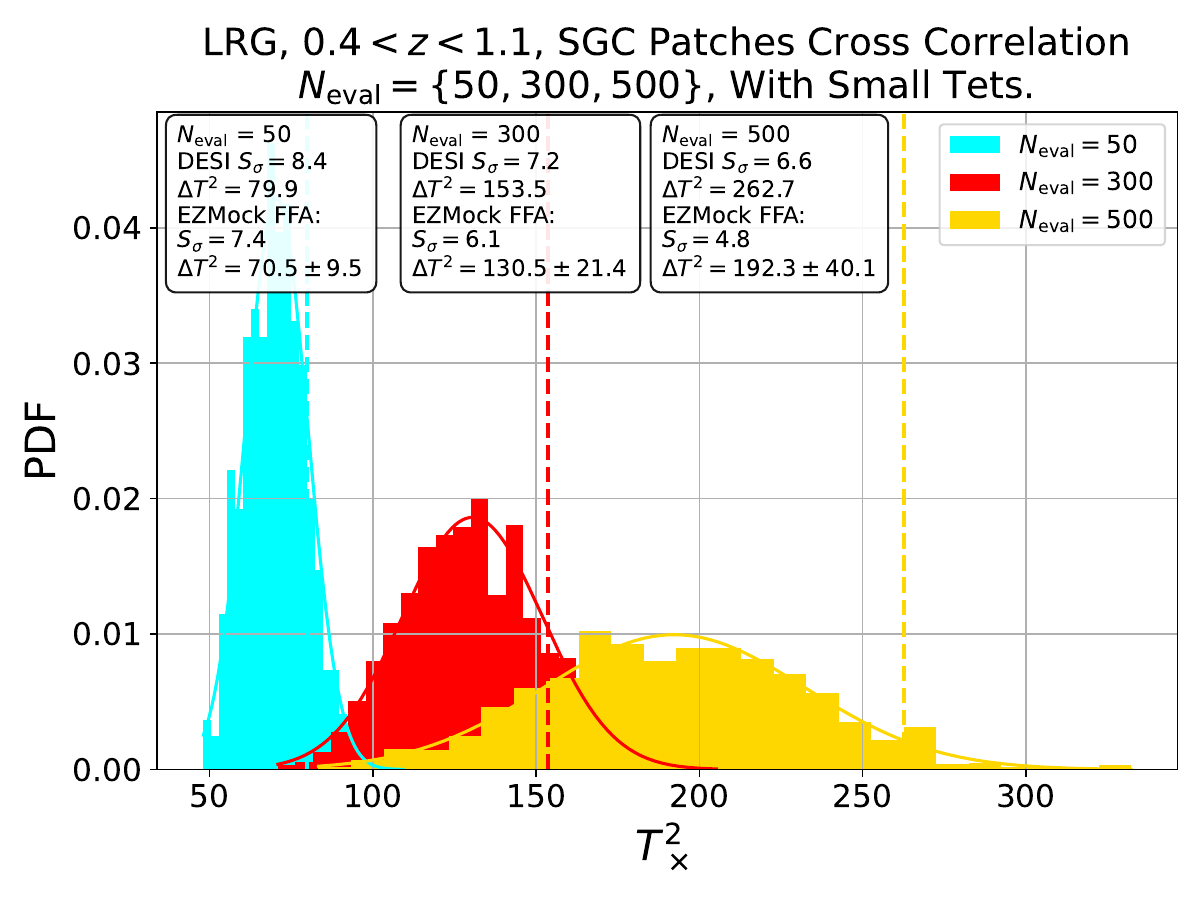}

    \caption{Same as \cref{fig:ST_and_NoST_Cross_Patches_T2_Auto}, but for the cross-correlation of each of the NGC and SGC patches ($0.4 < z < 1.1$) individually.}
    \vspace*{2ex}
    \label{fig:ST_NGC_and_SGC_Cross_Regions_T2_Auto}
\end{figure}

We now examine the detection significance of the connected 4PCF in the DESI LRG cross-correlation analysis using the $T_\times^2$ statistic. The $T^2$ test quantifies how strongly the measured 4PCF departs from the null distribution. \Cref{fig:ST_and_NoST_Cross_Patches_T2_Auto,fig:ST_NGC_and_SGC_Cross_Regions_T2_Auto} show the distributions of $T_\times^2$ for the \textsc{EZMocks} Fast Fiber Assignment (FFA) catalog alongside the value measured from the DESI data, allowing a visual assessment of the statistical significance and consistency of the detection. For brevity, only a representative subset of figures is included in this work.\footnote{%
    See \cite{zenodo_complete_results}.%
}

\subsection{4PCF Detection with Analytic Covariance}
\label{sec:Analytical_Cov_Results}

\begin{table}
    \centering
    {\textbf{\boldmath Detection Significance of the $\chi^2$ Analyses}}

    \smallskip

    % ---- COLUMN GROUP HEADERS ----
    \begin{tabular}{m{7em} @{} c @{\;} *{8}{S[table-format=2.1]}}
        \toprule
          \textbf{Sector} & \textbf{Analysis}
        & \multicolumn{2}{c}{\textbf{NGC}} 
        & \multicolumn{2}{c}{\textbf{SGC}} 
        & \multicolumn{2}{c}{\textbf{NGC $\times$ SGC}}
        & \multicolumn{2}{c}{\textbf{NGC $+$ SGC}}
        \\
        %& & & & & & \multicolumn{2}{c}{$N(N-1)/2$} \\
        & & {no-ST} & {ST} & {no-ST} & {ST} & {no-ST} & {ST} & {no-ST} & {ST} \\
        \midrule
        % ---- ROWS ----
        Full\newline($0.4 < z < 1.1$) & Auto & 0.2 & 6.9 & 3.2 & 5.2 & {---} & {---} & 3.2 & 8.6 \\
        Full\newline($0.4 < z < 1.1$) & Cross & {---} & {---} & {---} & {---}  & 0.7 & 16.7 & {---} & {---} \\
        Patches\newline($0.4 < z < 1.1$) & Cross & -1.4 & 9.8 & -0.9 & 5.0 & -1.4 & 12.3 & {---} & {---} \\
        Regions\newline($0.4 < z < 0.8$) & Cross &  -2.5 & 4.8 & {---} & {---} & -1.1 & 9.0 & {---} & {---} \\
        \bottomrule
    \end{tabular}
    \caption{Same as \cref{tab:T2_summary_Table}, but for the detection significance of the DESI dataset using the $\chi^2$ analysis for the different sectors of the Galactic Caps. Columns labeled ``ST'' include small tetrahedra, while those labeled ``no-ST'' exclude them. As in the $T^2$ analysis, \cref{tab:T2_summary_Table}, the inclusion of small tetrahedra systematically increases the detection significance across all sectors. Several no-ST measurements are consistent with low-significance detections, or even negative detection significances under our convention, whereas the corresponding ST measurements show clear detections. This indicates that the small-scale configurations included in the ST analysis carry a substantial fraction of the $\chi^2$ signal.}
    \label{tab:Chi2_summary_Table}
\end{table}

\begin{figure}
    \centering
    \includegraphics[width=0.5\textwidth]{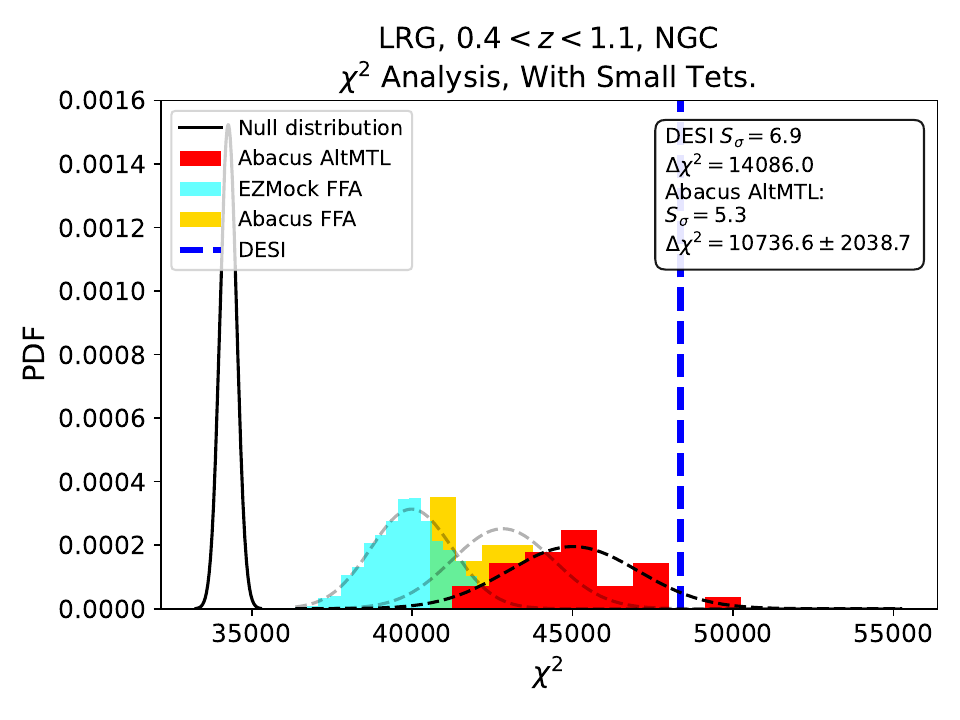}%
    \includegraphics[width=0.5\textwidth]{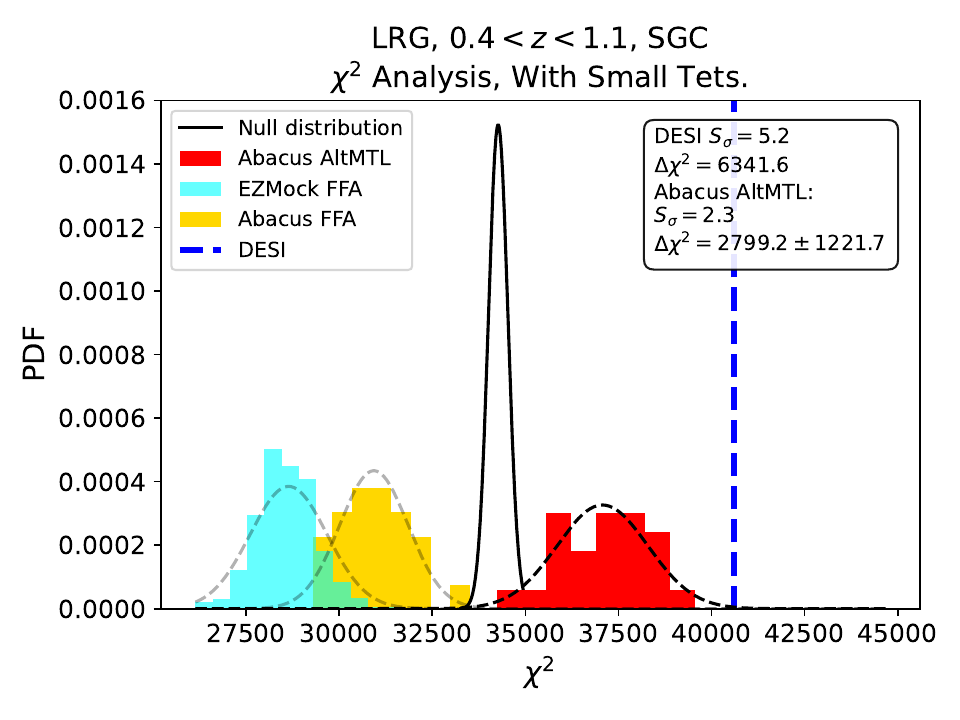}

    \caption{Detection significance of the 4PCF in the NGC (left image) and SGC (right image) for the LRG sample ($0.4 < z < 1.1$) using the $\chi^2$ test statistic, as explained in \cref{Sec:Cov-analytic}. The inset in each panel reports the corresponding detection significance $S_{\sigma}$ and the $\Delta \chi^2$ between DESI and the null distribution.}
    \label{fig:ST_NGC_Full_Chi2_Auto}
\end{figure}

We present the $\chi^2$ analysis, which directly inverts the analytic covariance to compute the detection significance. More details regarding the construction and properties of the covariance matrices can be found in \cref{Sec:Cov}. \Cref{tab:Chi2_summary_Table} shows the corresponding detection significances obtained from the $\chi^2$ analysis using the analytical covariance. The same qualitative trend observed in the $T^2$ analysis is present here: including small tetrahedra increases the detection significance in every sector. However, the contrast between the no-ST and ST measurements is even more striking for the $\chi^2$ statistic. In the no-ST case, several cross-correlation measurements have low or negative detection significances, indicating that these large-scale-only configurations do not produce a significant excess relative to the null hypothesis. Once small tetrahedra are included, the same sectors show strong detections, with the NGC$\times$SGC full cross-correlation increasing from $0.7\sigma$ to $16.7\sigma$, the patch-level NGC$\times$SGC cross-correlation increasing from $-1.4\sigma$ to $12.3\sigma$, and the region-level NGC$\times$SGC cross-correlation increasing from $-1.1\sigma$ to $9.0\sigma$. Further detection significances are in \cref{tab:chi2_det_table} (\cref{sec:Additional_Mst_significances}).

Across these $\chi^2$ analyses, we find that the distributions of the detection statistic obtained from the different mock ensembles are more widely separated than in the eigenmode-truncated $T^2$ analysis, with this behavior being particularly pronounced in the SGC; see \Cref{fig:ST_NGC_Full_Chi2_Auto} for an example. Specifically, the \textsc{Abacus} altMTL mocks produce a distribution at larger $\chi^2$ values, and closer to the DESI measurement, than the \textsc{Abacus} FFA and EZmock FFA distributions. One possible explanation for the stronger separation in the full-$\chi^2$ analysis is that it uses the complete 4PCF data vector, whereas the $T^2$ analysis retains only a limited number of covariance eigenmodes. Differences between the mock constructions that are carried by modes excluded from the truncated analysis may therefore contribute to the full $\chi^2$ and produce a larger separation between the distributions. The present analysis does not, however, determine whether this behavior is caused specifically by the treatment of fiber assignment, by other differences in the mock construction, or by the covariance model itself.

The behavior of \cref{tab:Chi2_summary_Table} suggests that the $\chi^2$ analysis is particularly sensitive to the smaller-scale 4PCF configurations, where nonlinear clustering enhances the signal. At the same time, it highlights the importance of accurately modeling the covariance in this regime, since small-scale configurations are also expected to receive larger non-Gaussian covariance contributions. A natural direction for future work is therefore to incorporate next-to-leading-order contributions to the 4PCF covariance, such as those theoretically modeled in \cite{ortola_cov_I, ortola_cov_II}.

\begin{figure}
    \centering
    \includegraphics[width=0.5\textwidth]{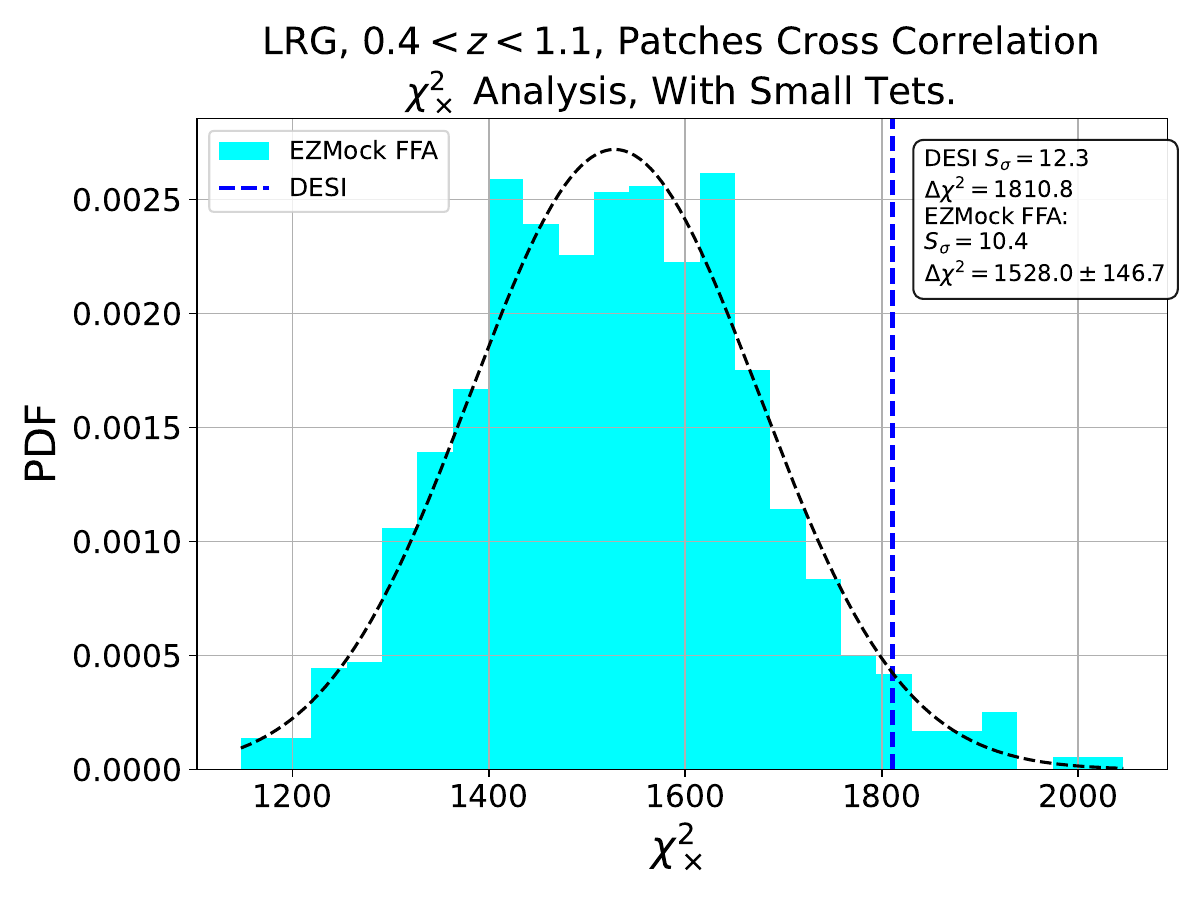}%
    \includegraphics[width=0.5\textwidth]{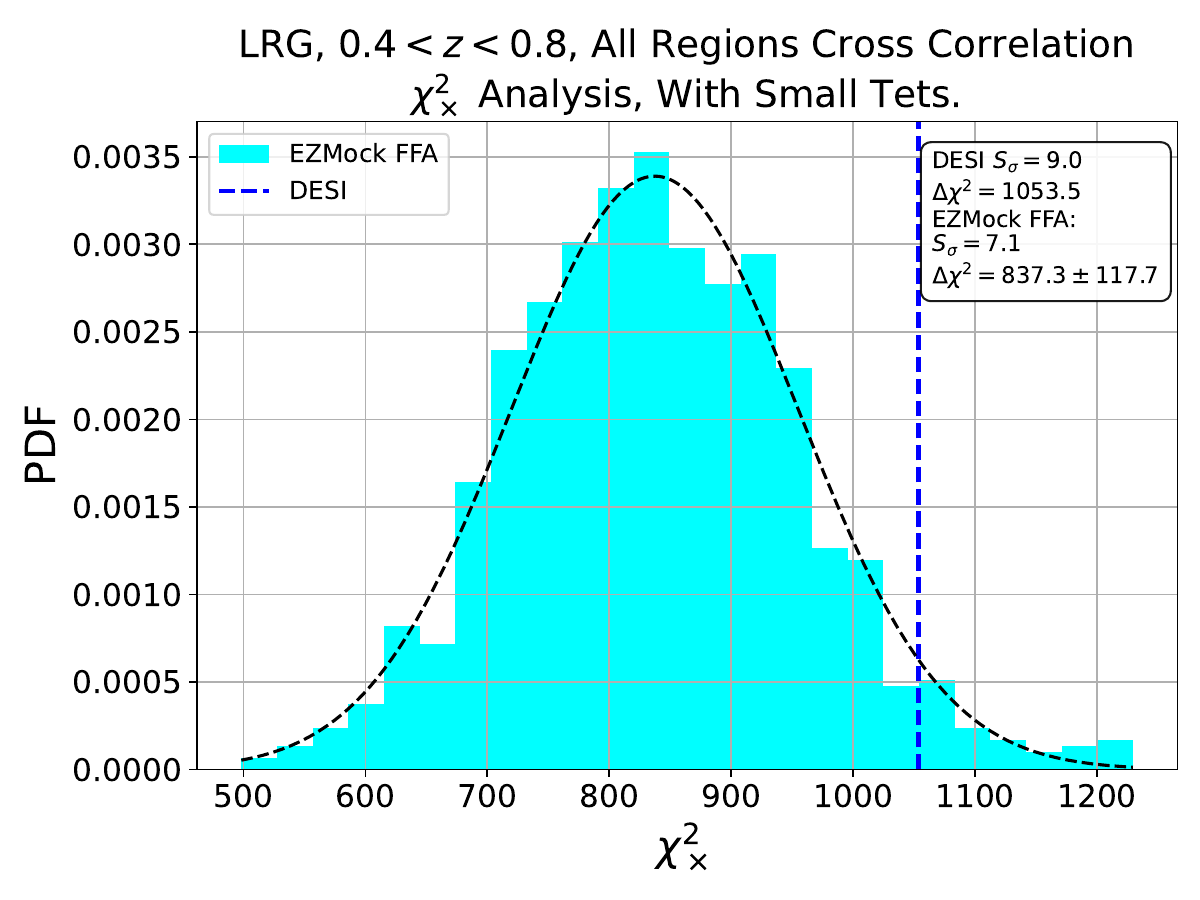}

    \caption{Detection significance of the 4PCF cross-correlation between all the NGC and SGC patches (left image) and regions (right image) of the LRG sample. We show the probability density of the cross-correlation test statistic $\chi_\times^2$ including all tetrahedra. The colored histogram represents results from the \textsc{EZMocks} FFA simulations, as indicated in the legend. Vertical dashed line marks the corresponding DESI measurement, while the inset reports the detection significance $S_{\sigma}$ and the $\Delta \chi^2$ relative to the null distributions.}
    \label{fig:ST_and_NoST_Cross_Patches_Chi2_Auto}

    \bigskip

    \centering
    \includegraphics[width=0.5\textwidth]{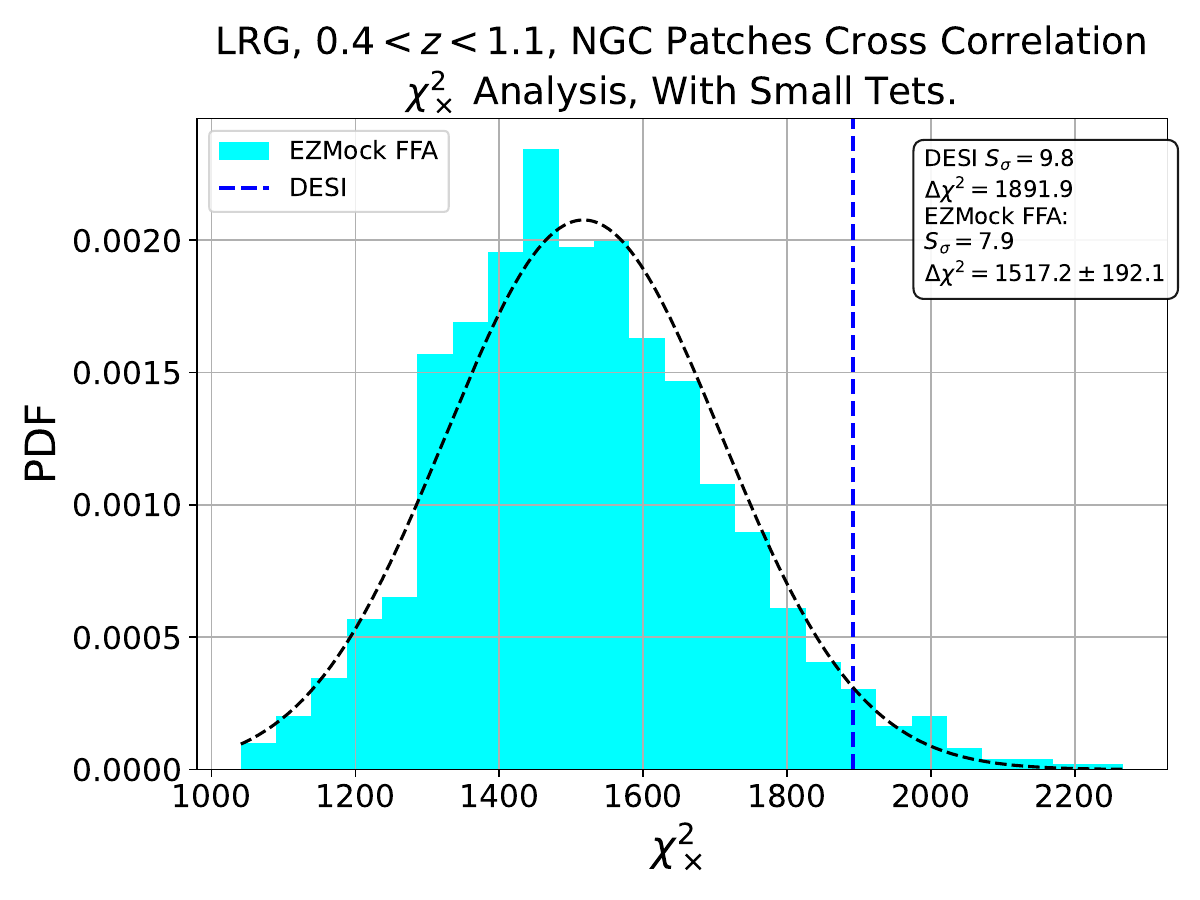}%
    \includegraphics[width=0.5\textwidth]{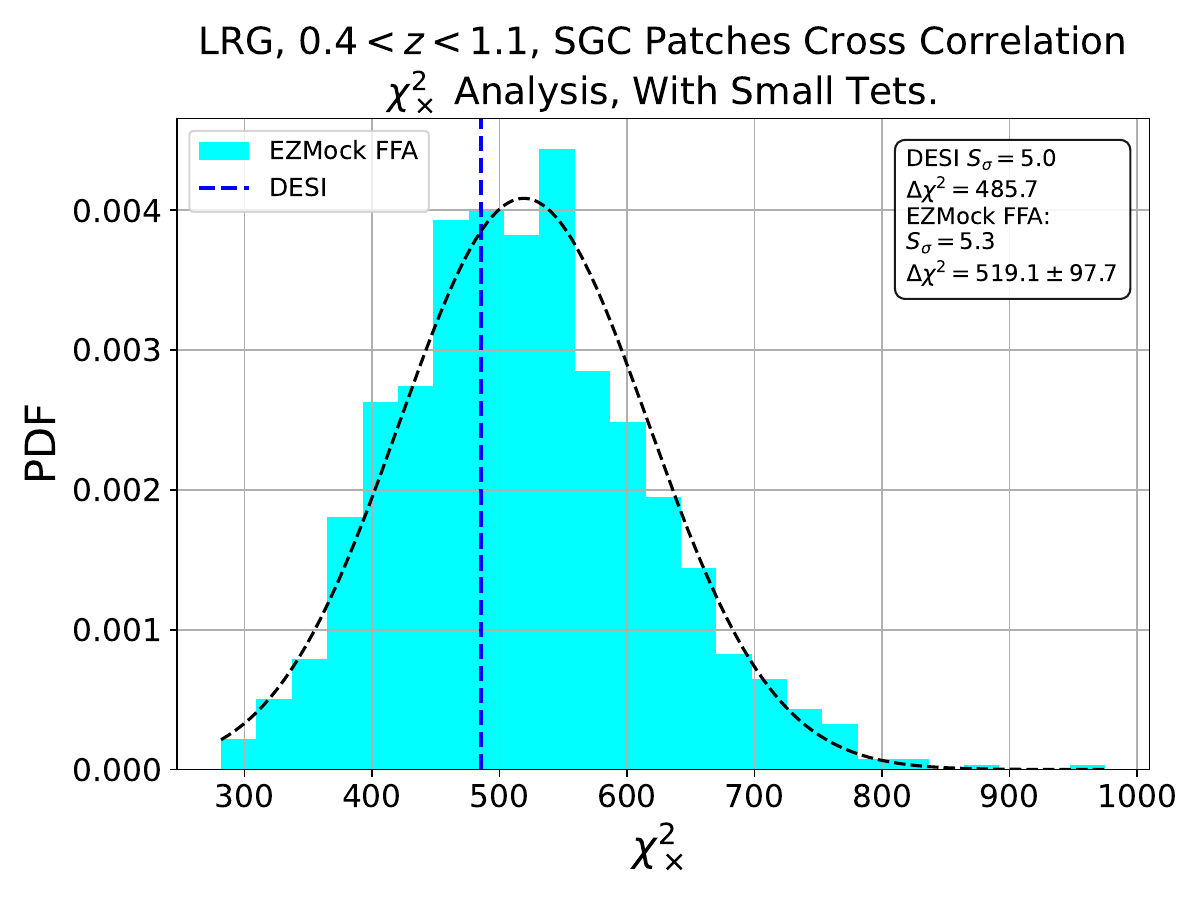}

    \caption{Same as \cref{fig:ST_and_NoST_Cross_Patches_Chi2_Auto}, but for the cross-correlation of each of the NGC and SGC patches ($0.4<z<1.1$) individually.}
    \label{fig:ST_NGC_and_SGC_Cross_Patches_Chi2}
\end{figure}

\subsubsection{Detection in Auto-Correlation}

In this subsection, we present the detection significance of the connected 4PCF in the DESI LRG auto-correlation analysis using the $\chi^2$ statistic. The $\chi^2$ test quantifies how strongly the measured 4PCF departs from the null distribution---see \cref{eq:Null_dist_def}. \Cref{fig:ST_NGC_Full_Chi2_Auto} shows the distributions of $\chi^2$ for the mock catalogs alongside the value measured from the DESI data. For brevity, only a representative subset of figures is included in this work.\footnote{%
    The complete set of results is made available online, see \cite{zenodo_complete_results}.%
}

\subsubsection{Detection in Cross-Correlation}

We now examine the detection significance of the 4PCF in the DESI LRG cross-correlation analysis using the $\chi_\times^2$ statistic. The $\chi^2$ test quantifies how strongly the measured 4PCF departs from the null distribution. \Cref{fig:ST_and_NoST_Cross_Patches_Chi2_Auto,fig:ST_NGC_and_SGC_Cross_Patches_Chi2} show the distributions of $\chi_\times^2$ for the \textsc{EZMock} FFA catalogs alongside the value measured from the DESI data. For brevity, only a representative subset of figures is included in this work.\footnote{%
    See \cite{zenodo_complete_results}.%
}

\section{Discussion \& Conclusions}
\label{Sec:Discussion}

In this work we have presented one of the first measurements of the connected 4PCF of DESI Year~1 LRGs, exploiting the isotropic basis of \cite{cahn_iso, iso_gen} and a GPU-accelerated implementation of the \textsc{encore} algorithm called \textsc{cadenza}. Our analysis uses the DESI Y1 LRG sample over $0.4 < z < 1.1$, split into North and South Galactic Caps and further subdivided into redshift-limited regions and equal-area patches (\cref{Sec:DESI_data_Y1}, \cref{tab:region-summary}). By working in the isotropic basis, we obtain a compact, rotationally-invariant description of tetrahedral clustering and cleanly separate the angular and radial dependence of the 4PCF. The use of \textsc{cadenza} allows us to evaluate the estimator in \cref{eq:4PCF_radial_def} $140\times$ faster than \textsc{encore}, enabling a high-fidelity measurement on the full DESI Y1 LRG footprint.

We find clear evidence for a non-zero connected 4PCF in the DESI data. Using the $T^2$ statistic with an empirical covariance constructed from the highest signal-to-noise ratio of the analytic covariance (\cref{sec:emp_cov}), we detect the 4PCF in the full NGC LRG sample at up to $\sim$17$\sigma$ when including small tetrahedra, and at a comparable level when combining NGC and SGC (\cref{tab:T2_summary_Table}). The corresponding SGC-only detections are more modest, reflecting the smaller volume and lower completeness of the southern footprint, but remain significant at the several-$\sigma$ level. A complementary $\chi^2$ analysis that directly inverts the analytic covariance yields qualitatively consistent results, with NGC, SGC, and the combined NGC+SGC all showing strong deviations from the null hypothesis when we include small tetrahedra (\cref{tab:Chi2_summary_Table}); \textit{i.\,e.}, we allow the side lengths of our tetrahedrons down to \SI{7.8}{\per\hHubble\Mpc}. The broad agreement between the $T^2$ and $\chi^2$ approaches, despite their different treatments of the covariance, gives us confidence that the detection is not an artifact of the chosen compression or test statistic.

Our multi-region and multi-patch analysis provides an important internal consistency check. The 4PCF measured in the full NGC and SGC footprints is in good agreement with both \textsc{EZMocks} and \textsc{Abacus} mocks on a mode-by-mode basis (\crefrange{fig:NGC_full_Meas}{fig:SGC_full_comp}), and the detection significance remains high when we restrict to high-completeness regions or to approximately equal-area patches. Cross-correlation measurements between NGC and SGC patches and regions further demonstrate that the signal is spatially coherent across the survey volume (\crefrange{fig:ST_and_NoST_Cross_Patches_T2_Auto}{fig:ST_NGC_and_SGC_Cross_Patches_Chi2}). This coherence disfavors explanations in terms of localized systematics confined to a single cap or particular region, and instead points to a genuine large-scale structure signal originating from nonlinear gravitational evolution and galaxy biasing.

The role of small tetrahedra is particularly revealing. While restricting to larger configurations (``no-ST'' columns in \cref{tab:T2_summary_Table,tab:Chi2_summary_Table}) yields only modest or even insignificant detection levels, the inclusion of small tetrahedra sharply increases the significance across essentially all configurations. This behavior is physically intuitive: the 4PCF is especially sensitive to nonlinear growth of structure and higher-order bias on quasi-linear and mildly nonlinear scales, and these are encoded in small tetrahedra. At the same time, the inclusion of small tetrahedra also underscores the importance of carefully modeling small-scale contributions and their covariance, as they dominate the overall signal-to-noise. A natural direction for future work is therefore to incorporate next-to-leading-order contributions to the 4PCF covariance, such as those theoretically modeled in \cite{ortola_cov_I, ortola_cov_II}.

Although our results indicate a strong detection of the connected galaxy 4PCF, we find some tensions between the DESI data and the mocks. We find the level of tension depends on the analysis configuration. In the $T^2$ analyses, the clearest discrepancies are found when the small tetrahedra are included, with configurations containing separations extending down to $7.8,h^{-1}{\rm Mpc}$. The fact that the tension becomes strongest when the maximum amount of small-scale information is retained provides a potentially important indication of its origin. On these scales, nonlinear evolution and the details of the galaxy--halo connection become increasingly important. The mock catalogs used here were primarily constructed and validated to reproduce the 2PCF rather than the connected 4PCF. In particular, higher-order correlations can contain additional sensitivity to the HOD and galaxy biasing \cite{Galaxy_Halo_DESI_DR2}. The connected 4PCF may therefore be probing aspects of the galaxy--halo connection, or more generally of the mock construction, to which the 2PCF is substantially less sensitive. These results should also be interpreted in light of the finite number of mock realizations, since a $\sim3\sigma$ fluctuation has a tail probability of order $10^{-3}$ and can therefore lie beyond all $\sim1000$ mocks without implying a substantially larger tension. In contrast, the $\chi^2$ analysis shows no significant discrepancies between the data and the mocks, although differences are observed among the mock catalogs themselves, with the Abacus FFA and \textsc{EZmock} distributions lying close together and the Abacus altMTL distribution shifted toward larger $\chi^2$ values.

Naively, one might expect that a mismatch between DESI and the mocks at the connected 4PCF level could also imply that the 2PCF covariance is not perfectly reproduced, since the 2PCF covariance is itself related to a 4PCF. However, it is important to recall that the connected 4PCF is not the leading contribution in the perturbative expansion. Rather, the leading-order 4PCF contribution is built from four linear Gaussian density contrasts, which, through Wick's theorem, reduce to products of two 2PCFs \cite{Bernardeau_2002}. Therefore, a mismatch between the data and the mocks at the connected 4PCF level would enter the 2PCF covariance only at next-to-leading order in perturbation theory.

From this perspective, the small-scale sensitivity observed here may also illustrate an additional use of the connected 4PCF beyond cosmological measurements. Higher-order statistics might probe aspects of the galaxy--halo connection that are not fully constrained by two-point clustering, making the connected 4PCF a potentially useful diagnostic for the construction and validation of future high-fidelity mock catalogs. Determining whether the discrepancies identified here can be associated with particular ingredients of the galaxy--halo model will require dedicated mock comparisons and is left for future work.

There are also several limitations that should be borne in mind. Our analytic covariance model necessarily involves approximations (\textit{e.\,g.}\ truncation at linear order in perturbation theory and linear bias only), and the number of available mocks is finite. These limitations motivate our use of eigenmode compression and hybrid empirical covariances, but these limitations also mean that the reported significances should be viewed as approximate. Finally, our current analysis is performed on DESI Y1, which, while already delivering volumes substantially larger than previous surveys, still represents only a fraction of the final DESI dataset. Statistical errors, particularly for cross-correlation configurations and sub-regions, will shrink further with Y3 and Y5 data.

Despite these caveats, our results demonstrate that the 4PCF is now a practical and powerful observable for modern spectroscopic surveys. The significant detection we obtain in DESI Y1 LRGs opens several avenues for future work. On the observational side, applying the same methodology to the DESI BGS, QSO and ELG samples will enable both multi-tracer and cross-correlation 4PCF analyses, improve constraints on higher-order bias parameters, and sharpen tests of gravitational physics. Combining the 4PCF with lower-order statistics (2PCF and 3PCF) \cite{gualdi_joint_bi_tri} will allow joint fits that exploit their complementary dependence on cosmology and biasing, and may help break degeneracies that limit current analyses.

In summary, we have measured the connected 4PCF of DESI Y1 LRGs over a range of survey geometries, redshift cuts, and angular configurations, and have demonstrated a robust, high-significance detection in both auto- and cross-correlation. The agreement with state-of-the-art mock catalogs and the consistency between independent statistical tests validate both our measurement and our covariance modeling at the level required for Y1. These results establish the 4PCF as a mature tool for large-scale structure analysis and lay the groundwork for exploiting higher-order statistics to extract cosmological information.

\section*{Acknowledgments}

WOL acknowledges support from Grant 63041 from the John Templeton Foundation. The opinions expressed in this publication are those of the author(s) and do not necessarily reflect the views of the John Templeton Foundation. This material is based upon work supported by the National Science Foundation Graduate Research Fellowship under Grant No.\ DGE-2236414. Any opinions, findings, and conclusions or recommendations expressed in this material are those of the author(s) and do not necessarily reflect the views of the National Science Foundation. AK was supported as a CITA National Fellow by the Natural Sciences and Engineering Research Council of Canada (NSERC), funding reference \#DIS-2022-568580. AG and ZS acknowledge funding from NASA grant number 80NSSC24M0021. The authors acknowledge UFIT Research Computing for providing computational resources on HiPerGator and support that have contributed to the research results reported in this publication.

This material is based upon work supported by the U.S. Department of Energy (DOE), Office of Science, Office of High-Energy Physics, under Contract No. DE–AC02–05CH11231, and by the National Energy Research Scientific Computing Center, a DOE Office of Science User Facility under the same contract. Additional support for DESI was provided by the U.S. National Science Foundation (NSF), Division of Astronomical Sciences under Contract No. AST-0950945 to the NSF’s National Optical-Infrared Astronomy Research Laboratory; the Science and Technology Facilities Council of the United Kingdom; the Gordon and Betty Moore Foundation; the Heising-Simons Foundation; the French Alternative Energies and Atomic Energy Commission (CEA); the Secretariat of Science, Humanities, Technology and Innovation (SECIHTI) of Mexico; the Ministry of Science, Innovation and Universities of Spain (MICIU/AEI/10.13039/501100011033), and by the DESI Member Institutions: \url{https://www.desi.lbl.gov/collaborating-institutions}. Any opinions, findings, and conclusions or recommendations expressed in this material are those of the author(s) and do not necessarily reflect the views of the U. S. National Science Foundation, the U. S. Department of Energy, or any of the listed funding agencies.

The authors are honored to be permitted to conduct scientific research on I'oligam Du'ag (Kitt Peak), a mountain with particular significance to the Tohono O’odham Nation.

\appendix

\section{Additional Measurements \& Detection Significances}
\label{sec:Additional_Mst}

In this Appendix, we present a set of selected supplementary figures showing the measured 4PCF in various sky sectors and comparisons to the mock catalogs. These include measurements for the individual patches and regions of both the NGC and SGC, allowing a more detailed view of the consistency between the DESI data and the different mocks beyond what is shown in the main text. By displaying the 4PCF across these smaller angular subdivisions, we provide an additional check on the stability of the signal across the survey footprint.

We also include extended detection significance information in the form of two summary tables, \cref{tab:T2_det_table,tab:chi2_det_table}. These report the detection significances obtained from the $T^2$ and $\chi^2$ analyses, respectively, for the auto-correlation measurements in every NGC and SGC patch and region. To complement the tabulated results, we also provide a selection of additional figures illustrating these detection significances.

\Cref{tab:T2_det_table} shows the $T^2$ auto-correlation detection significance for the individual patches and regions of the NGC and SGC. As in the full-cap results, the inclusion of small tetrahedra generally increases the detection significance relative to the no-ST measurements. This is consistent with the expectation that smaller-scale configurations contain additional 4PCF signal, due to the larger clustering amplitude and stronger nonlinear contributions on these scales.

There are, however, two features of the table that are not fully understood. First, SGC Patch~1 shows the opposite behavior from the general trend: the detection significance decreases from $4.1\sigma$ in the no-ST case to $3.7\sigma$ when small tetrahedra are included. This decrease is also present in the corresponding $\chi^2$ analysis, shown in \cref{tab:chi2_det_table}, suggesting that it is not specific to the empirical-covariance $T^2$ statistic. Second, although the patches were selected to have approximately similar volumes, the detection significances are not uniform across patches. For example, in the NGC ST measurements the values range from $3.4\sigma$ to $6.2\sigma$, while in the SGC ST measurements they range from $2.6\sigma$ to $5.8\sigma$. We would therefore have expected the patch-level significances to be more comparable, but the measured values show noticeable patch-to-patch variation. The origin of this variation is unclear. Based on the tests in \cref{Sec:Systematics}, however, we do not find evidence that the observed behavior is driven by the systematics considered in this work.

\subsection{Measurements}

\begin{figure}
    \centering
    \includegraphics[width=0.83\textwidth]{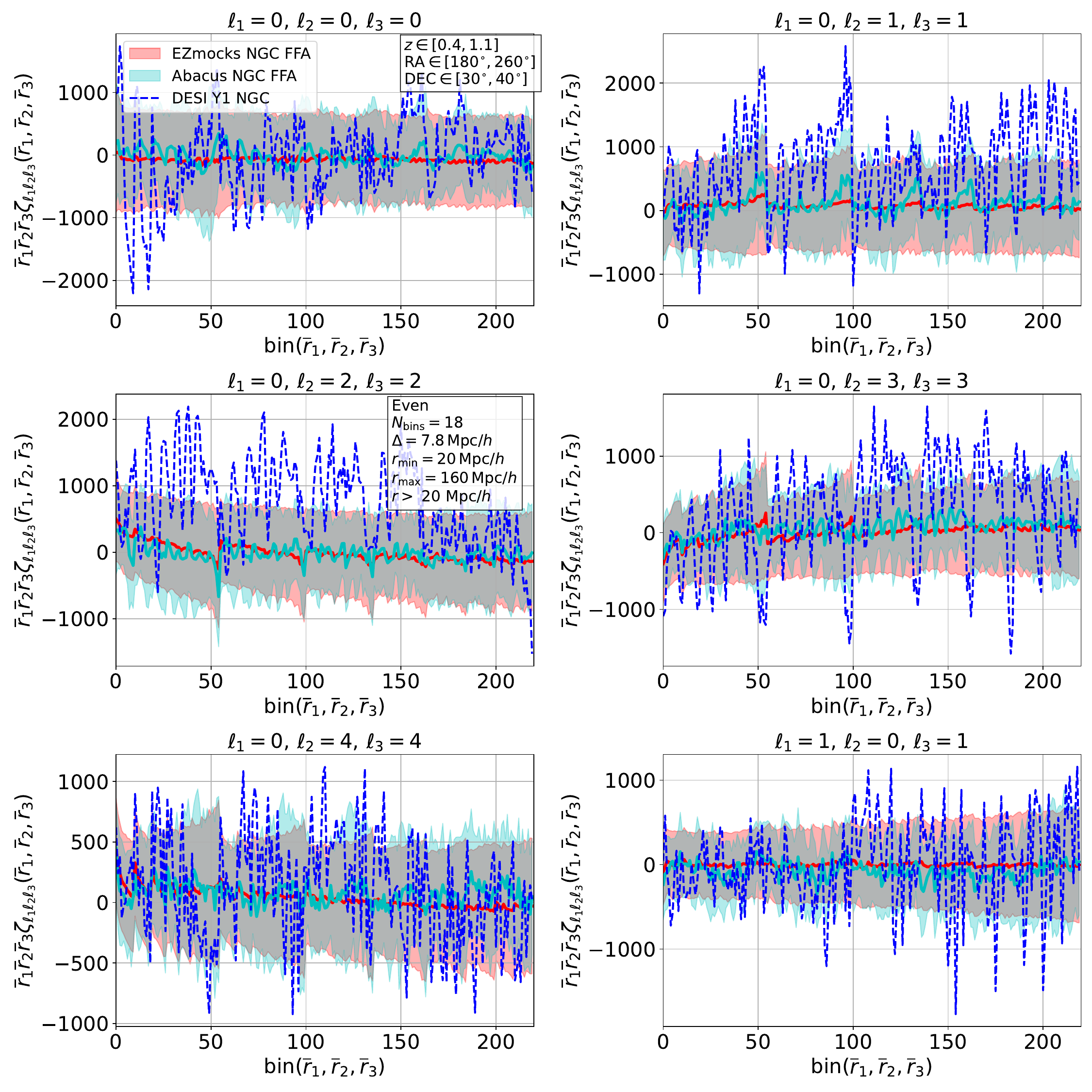}

    \caption{Measurement of the connected 4PCF modes $\ell_1, \ell_2, \ell_3$ from DESI Y1 NGC Patch~1 (blue dashed line) as a function of radial index, compared to mock catalogs. The red solid line and pink shaded region represent the mean and $1\sigma$ deviations from \textsc{EZMocks} realizations, while the light blue line and shaded region show results from \textsc{Abacus} mocks. The measurement is restricted to galaxies in the redshift range z $ \in [0.4, 1.1]$ and the angular region defined by RA $\in [180^\circ, 260^\circ]$ and DEC $\in [30^\circ, 40^\circ]$, as indicated in the top-left panel inset. Radial binning and scale cuts are defined in the inset of the middle-left panel; $\Delta$ represents the bin separation. The consistency of the DESI data with the mock predictions across multiple $\ell$ combinations shows we have robustly detected a non-zero connected 4PCF signal. In the $\ell_1=0$, $\ell_2=\ell_3=2$ panel, the data fall outside the $1\sigma$ region of the mocks due to strong correlations in the error bars. Nevertheless, \cref{fig:NGC_P1_comp} demonstrates the data and mocks remain consistent.}
    \label{fig:NGC_P1_meas}
\end{figure}

\begin{figure}
    \centering
    \includegraphics[width=\textwidth]{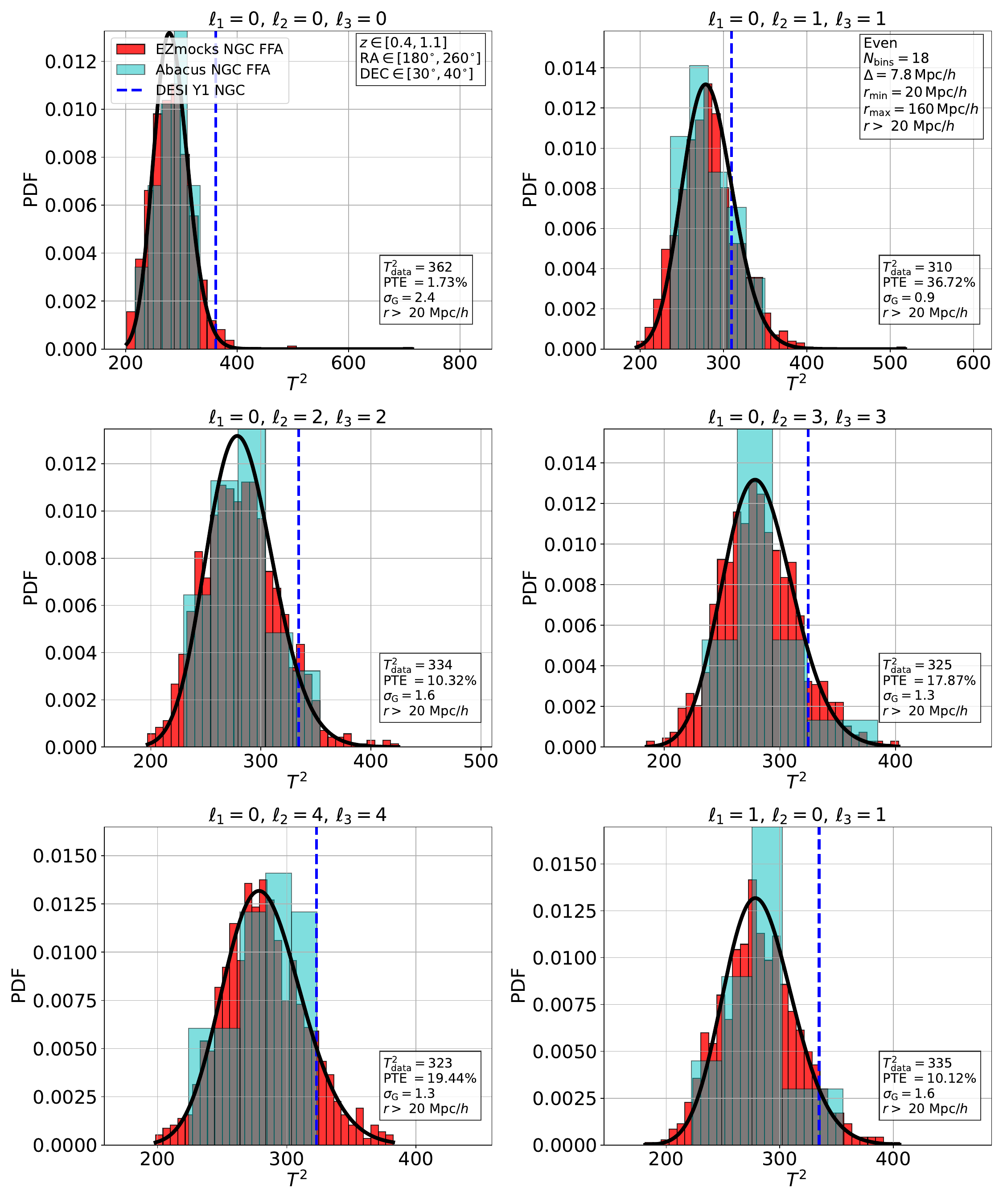}

    \caption{PDFs of the $T^2$ statistics for the 4PCF modes measured from \textsc{EZMocks} realizations (red) and \textsc{Abacus} mocks (light blue) in DESI Y1 NGC Patch~1. The black curve shows a $T^2$ distribution. The vertical blue dashed line indicates the value of $T^2$ measured from DESI Y1 data. Each panel corresponds to a different $(\ell_1, \ell_2, \ell_3)$ combination. The PTE and the corresponding number of $\sigma$ (denoted $\sigma_{\mathrm{G}}$) on a Gaussian are reported in each panel inset. The binning scheme and scale cuts applied are indicated in the inset. The mocks closely follow the analytic $T^2$ distribution (black curve), peaking at $T^2 \simeq 270$, with the data lying well within the distributions, indicating good agreement between the data and mocks.}
    \label{fig:NGC_P1_comp}
\end{figure}

\begin{figure}
    \centering
    \includegraphics[width=\textwidth]{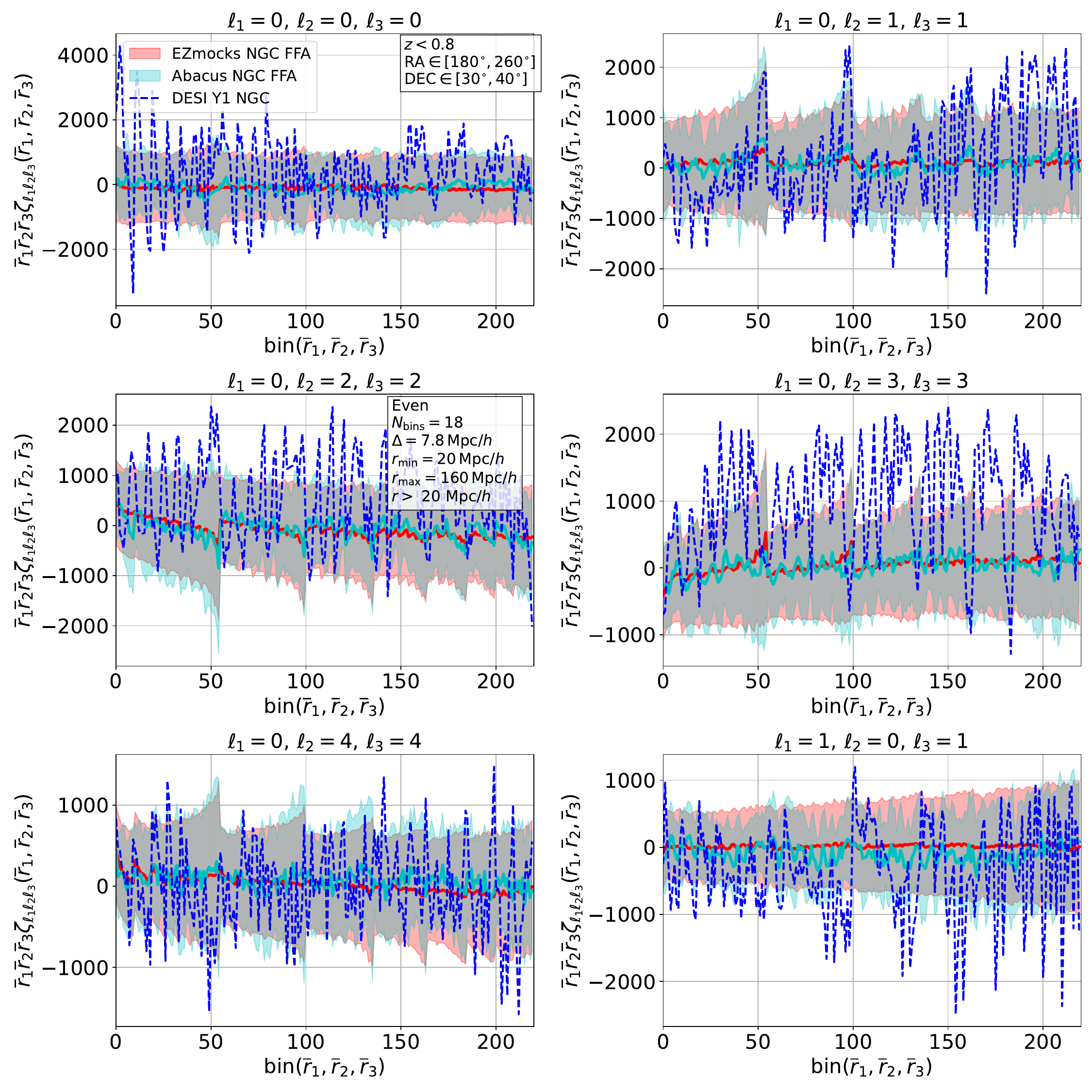}

    \caption{Same as \cref{fig:NGC_P1_meas}, but for DESI Y1 NGC $z$-cut Region~1 ($z < 0.8$). The DESI data and mocks show consistency across multiple $\ell$ combinations. In the two lower panels in the right column, part of the data fall outside of the $1\sigma$ region of the mocks. However, \cref{fig:NGC_R1_comp} demonstrates the data and mocks remain consistent.}
    \label{fig:NGC_R1_meas}
\end{figure}

\begin{figure}
    \centering
    \includegraphics[width=\textwidth]{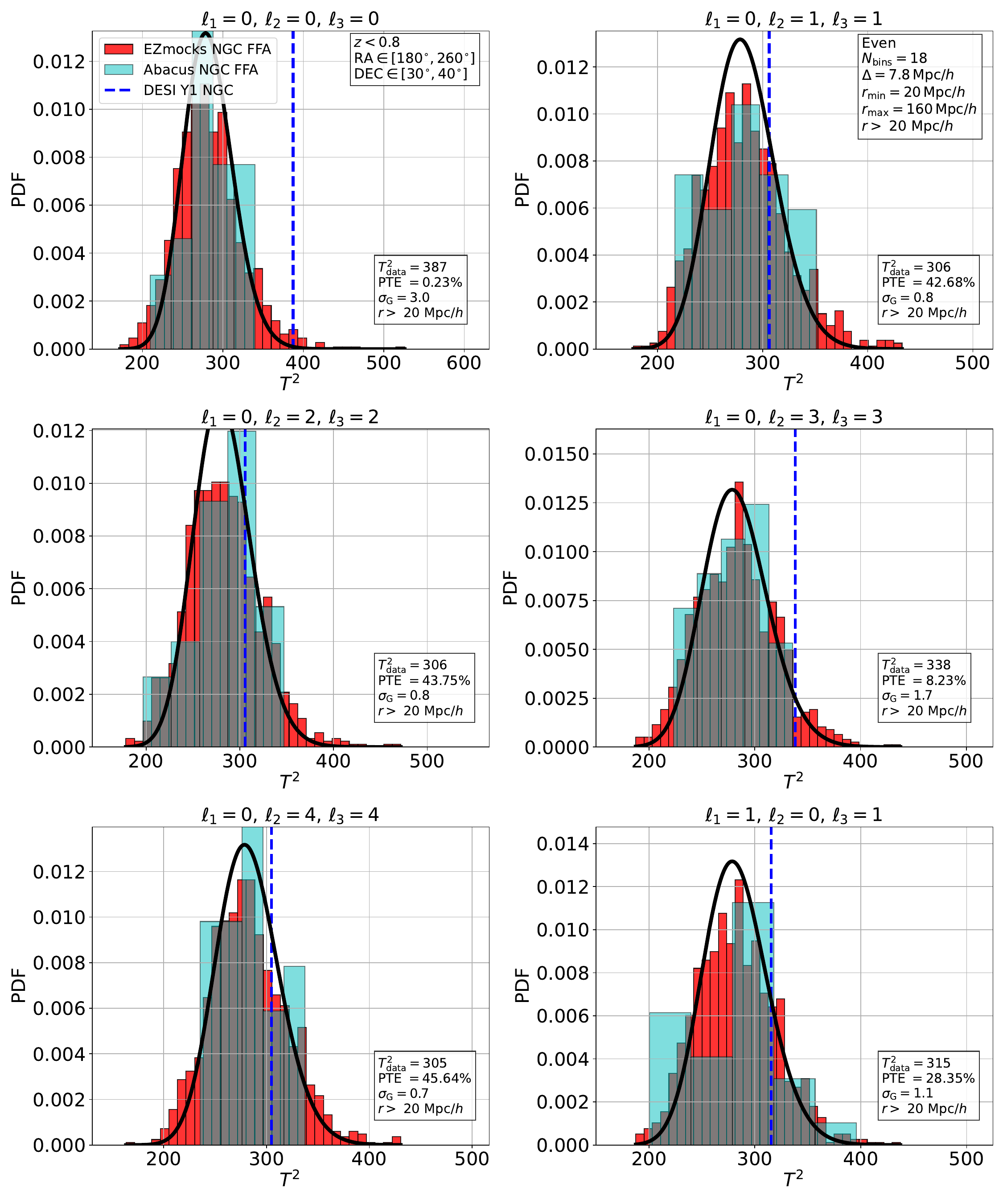}

    \caption{Same as \cref{fig:NGC_P1_comp}, but for a DESI Y1 NGC $z$-cut Region~1 ($z < 0.8$). The mocks closely follow the analytic $T^2$ distribution (black curve), peaking at $T^2 \simeq 270$, with the data lying well within the distributions, indicating good agreement between the data and mocks.}
    \label{fig:NGC_R1_comp}
\end{figure}

\begin{figure}
    \centering
    \includegraphics[width=\textwidth]{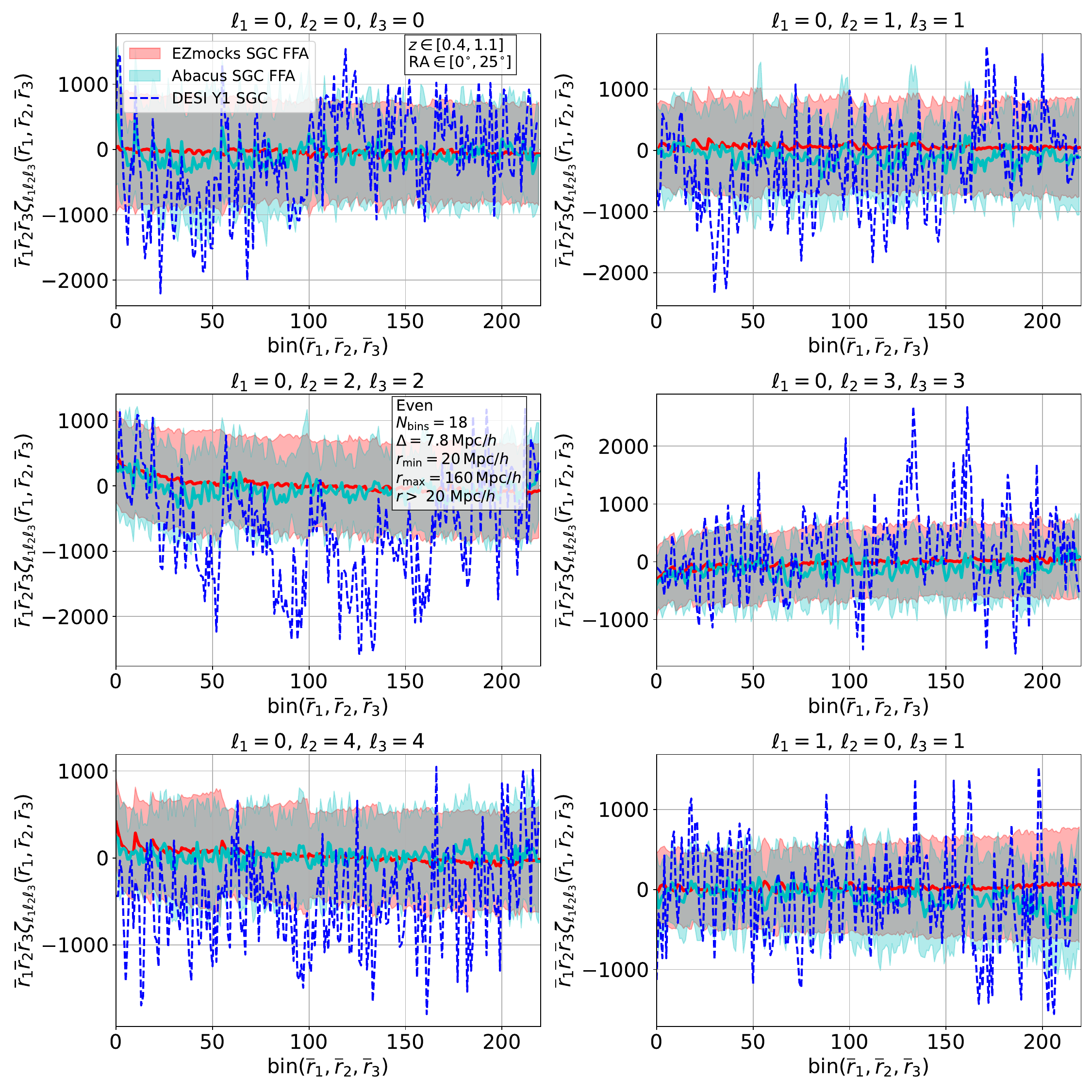}

    \caption{Same as \cref{fig:NGC_P1_meas}, but for SGC Patch~2. This SGC Patch uses the full available DEC range. The DESI data and mocks show consistency across the multiple $\ell$ combinations. In the two lower panels in the left column, part of the data fall outside of the $1\sigma$ region of the mocks. However, \cref{fig:SGC_P2_comp} demonstrates the data and mocks remain consistent.}
    \label{fig:SGC_P2_meas}
\end{figure}

\begin{figure}
    \centering
    \includegraphics[width=\textwidth]{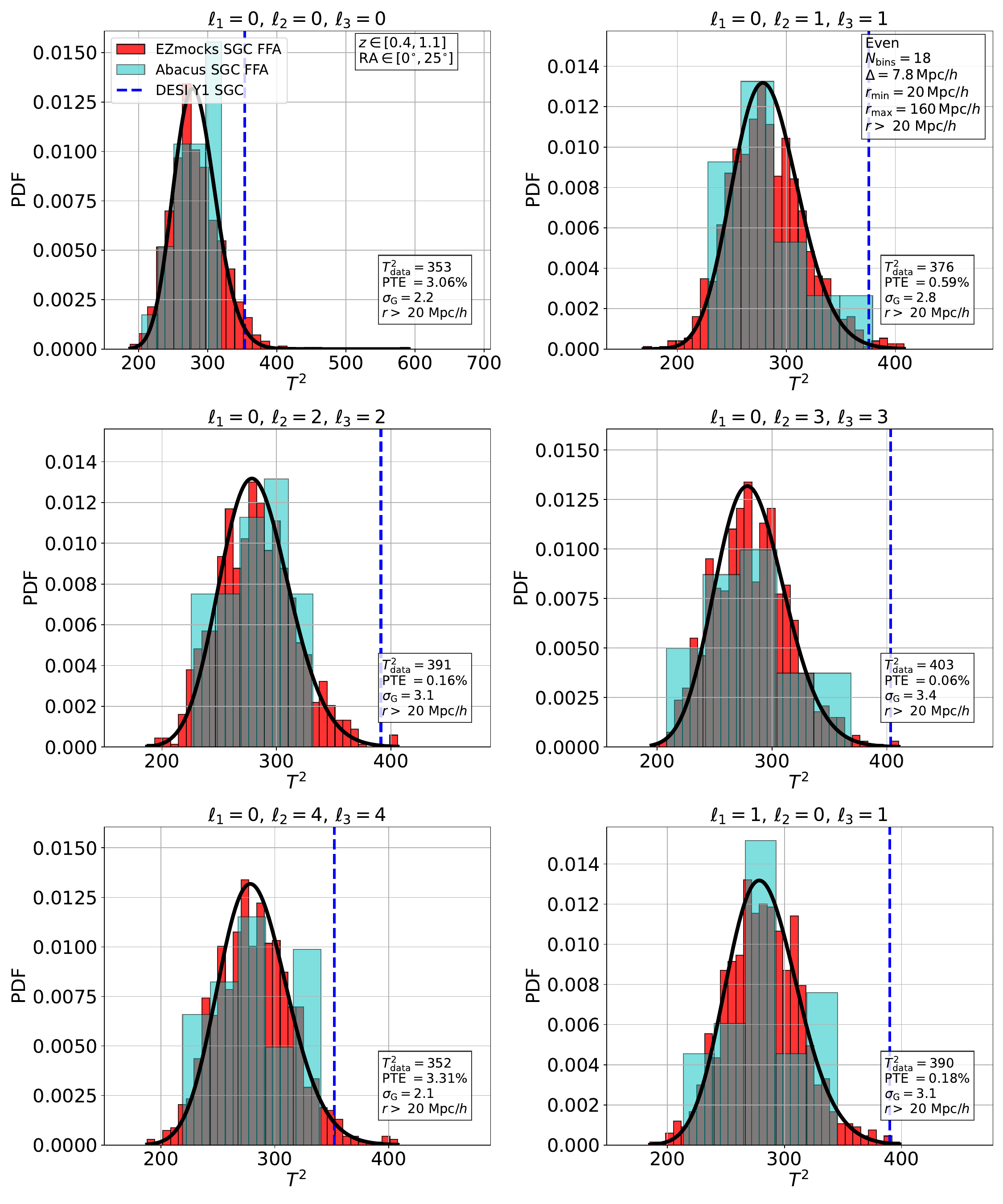}

    \caption{Same as \cref{fig:NGC_P1_comp}, but for SGC Patch~2. The mocks closely follow the analytic $T^2$ distribution (black curve), peaking at $T^2 \simeq 270$. Some panels show the data lying towards the end of the distributions, possibly indicating a covariance under estimation in the SGC, but distributions still demonstrate consistency between data and mocks. We discard systematic effects based on the results of \cref{Sec:Systematics}.}
    \label{fig:SGC_P2_comp}
\end{figure}

\begin{figure}
    \centering
    \includegraphics[width=\textwidth]{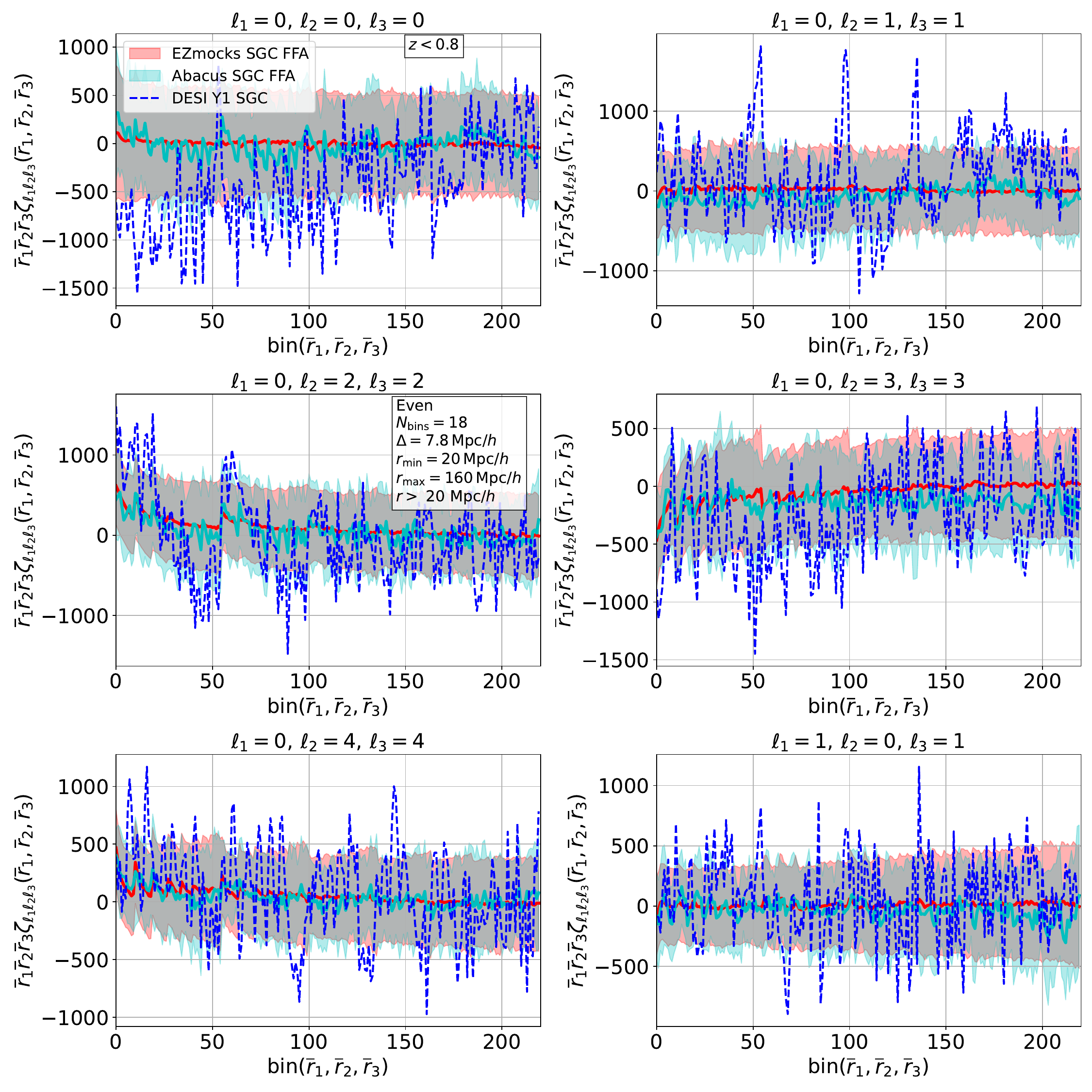}

    \caption{Same as \cref{fig:NGC_P1_meas}, but for the DESI Y1 SGC $z$-cut Region ($z < 0.8$). The SGC Region uses the full available RA and DEC range. The DESI data and mocks show consistency across the multiple $\ell$ combinations. In the two panels of the first row, part of the data fall outside of the $1\sigma$ region of the mocks. However, \cref{fig:SGC_R1_comp} demonstrates the data and mocks remain consistent. }
    \label{fig:SGC_R1_meas}
\end{figure}

\begin{figure}
    \centering
    \includegraphics[width=\textwidth]{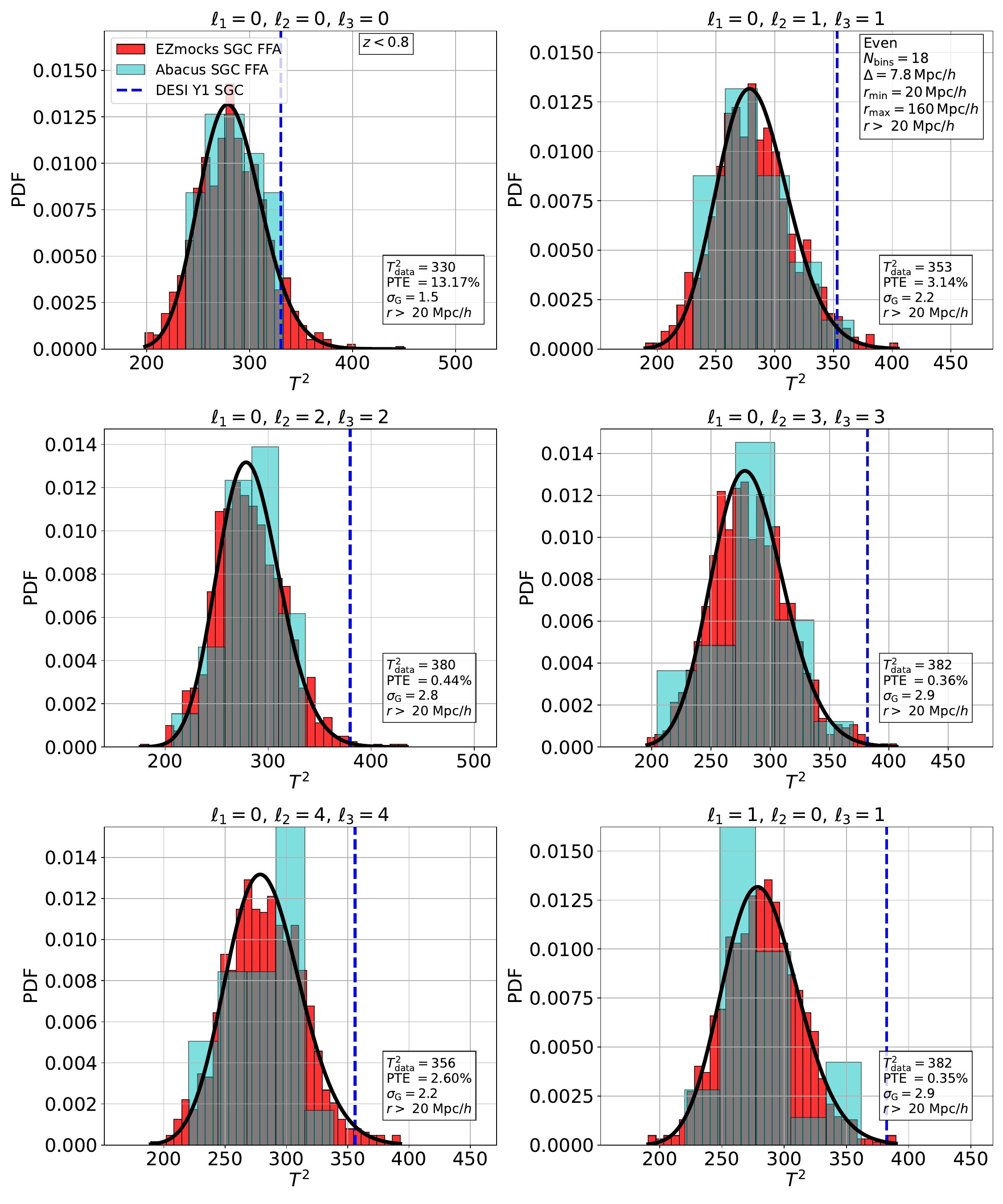}

    \caption{Same as \cref{fig:NGC_P1_comp}, but for the DESI Y1 SGC $z$-cut Region ($z < 0.8$). The mocks closely follow the analytic $T^2$ distribution (black curve), peaking at $T^2 \simeq 270$. Some panels show the data lying towards the end of the distributions, possibly indicating an under-estimation of the covariance in this SGC Region, but distributions still demonstrate consistency between data and mocks. We discard systematic effects based on the results of \cref{Sec:Systematics}.}
    \label{fig:SGC_R1_comp}
\end{figure}

In this subsection we present supplementary figures showing the measured 4PCF in each survey subdivision, and comparisons against the mock catalogs. The NGC is in \crefrange{fig:NGC_P1_meas}{fig:NGC_R1_comp}, which include both the data measurement and the mocks. The SGC is in \crefrange{fig:SGC_P2_meas}{fig:SGC_R1_comp}.

\clearpage

\subsection{Detection Significances}
\label{sec:Additional_Mst_significances}

\begin{table}
    \centering
    {\textbf{\boldmath Detection Significance of the $T^2$ Auto Analysis}}

    \smallskip

    \begin{tabular}{l *{4}{S[table-format=1.1]}}
        \toprule
        \textbf{Sector}
        & \multicolumn{2}{c}{\textbf{NGC}} 
        & \multicolumn{2}{c}{\textbf{SGC}}  
        \\
        & {no-ST} & {ST} & {no-ST} & {ST}  \\
        \midrule
        Patch 1  & 1.9 & 4.1 & 4.1 & 3.7 \\
        Patch 2  & 1.7 & 6.2 & 2.4 & 5.8 \\
        Patch 3  & 2.6 & 3.4 & 2.3 & 2.6 \\
        Patch 4  & 3.1 & 4.2 & {---} & {---} \\
        Region 1 & 0.1 & 2.5 & 4.2 & 5.3 \\
        Region 2 & 1.7 & 5.8 & {---} & {---}  \\
        \bottomrule
    \end{tabular}

    \caption{Detection significance of DESI Y1 LRGs using the $T^2$ auto-correlation analysis for different patches/regions and configurations. Columns labeled ``ST'' include small tetrahedra, while those labeled ``no-ST'' exclude them. We list the detection-significance values corresponding to an empirical covariance (\cref{sec:emp_cov}) constructed with 300 eigenvalues. For the SGC only one region exists and is labeled Region~1. We obtain a lower detection significance in the SGC Patch~1 for the ``ST'' than the ``no-ST'', contrary to expectation. The same behavior occurs when using the $\chi^2$ analysis, as shown in \cref{tab:chi2_det_table}. We discard systematics effects on the SGC Patch~1, based on the result of \cref{Sec:Systematics}. Although the patches were selected to have approximately similar volumes, the detection significances show noticeable patch-to-patch variation, both for the ``no-ST'' and ``ST'' measurements. We would have expected the four patches to yield more comparable detection significances, and the origin of these differences is not fully understood. Similarly, the decrease observed in SGC Patch~1 when going from ``no-ST'' to ``ST'' does not follow the general trend seen in the other sectors, and its origin is also not fully understood.}
    \label{tab:T2_det_table}

    \bigskip

    \centering
    {\textbf{\boldmath Detection Significance of the $\chi^2$ Auto Analysis}}

    \smallskip

    \begin{tabular}{l *{4}{S[table-format=1.1]}}
        \toprule
        \textbf{Sector}
        & \multicolumn{2}{c}{\textbf{NGC}} 
        & \multicolumn{2}{c}{\textbf{SGC}}  
        \\
        & {no-ST} & {ST} & {no-ST} & {ST} \\
        \midrule
         Patch 1  & 0.9 & 1.8 & 2.5 & 2.4 \\
         Patch 2  & 1.1 & 2.3 & 0.5 & 0.8 \\
         Patch 3  & 1.0 & 1.5 & 0.4 & 0.8 \\
         Patch 4  & 2.7 & 3.3 & {---} & {---}  \\
         Region 1 & 2.0 & 2.4 & 3.7 & 4.1 \\
         Region 2 & 0.7 & 2.4 & {---} & {---}  \\
        \bottomrule
    \end{tabular}

    \caption{Detection significance of the DESI Y1 LRGs using the $\chi^2$ auto-correlation analysis for different patches/regions and configurations. Columns labeled ``ST'' include small tetrahedra, while those labeled ``no-ST'' exclude them. For the SGC only one region exists and is labeled Region~1.}
    \label{tab:chi2_det_table}
\end{table}

\begin{figure}
    \vspace*{-8ex}
    \centering
    \includegraphics[width=0.9\textwidth]{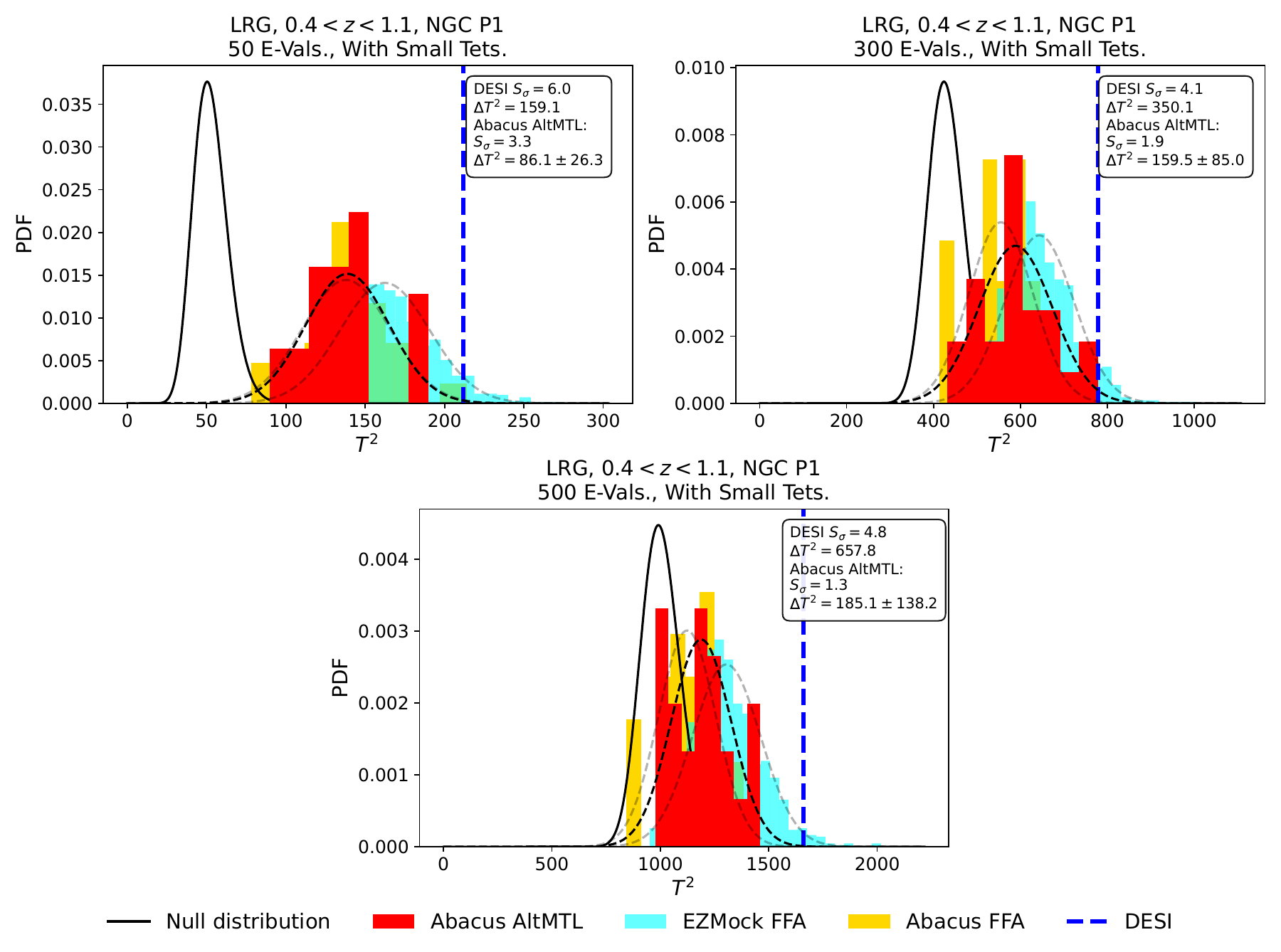}

    \vspace*{-2ex}
    \caption{Detection significance of the 4PCF in the NGC Patch~1 ($0.4 < z < 1.1$). Each panel shows the distribution of the test statistic $T^2$ computed using \numlist{50; 300; 500} eigenvalues (from left to right) of the covariance, as explained in \cref{sec:emp_cov}. The inset in each panel reports the corresponding detection significance in $\sigma$, $S_{\mathrm{\sigma}}$. It also gives the $\Delta T^2$ between DESI and the null distribution, quantifying the robustness of the 4PCF detection. Contrary to expectation, the detection significance decreases when increasing from 50 to 300 eigenvalues.}
    \label{fig:ST_NGC_P1_T2_Auto}

    \bigskip

    \centering
    \includegraphics[width=0.9\textwidth]{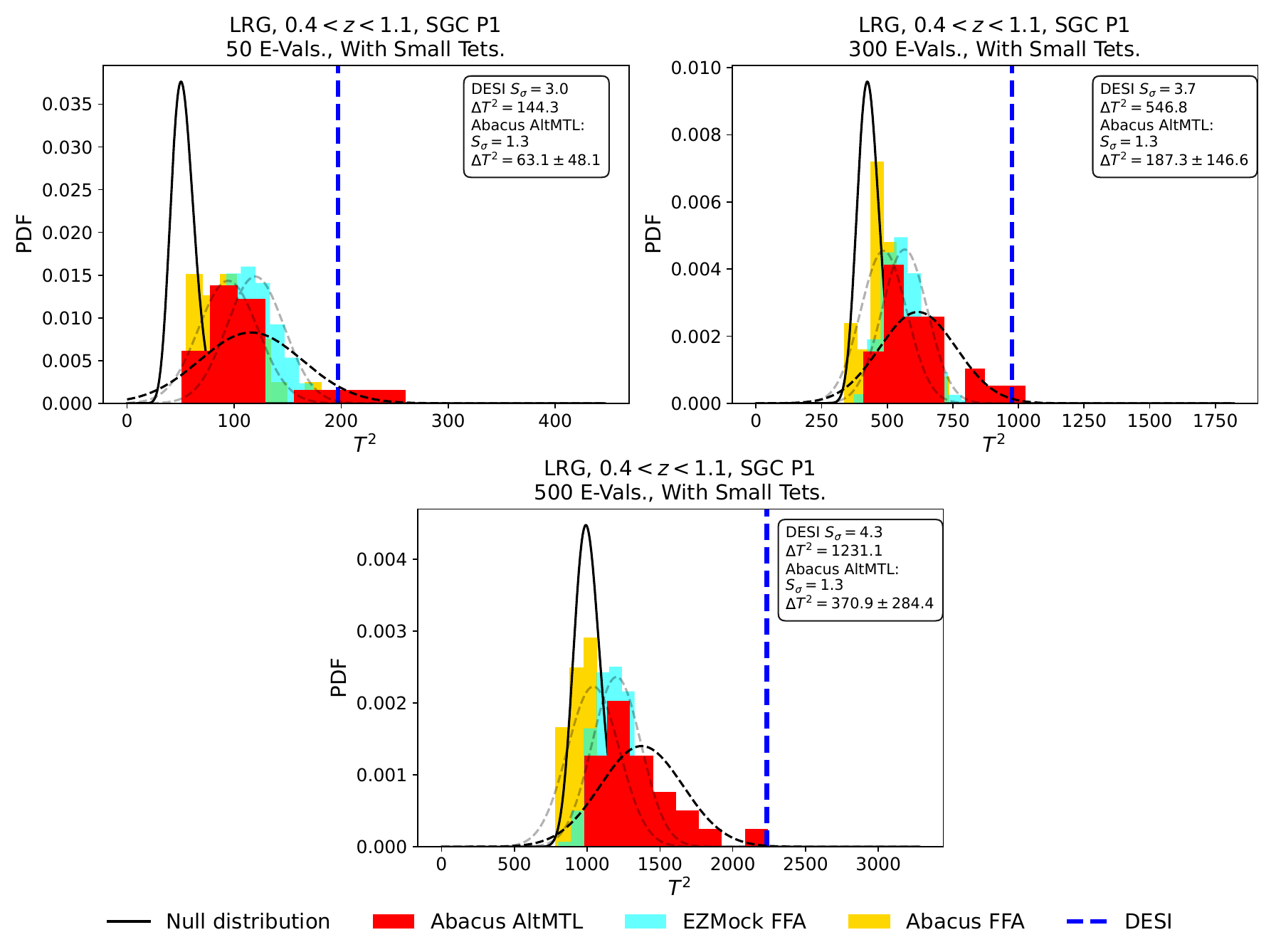}

    \vspace*{-2ex}
    \caption{Same as \cref{fig:ST_NGC_P1_T2_Auto}, but for SGC Patch 1 ($0.4 < z < 1.1$). As expected, the detection significance increase when increasing from 50 to 300 to 500 eigenvalues. }
    \label{fig:ST_SGC_P1_T2_Auto}
\end{figure}

\begin{figure}
    \vspace*{-2ex}
    \centering
    \includegraphics[width=0.9\textwidth]{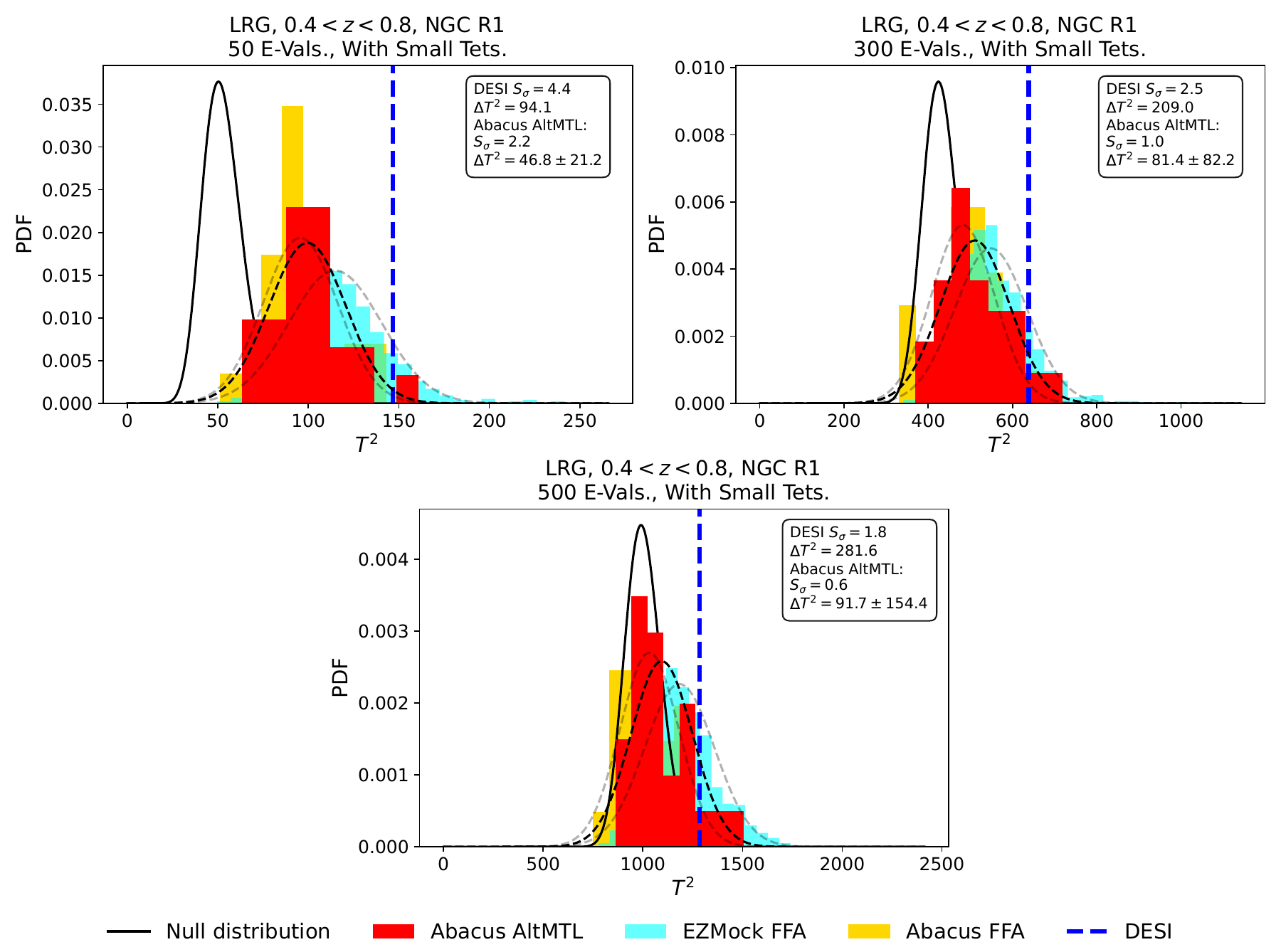}

    \vspace*{-2ex}
    \caption{Same as \cref{fig:ST_NGC_P1_T2_Auto}, but for NGC Region 1 ($0.4 < z < 0.8$). Contrary to expectation, the detection significance decreases when increasing from 50 to 300 to 500 eigenvalues.}
    \label{fig:ST_NGC_R1_T2_Auto}

    \bigskip

    \centering
    \includegraphics[width=0.9\textwidth]{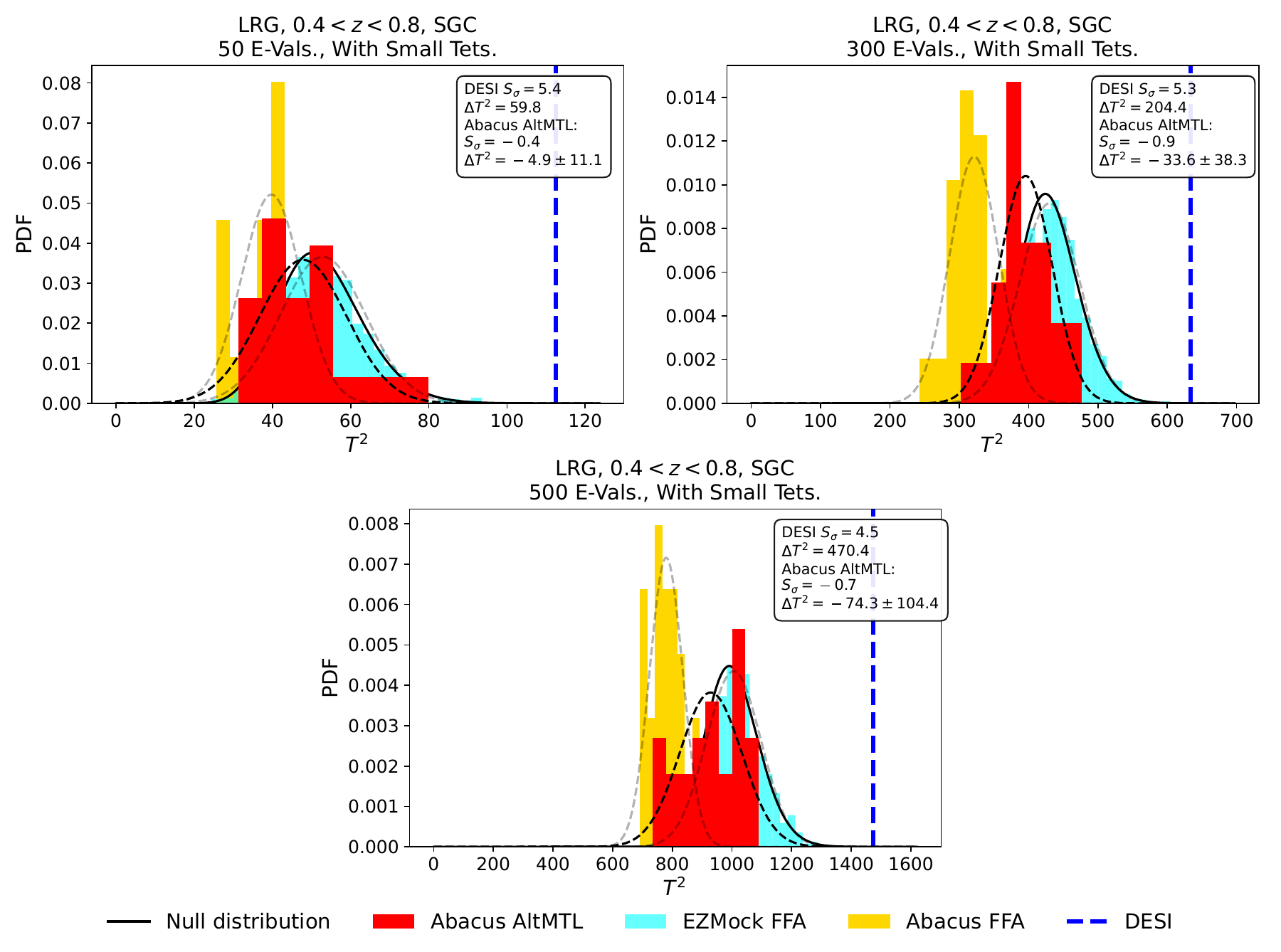}

    \vspace*{-2ex}
    \caption{Same as \cref{fig:ST_NGC_P1_T2_Auto}, but for SGC Region~1 ($0.4 < z < 0.8$). Contrary to expectation, the detection significance decreases when increasing from 50 to 300 to 500 eigenvalues. The mock distributions also lie below, or very close to, the null distribution, suggesting that these mocks may not provide a reliable representation of the connected 4PCF signal in this region, since they predict little separation from the purely Gaussian null expectation. }
    \label{fig:ST_SGC_Region_T2_Auto}
\end{figure}

\begin{figure}
    \centering
    \includegraphics[width=0.5\textwidth]{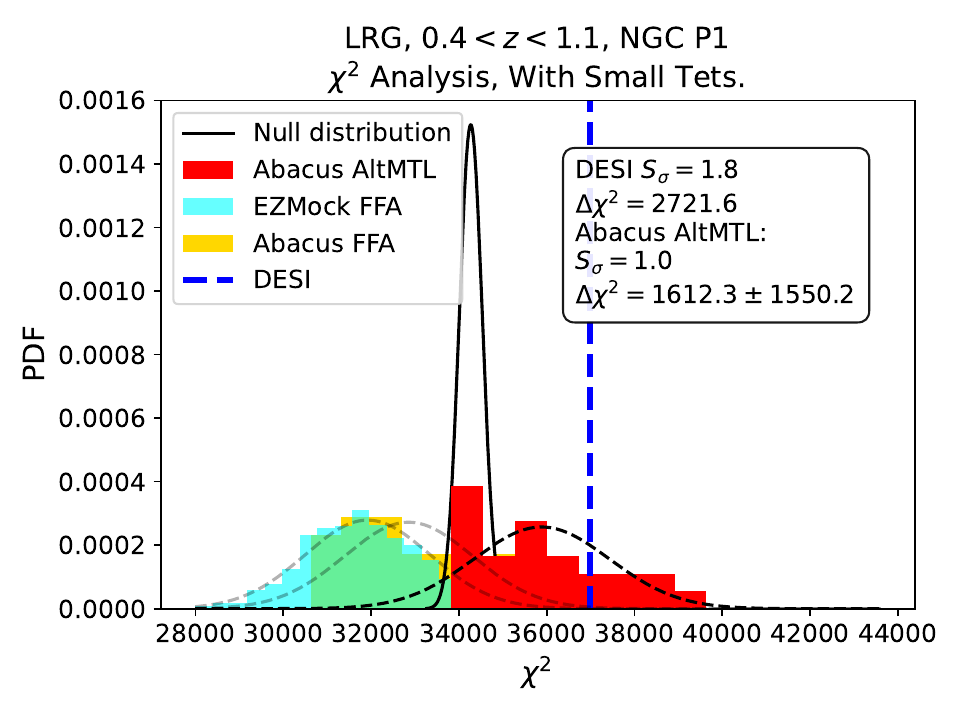}%
    \includegraphics[width=0.5\textwidth]{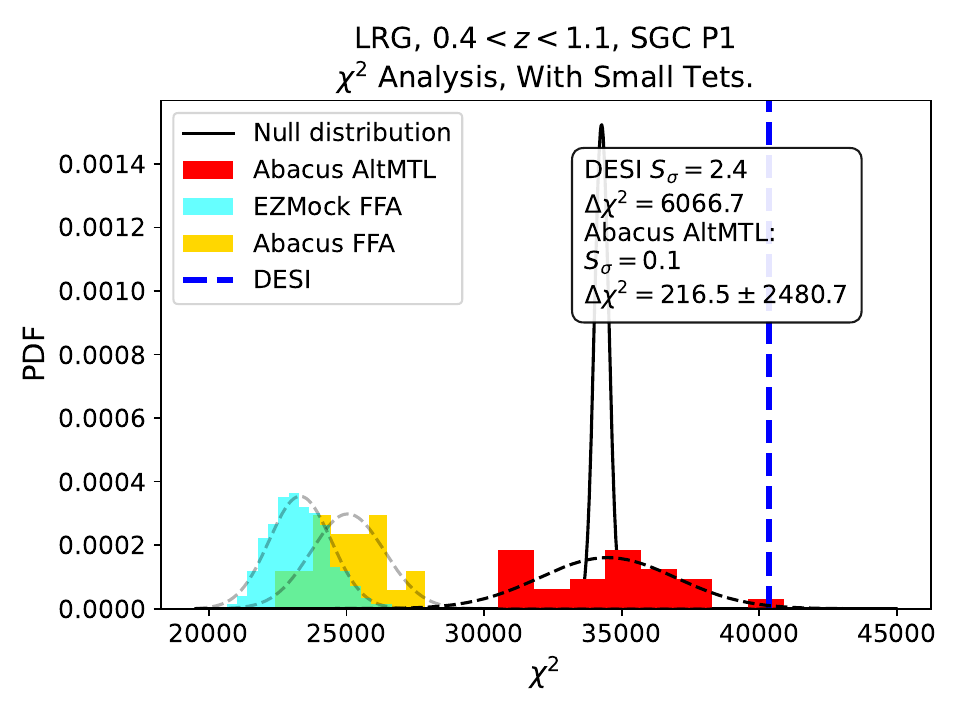}

    \caption{Detection significance of the 4PCF in the NGC (left) and SGC (right) Patch~1 for the LRG sample ($0.4 < z < 1.1$) using the $\chi^2$ test statistic, as explained in \cref{Sec:Cov-analytic}. The inset in each panel reports the corresponding detection significance $S_{\sigma}$ and the effective $\Delta \chi^2$ between DESI and the null distribution, quantifying the robustness of the 4PCF detection.}
    \label{fig:ST_NGC_P1_Chi2_Auto}
\end{figure}

\begin{figure}
    \centering
    \includegraphics[width=0.5\textwidth]{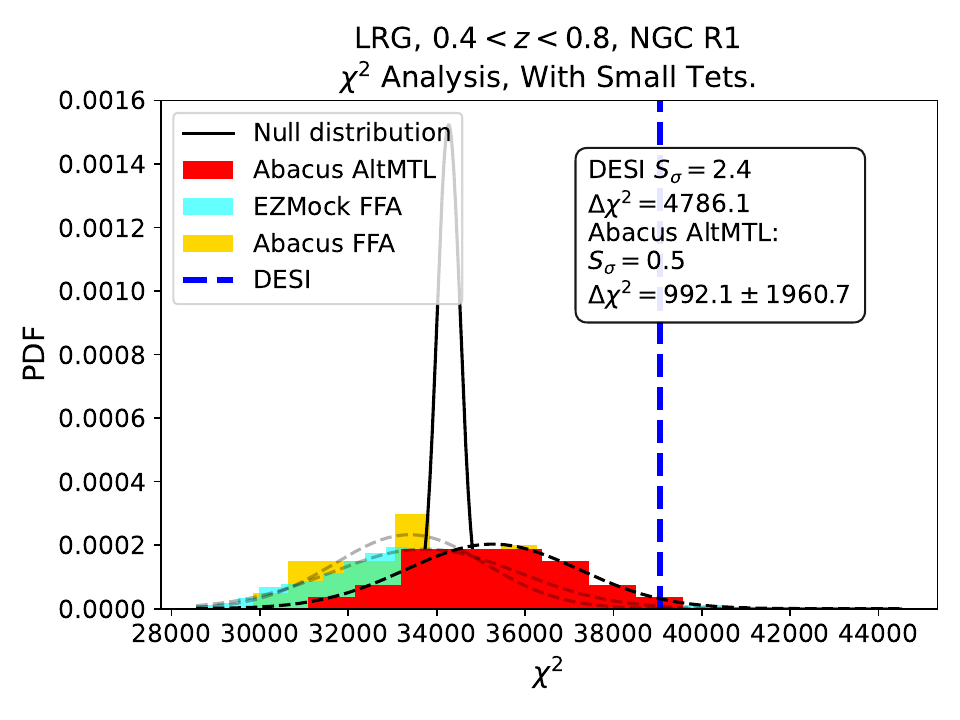}%
    \includegraphics[width=0.5\textwidth]{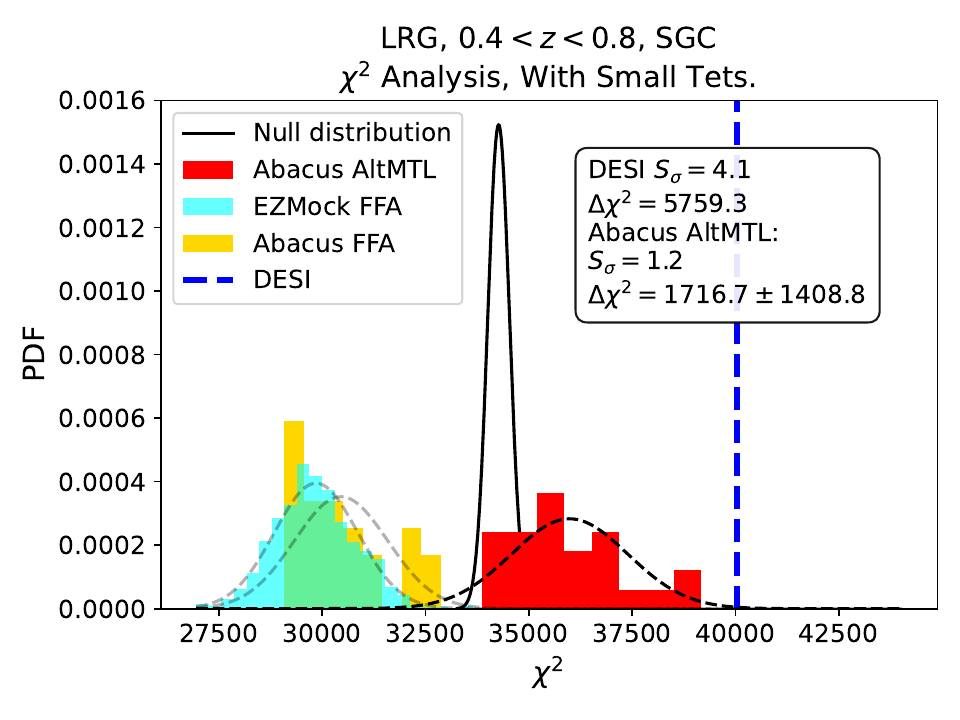}

    \caption{Same as \cref{fig:ST_NGC_P1_Chi2_Auto}, but for NGC Region~1 (left) and the sole SGC Region (right) (both for $0.4<z<0.8)$. Although one might naively expect a larger detection significance in the NGC than in the SGC, we find the opposite for this setup. This arises because, in the $z$-cut analysis, the SGC includes its full footprint, whereas the NGC is restricted to a smaller area than the full NGC.}
    \label{fig:ST_NGC_R1_Chi2_Auto}
\end{figure}

In this subsection we present the supplementary figures associated with the detection significance of the 4PCF, as obtained from both the $T^2$ and $\chi^2$ analyses. These complement the summary of detection significance results listed in \cref{tab:T2_det_table,tab:chi2_det_table}. The supplementary $T^2$ analyses are \crefrange{fig:ST_NGC_P1_T2_Auto}{fig:ST_SGC_Region_T2_Auto}, covering all auto-correlation measurements in the NGC and SGC and corresponding directly to the detection significances summarized in \cref{tab:T2_det_table}. The $\chi^2$ analyses are \crefrange{fig:ST_NGC_P1_Chi2_Auto}{fig:ST_NGC_R1_Chi2_Auto}, and correspond to the results reported in \cref{tab:chi2_det_table}. These supplementary figures present the reduced analyses performed either over the full redshift range within individual sky patches or over narrower redshift ranges within smaller sky regions, as defined in \cref{Sec:DESI_data_Y1}. Overall, these reduced analyses yield lower detection significances than the corresponding results presented in \cref{sec:Results}. This is expected because the smaller survey volumes contain fewer galaxies and configurations, and therefore provide a weaker total 4PCF signal.

For the $T^2$ analysis, some results exhibit a decrease in detection significance as the number of eigenmodes increases, contrary to the naive expectation that including additional high-SNR modes should monotonically increase the signal. Since this signal-to-noise is estimated from a finite set of mocks (in each of the regions and patches described in \cref{tab:region-summary}), finite-mock noise can contaminate the inferred SNR and the associated eigenmodes, and the inclusion of additional (increasingly noise-sensitive) modes can in some cases reduce the stability of the inverse covariance and dilute the detection significance. 

As a reminder to the reader and as described in \cref{sec:emp_cov}, we construct the empirical covariance using three choices of retained eigenmodes (the highest-SNR modes), $N_{\mathrm{eval}} \in \{50, 300, 500\}$. Consequently, \crefrange{fig:ST_NGC_P1_T2_Auto}{fig:ST_SGC_Region_T2_Auto} for the $T^2$ analysis each display three panels, one per $N_{\mathrm{eval}}$, with the corresponding detection significance reported in the inset of each panel. In contrast, \crefrange{fig:ST_NGC_P1_Chi2_Auto}{fig:ST_NGC_R1_Chi2_Auto} show the $\chi^2$ analyses performed using the analytic covariance, which has approximately \num{34000} degrees of freedom; therefore, only a single panel is shown for each case. The detection significance for each $\chi^2$ analysis is likewise reported in the inset.

\section{Covariance Matrix Treatments}
\label{Sec:Cov}

We adopt two basic approaches for the covariance matrix in this work. First, we use a purely empirical covariance obtained from the mocks. However, we have at most \num{1000} mocks (\textsc{EZMocks}) and many fewer (25) \textsc{Abacus} mocks. That means we cannot obtain an invertible matrix from the mocks for the full number of degrees of freedom of our 4PCF, of order \numrange{9000}{34000} depending on the choice of analysis (with or witout small tetrahedrons). However, at each individual $(\ell_1, \ell_2, \ell_3)$ mode, we have of order 250 degrees of freedom, given our radial binning, and hence can obtain covariances for each mode individually from mocks. We describe this further below, after first continuing here to outline our overall covariance matrix strategy.

Our second approach to the covariance is to use an analytical template based on assuming the density field is a Gaussian Random Field (GRF). While this is clearly not correct in detail---if it were, then there would be no connected 4PCF for us to measure!---the GRF contribution is always the leading-order term in the connected 4PCF covariance. It enables us to write down analytical expressions for the covariance in terms of integrals of the galaxy power spectrum. Since this template is from an analytical calculation, it is smooth and of arbitrarily good ``precision'' (though it can be less accurate since the density field is not actually a GRF), and thus the covariance is smoothly invertible. 

This template has two parameters, the volume and the shot noise, which we calibrate by matching to a noisy, empirical, full covariance from the mocks, where the matching can be done without inverting this latter. We also describe this more fully in what follows.

\subsection{Empirical Covariance}
\label{sec:emp_cov}

\subsubsection{Block Diagonal Assumption}
At each $(\ell_1, \ell_2, \ell_3)$ mode, we compute the covariance purely from the mocks as follows:
\begin{align}
    \label{eq:def_Cov}
    &\mathrm{Cov}\mleft(\zeta_{\ell_1 \ell_2 \ell_3}(r_1, r_2, r_3),
    \zeta_{\ell_1 \ell_2 \ell_3}(r_1', r_2', r_3')\mright) =
    \frac{1}{N_{\mathrm{mocks}} - 1}
    \\
    &\qquad\times \sum_{n = 1}^{N_{\mathrm{mocks}}}
    \mleft[\zeta^{(n)}_{\ell_1 \ell_2 \ell_3}(r_1, r_2, r_3) - \bar{\zeta}_{\ell_1 \ell_2 \ell_3}(r_1, r_2, r_3)\mright]
    \mleft[\zeta^{(n)}_{\ell_1 \ell_2 \ell_3}(r_1', r_2', r_3') - \bar{\zeta}_{\ell_1 \ell_2 \ell_3}(r_1', r_2', r_3')\mright],
    \nonumber
\end{align}
where superscript $n$ indicates the measured 4PCF coefficient from the $n^{\mathrm{th}}$ mock, and bar the mean over all of the mocks. The $N_{\mathrm{mocks}} - 1$ in the denominator in front is because we estimated the mean from the mocks as well, effectively stealing one piece of information from what we then use to compute the covariance, and making the averaging be over $\left(N_{\mathrm{mocks}} - 1\right)$ independent pieces of information. With $N_{\mathrm{mocks}} = \num{1000}$, this is clearly not a significant difference from a naive averaging (if we were to obtain an empirical covariance from the 25 \textsc{Abacus} mocks, then it would be a few percent correction).

\textbf{In the relation above, the matching $\ell_i$ but differing $r_i$ (primed vs.\ unprimed) are intentional}, as we are estimating the covariance of the radial bins with each other but at fixed mode $(\ell_1, \ell_2, \ell_3)$. This approach corresponds to taking the covariance as block diagonal, where each block captures the covariance of the different sets of radial bins at that one $\ell$-mode.

\subsubsection{Eigenmode Compression Prior to Empirical Covariance Estimation}

Before constructing the empirical covariance from mocks, it is essential to ensure that the resulting covariance matrix is invertible given the finite number of available mocks.  The full 4PCF data vector contains thousands of elements, vastly exceeding the number of \textsc{EZMock} realizations ($N_{\mathrm{mocks}} = \num{1000}$). Direct inversion of the full empirical covariance would therefore be impossible, as the matrix would be singular.

To address this, we first compute the eigenvalues and eigenvectors of the \emph{analytic} covariance matrix and rank the modes by highest SNR following \cite{phil_4pcf, Scoccimarro_Theory_To_observation_Bisp, Cov_Accuracy_Taylor}. This provides a well-defined, theory-motivated orthonormal basis in which the 4PCF modes are naturally ordered by decreasing information content.

We then project each mock 4PCF measurement onto this eigenbasis but only retain a restricted number of modes,
\[
    N_{\mathrm{eval}} \in \{50, 300, 500\},
\]
corresponding to the highest-SNR eigenvalues of the analytic covariance. The empirical covariance is subsequently computed \emph{only} within this reduced subspace. Because $N_{\mathrm{eval}} \ll N_{\mathrm{mocks}}$, the resulting empirical covariance will be invertible.

This hybrid approach --- analytic-eigenmode compression followed by mock-based covariance estimation --- allows us to retain the modes with the highest signal to noise.

\subsubsection{Shape and Center of the Test Statistic's Distribution}
Now, when each $\ell$-mode is analyzed using this covariance, the expected distribution will not be a $\chi^2$ distribution. Let us briefly review why.

\paragraph{Simplified Case: Independent Radial-Bin Sets.}
Consider the limit where each radial-bin set is independent of the others; this corresponds to a diagonal covariance at the given $\ell$-mode. If each radial-bin-set's coefficient is distributed as a Gaussian around its mean, then evaluating the overall significance by summing up their squares weighted by inverse variance, as is the optimal weight, would result in a $\chi^2$ distribution, as long as the variance were perfectly known. However, if we consider this idealized case above, but where the variance of each mode is estimated from a finite set of mocks, this variance itself will have noise. While the noise on the covariance estimator averages to zero in the limit of infinite mocks (it is an unbiased estimator), the noise on the inverse covariance will not. Thus, adding up squares of variables with Gaussian-distributed noise (the signal) weighted by an inverse variance that also has noise, does not result in a $\chi^2$ distribution. The distribution one does get, instead, will be both biased in central value, and different in shape, from a $\chi^2$. It turns out that the correct distribution to which to compare the summed, inverse variance weighted squares, in this case, is the $T^2$ distribution \cite{T2_statistic_Paper}.

\paragraph{Generalization to Correlated Radial-Bin Sets.}
Now, let us generalize this to the case where each radial-bin set is not independent. In this case, the basic picture remains the same---the noise on the covariance (coming from its being estimated from a finite number of mocks) creates a biased inverse covariance, leading to both a different shape and center of the resulting distribution. 

The noise bias of the center was first identified in \cite{hartlap}, and detailed analysis of the distribution shape carried out by \cite{sellentin}, also reviewed in \cite{gouyou}. We can view weighting by the inverse covariance as simply rotating to the basis where each resulting linear combination of modes is independent, weighting these by inverse variance, and summing. 

This picture is appropriate because, by the spectral theorem, the covariance matrix $\mathbf{C}$, which is by construction symmetric, is also diagonalizable, as $\mathbf{C} = \mathbf{S} \mathbf{D} \mathbf{S}^{-1}$. $\mathbf{D}$ is a diagonal matrix whose elements are the eigenvalues, and $\mathbf{S}$ is a rotation matrix whose columns are the eigenvectors; thus $\mathbf{S}^{-1}$ is also a rotation matrix, and $\mathbf{S}^{-1} = \mathbf{S}^{\mathrm{T}}$. Thus the $\mathbf{S}$ matrices do change the basis as outlined above.

%ZS Note to self---has anyone considered how the rotation matrix also gets messed up by this, not just the varaiance in the new basis? Does sellentin and heavens address this?

It turns out that,  as one might guess from the arguments above, in the correlated-error case, the appropriate distribution to which to compare the summed, inverse-covariance-weighted radial-bin-set coefficients is a $T^2$ distribution. Thus, this is what we do.

\subsubsection{“Leave One Out” Test}

However, as a test, we also take each mock, one by one, as if it is the real DESI data, use the other $N_{\mathrm{mocks}} - 1$ mocks to compute a covariance, and compute the $T^2$ of the excluded mock. We do this for each mock and thus obtain the empirical PDF of their test statistic values. This turns out to match well the expected $T^2$ distribution, showing us that we have sufficiently understood the impact of the noise on the empirical covariance. We term this test the ``leave-one-out'' test.

We highlight an important subtlety here. One might naively think that it does not matter if one leaves one mock out in estimating the covariance that is then used to assess it, as this should be an order $1/N_{\mathrm{mocks}}$ change, where $N_{\mathrm{mocks}}$ for the \textsc{EZMocks} is \num{1000}---hence, \SI{0.1}{\percent}. One might think not leaving one out would then be the more computationally efficient choice, as then one could compute a single \textsc{EZMocks} covariance matrix (using all \textsc{EZMocks}) and use it for the test on all \num{1000} mocks, rather than recomputing a new covariance matrix from 999 mocks for each ``excluded'' mock being treated as the ``fake data'', as we in fact do.

However, ``leaving one out'' actually turns out to be essential. This is because, if the mock being analyzed is included in the estimate of the covariance, then by construction the largest deviation that mock can ever have from the mean will be $1$ in $N_{\mathrm{mocks}}$ rare; for the \num{1000} \textsc{EZMocks}, this corresponds to 1 in \num{1000}, or just over $3\sigma$. Thus, failure to ``leave one out'' will lead to a test result that is very consistent (by inadvertent construction), and thus not useful.

\subsubsection{“Leave One Out” Test and the \textsc{Abacus} Mocks}

One further note on this point is in order. The \textsc{Abacus} mocks are far too few in number (25) ever to estimate an empirical covariance, even in the restricted case we discussed above where we treat each $\ell$-mode separately (each $\ell$-mode has of order 250 degrees of freedom). Thus, to assess their significance with an empirical covariance matrix, we use the empirical covariance matrix from the \textsc{EZMocks}, just as we did in the \textsc{EZMocks} case. But here, the \textsc{Abacus} mocks were not included in computation of the covariance, and so we do not need to worry about leaving any \textsc{EZMocks} out; we can use the same \textsc{EZMocks}-estimated covariance to test all 25 \textsc{Abacus} mocks.

One might worry that this slight difference in how the test was performed on the \textsc{EZMocks}, vs. on the \textsc{Abacus}, could cause discrepant results. However, in this case, the use of all \num{1000} \textsc{EZMocks} in computing the covariance matrix used on \textsc{Abacus}, versus the use of 999 to get the covariance for each ``excluded'' \textsc{EZMocks}, really should be a \SI{0.1}{\percent} effect, and thus is not significant.

\subsubsection{Replication and the \textsc{Abacus} Mocks}

To construct the \textsc{Abacus} mocks, the underlying $N$-body simulation volume is smaller than the region spanned by the DESI footprint. The simulation box is therefore replicated in three dimensions before the survey mask is applied. While this procedure yields a volume large enough to match the survey geometry, it does not introduce new independent large-scale modes: each replicated region is drawn from the same realization of the density field. Consequently, the effective number of modes sampled by the mocks is determined by the volume of the original, non-replicated simulation box, which we denote as $V_{\mathrm{unique}}$.

By contrast, we use $V_{\mathrm{mock}}$ to denote the full volume after all replications have been included, \textit{i.\,e.}, the volume spanned by the tiled boxes prior to imposing the DESI mask. Since the ratio $V_{\mathrm{mock}}/V_{\mathrm{unique}}$ measures how much larger the apparent mock volume is compared to the number of independent modes it contains, we rescale the covariance matrix by this factor. For the DESI NGC (SGC) across the full redshift range $0.4 < z < 1.1$, this ratio is $1.25$ ($1.15$).

This effect has been examined in the context of the \textsc{Uchuu} mocks produced for SDSS BOSS, which were made with a similar (though not exactly the same) replication scheme. In the 2PCF, \cite{uchuu_rep} found that at lowest order, one could correct the difference in covariances to which this mode-counting leads by multiplying the covariance matrix by the ratio $V_{\mathrm{full}} / V_{\mathrm{single\ box}}$, where $V_{\mathrm{full}}$ is the full volume used in the non-replicated approach (\textit{e.\,g.}\ it could be either the full survey volume, or some $V_{\mathrm{eff}}$ derived from the typical formulae for this, such as \cite{hou_covar}). $V_{\mathrm{single\ box}}$ is the volume of the single, initial box that was replicated.
%ZS---need to check this against Appendix B.

However, \cite{uchuu_rep} found that this rescaling still under-estimated the covariance of the 2PCF appropriate for the replicated boxes by \SIrange{15}{20}{\percent} (see their Appendix~B). \cite{phil_uchuu} used the \textsc{Uchuu} mocks to argue that the BOSS parity-violation results were due to under-estimated covariances, as that work found the \textsc{Uchuu}-based covariances to be larger than the \textsc{patchy}-mock-based covariance of the original analysis, even after making the volume correction. However, the 4PCF covariance should behave roughly (on dimensional grounds) as the 2PCF covariance squared, so a \SIrange{15}{20}{\percent} over-estimate by the Uchuu mocks, even after the volume correction, would translate to a \SIrange{30}{40}{\percent} over-estimate of the 4PCF covariance by \textsc{Uchuu}. We regard this additional, uncorrected replication effect as an alternative possible explanation for the results of \cite{phil_uchuu}.

\subsection{Analytic Covariance Template}
\label{Sec:Cov-analytic}

Using the public \textsc{Analytic4PC} code described in \cite{Krowleski_No_PV_det}, which includes parallelization of the calculation to multiple cores, we compute a large suite of analytic covariance matrices using the formalism of \cite{hou_covar}. The input required for this is the galaxy power spectrum, which we obtain from the data for both redshift regions as described further below. Following \cite{Krowleski_No_PV_det}, the analytic covariance is computed in 18 equally sized radial bins between lengths of \SI{20}{\per\hHubble\Mpc} and \SI{160}{\per\hHubble\Mpc}. The number of radial bins is the main factor driving the cost of the computation. Additional numerical parameters, \textit{e.\,g.}\ concerning the numerical integration, are similarly left as described in \cite{hou_covar,Krowleski_No_PV_det}. We note that recent work \cite{ortola_cov_I, ortola_cov_II} has derived the next-to-leading order contributions to the 4PCF covariance matrix; exploring how the inclusion of these higher-order terms affects the results presented here will be an interesting direction for future work.

\subsubsection{Power Spectrum}

We measured the Y1 LRG power spectrum on both the $0.4 < z < 0.8$ and the $0.4 < z < 1.1$ regions, in the NGC. We had to do this measurement because the DESI Y1 cosmology papers used a $\theta$-cut to reduce the impact of fiber assignment \cite{Pinon24}. However, this cut depends on the angular separation of each galaxy pair on the sky, and therefore requires knowledge of the absolute orientation of the pair with respect to the observer. We cannot track the absolute orientation of the tetrahedron in our current 4PCF algorithm, so we cannot impose an analogous cut in our measurement. Thus we want the power spectrum with no $\theta$-cut, and so we must remeasure it.

However, the measured power spectrum is noisy, and we would like a smooth model power spectrum as an input to the covariance (otherwise the noise would cause problems in the matrix inversion). We therefore fit an EFT model to the measured power spectrum using \textsc{velocileptors}\footnote{\url{https://github.com/sfschen/velocileptors}} \cite{Chen20a,Chen20b} wrapped through \textsc{desilike}\footnote{\url{https://github.com/cosmodesi/desilike}}. However, the EFT terms blow up at high $k$, and so we fit only on the region up to $k_{\mathrm{max}} = \SI{0.5}{\hHubble\per\Mpc}$ (see Fig.~15 of \cite{Slepian_Parity_Meas_DESI}). 

We then fit a power law to the average of the \textsc{Abacus} AltMTL mocks above this $k_{\mathrm{max}}$, as the average of the mocks was sufficiently smooth and should be faithful to the high-$k$ behavior, as they are $N$-body and also have the most accurate fiber assignment method applied to them.

\subsubsection{Shot Noise}

The analytic covariance contains two unknown parameters: the shot noise $1/\bar{n}$, with $\bar{n}$ the number density, and the effective volume $V_{\mathrm{eff}}$. One may choose a shot noise value other than the nominal one to better allow the analytical template to capture structure in the (noisy and non-invertible, but presumably more accurate) empirical covariance. This method was first suggested in \cite{xu_2012} for the 2PCF, and then applied to  the 3PCF in \cite{se_3pt_alg, se_boss_3pcf, se_3pcf_bao} and to the 4PCF of data in \cite{phil_4pcf} in the even-parity sector and \cite{hou_covar} in the odd sector. We do this by optimizing the likelihood $\mathcal{L}$, in terms of $\bar{n}$ and $V_{\mathrm{eff}}$:
\begin{equation}
    - \log\mleft(\mathcal{L}_1(\bar{n}, V_{\mathrm{eff}})\mright) = \frac{N_{\mathrm{mock}}}{2} \mleft[\mathrm{Tr}\mleft( \mathbf{C}_{\mathrm{model}}^{-1}(\bar{n}, V_{\mathrm{eff}})\,\mathbf{C}_{\mathrm{mock}} \mright) - \log\mleft( \det\mleft(\mathbf{C}_{\mathrm{model}}^{-1}(\bar{n}, V_{\mathrm{eff}})\mright) \mright) \mright] + \dots
\end{equation}
with
\begin{equation}
    V_{\mathrm{eff}} \equiv
    \frac{V_{\mathrm{fid}}\, N_{\mathrm{dof}}}{\mathrm{Tr}\mleft( \mathbf{C}_{\mathrm{ana}}^{-1} \mathbf{C}_{\mathrm{mock}} \mright)},
\end{equation}
where $V_{\mathrm{fid}}$ is the fiducial survey volume and $N_{\mathrm{dof}}$ is the dimension of the compressed data vector. The optimization is performed with $\bar{n}$ ranging from \SIrange{0.5e-4}{3e-4}{\hHubble\cubed\per\Mpc\cubed}, using at each $\bar{n}$ the optimal $V_{\mathrm{eff}}$. We also optimize each of the patches, regions, and full galactic caps to match the \textsc{Abacus} AltMTL distribution; the NGC and SGC optimization is done separately given their independent level of completeness.

\section{Computational Cost}
\label{Sec:Comp_Cost}

We now summarize the computational cost of the results reported in this work. For each of the data subsets described in \cref{tab:region-summary}, we ran on \num{1000} \textsc{EZMocks} and 25 \textsc{Abacus} mocks; these latter had both FFA and AltMTL, so they count as 50 rather than 25. 

Including the data, we thus have \num{1051} catalogs. Each 4PCF run takes about 45 minutes of wall time on one A100 GPU, including all of the randoms required; thus, overall the cost was 800 hours for each subset. Since we have 12 subsets, the total cost was roughly \num{10000} GPU hours. These computations were all performed on the University of Florida's HiPerGator 3.0, and as noted already, using NVIDIA A100 GPUs.

Computing the analytic covariance (see \cref{Sec:Cov-analytic}), including both the parity-odd and parity-even parts, required around \num{5000} CPU hours for each set of inputs using the public \textsc{Analytic4PC} code introduced in \cite{Krowleski_No_PV_det}.
As the current implementation of the code focuses on clarity and simplicity of implementation (although it does provide for scalable computations across many CPU cores), the cost could easily be reduced by exploiting symmetries in the indices of the analytic covariance.
However, since the absolute cost of calculating the analytic covariance was low enough not to be an impediment, we used the existing code as-is without modifications.

\section{Variance Scaling in Auto and Cross}

\begin{figure}
    \centering
    \includegraphics[width=0.5\textwidth]{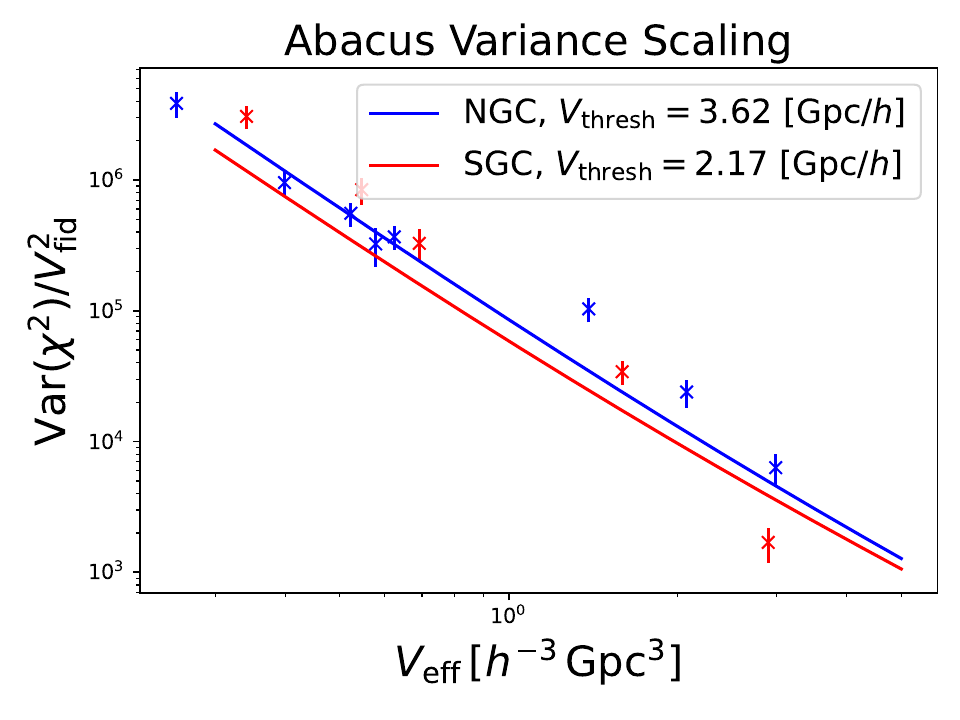}%
    \includegraphics[width=0.5\textwidth]{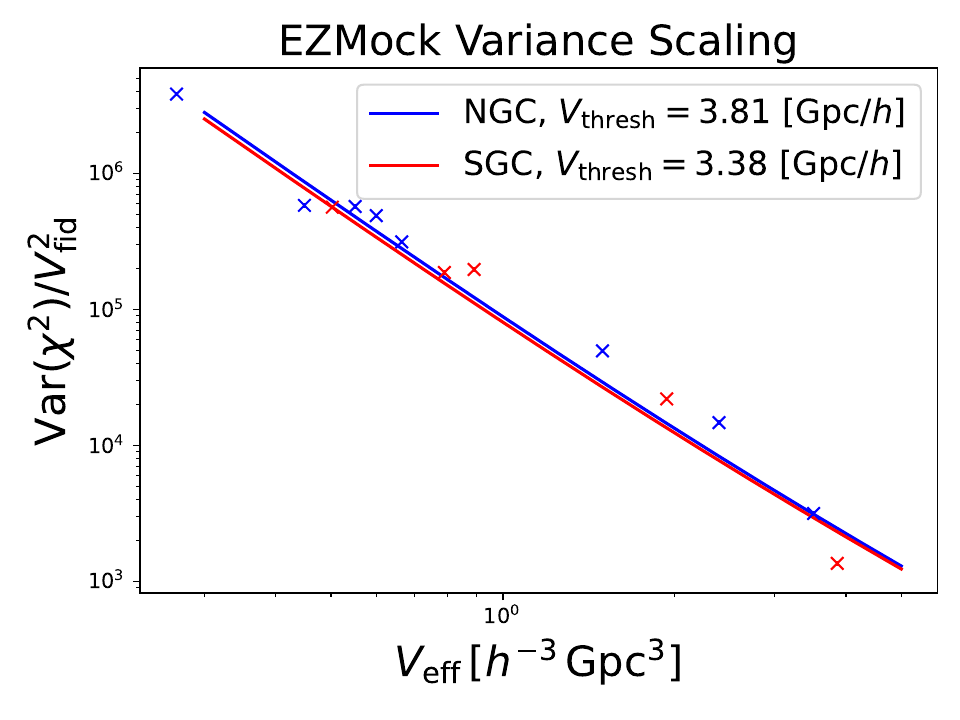}

    \caption{Scaling of the variance of the $\chi^2$ distribution with effective volume, $V_{\mathrm{eff}}$, for the DESI Y1 LRG sample. The left panel shows results obtained from the \textsc{Abacus} mocks, while the right panel shows the corresponding results from the \textsc{EZMocks}. The crosses in each panel correspond to the mocks' measurement, and solid curves show the theoretical scaling relation of \cite{Krowleski_No_PV_det}, which includes a single free parameter, $V_{\mathrm{thresh}}$. Error-bars in the \textsc{Abacus} panel indicate the $1\sigma$ uncertainty arising from the finite number of mocks. Error-bars become small compared to the size of the \textsc{EZMocks}' crosses due to the large number of realizations.}
    \label{fig:Variance_Scailing}
\end{figure}

The variance of the $\chi^2$ statistic across the mock realizations is not determined solely by the analytic covariance used to construct the statistic. It also depends on survey geometry, nonlinear mode coupling, window-function effects, and finite-mock noise. \cite{Krowleski_No_PV_det} derived a phenomenological model for the dependence of this variance on the effective survey volume, $V_{\mathrm{eff}}$, defined in their Eq.~(B.1). Their result can be written as
\begin{equation}
    \frac{\mathrm{Var}(\chi^2)}{V_{\mathrm{fid}}^2}
    \simeq
    \frac{2N_{\mathrm{dof}}}{V_{\mathrm{eff}}^2}
    \left(
        1+\frac{V_{\mathrm{thresh}}}{V_{\mathrm{eff}}}
    \right),
\end{equation}
where the first term represents the Gaussian contribution, which scales as $V_{\mathrm{eff}}^{-2}$, while the second represents a non-Gaussian finite-volume correction scaling as $V_{\mathrm{eff}}^{-3}$. The free parameter $V_{\mathrm{thresh}}$ specifies the effective volume at which the Gaussian and non-Gaussian contributions are equal. Consequently, the non-Gaussian contribution becomes increasingly important for $V_{\mathrm{eff}}<V_{\mathrm{thresh}}$.

In \cref{fig:Variance_Scailing}, each point shows the variance of the $\chi^2$ distribution measured from the mock realizations for one survey region, while the solid curves show fits to the scaling relation above. Both the \textsc{Abacus} and \textsc{EZMock} measurements exhibit the expected decrease in variance with increasing effective volume. The fitted values of $V_{\mathrm{thresh}}$ are of order a few \si{\per\hHubble\cubed\Gpc\cubed}, indicating that non-Gaussian finite-volume contributions remain relevant over the range of volumes considered. The differences between the fitted values for the mock families and sky regions quantify differences in the effective strength of these non-Gaussian contributions, with \textsc{Abacus} preferring a smaller $V_{\mathrm{thresh}}$ than \textsc{EZMocks}, particularly in the SGC.

\section{Systematics Tests}
\label{Sec:Systematics}

To assess the robustness of our measurements to known observational systematics, we repeat both the $T^2$ and $\chi^2$ analyses after selectively disabling two key DESI weights: the imaging systematics weights and the redshift failure ($z$-fail) weights. For each of these systematics, we recompute the detection significance in the full NGC and SGC samples while keeping all other aspects of the analysis fixed, and we report the resulting values in \cref{tab:T2_Sys_Tests_Summary,tab:chi2_Sys_Tests_Summary}. As in the fiducial analysis, we present results both including small tetrahedra (``ST'') and excluding them (``no-ST''). Any substantial deviation from the fiducial detection significance values (\cref{tab:T2_summary_Table,tab:Chi2_summary_Table}) indicates that the removed weight contributes notably to the measured signal. In contrast, impact of the detection significance implies that the corresponding systematic does not bias our conclusions. By conducting the same suite of systematic toggles for both estimators, we ensure that any claimed detection is not an artifact of residual imaging variations or redshift measurement failures.\footnote{%
    Figures of results are made available online, see \cite{zenodo_complete_results}.%
}

\begin{table}
    \centering
    {\textbf{\boldmath Tests of Systematics: Detection Significance for the $T^2$ Analyses}}

    \smallskip

    % ---- COLUMN GROUP HEADERS ----
    \begin{tabular}{l @{\hspace*{6pt}} c *{8}{S[table-format=2.1]}}
        \toprule
        \textbf{Systematic} & \textbf{Analysis}
        & \multicolumn{2}{c}{\textbf{NGC}} 
        & \multicolumn{2}{c}{\textbf{SGC}} 
        & \multicolumn{2}{c}{\textbf{NGC $\times$ SGC}}   & \multicolumn{2}{c}{\textbf{NGC $+$ SGC}}
        \\
        %& & & & & & \multicolumn{2}{c}{$N(N-1)/2$} & \\
        & & {no-ST} & {ST} & {no-ST} & ST & {no-ST} & {ST} & {no-ST} & {ST} \\
        \midrule
        % ---- ROWS ----
        Imaging & Auto & 3.7 & 16.3 & 5.0 & 3.2 & {---} & {---} & 6.2 & 16.6 \\
        Imaging & Cross & {---} & {---} & {---} & {---} & 2.8 & 8.6 & {---} & {---} \\
        $z$-Fail & Auto & 3.8 & 16.5 & 4.4 & 3.1 & {---} & {---} & 5.8 & 16.8 \\
        $z$-Fail & Cross & {---} & {---} & {---} & {---} & 2.7 & 8.6 & {---} & {---} \\
        \bottomrule
    \end{tabular}

    \caption{Detection significance of the $T^2$ analysis when the imaging systematics weights and redshift failure ($z$-Fail) weights are individually turned off, used as a diagnostic test of potential systematics in the Full NGC and SGC samples. Columns labeled ``ST'' include small tetrahedra, while ``no-ST'' excludes them. For comparison, the corresponding fiducial detection significances (with all weights applied) are listed in \cref{tab:T2_summary_Table}. We find the detection significances reported here to be consistent, within $1\sigma$, of those in the fiducial analysis.}
    \label{tab:T2_Sys_Tests_Summary}

    \bigskip

    \centering
    {\textbf{Tests of Systematics: Detection Significance for the $\chi^2$ Analyses}}

    \smallskip

    % ---- COLUMN GROUP HEADERS ----
    \begin{tabular}{l @{\hspace*{6pt}} c *{8}{S[table-format=2.1]}}
        \toprule
        \textbf{Systematic} & \textbf{Analysis}
        & \multicolumn{2}{c}{\textbf{NGC}} 
        & \multicolumn{2}{c}{\textbf{SGC}} 
        & \multicolumn{2}{c}{\textbf{NGC $\times$ SGC}}   & \multicolumn{2}{c}{\textbf{NGC $+$ SGC}}
        \\
        %& & & & & & \multicolumn{2}{c}{$N(N-1)/2$} & \\
        & & {no-ST} & {ST} & {no-ST} & {ST} & {no-ST} & {ST} & {no-ST} & {ST} \\
        \midrule
        % ---- ROWS ----
        Imaging & Auto & -1.0 & 6.4 & 2.9 & 5.1 & {---} & {---} & 3.1 & 8.2 \\
        Imaging & Cross & {---} & {---} & {---} & {---}  & 1.4 & 17.6 & {---} & {---} \\
        $z$-Fail & Auto & 0.2 & 7.1 & 2.4 & 4.7 & {---} & {---} & 2.4 & 8.5 \\
        $z$-Fail & Cross & {---} & {---} & {---} & {---}  & 1.3 & 17.2 & {---} & {---} \\
        \bottomrule
    \end{tabular}

    \caption{Detection significance of the $\chi^2$ analysis when the imaging systematics weights and redshift failure ($z$-Fail) weights are individually turned off, used as a diagnostic test of potential systematics in the Full NGC and SGC samples. Columns labeled ``ST'' include small tetrahedra, while ``no-ST'' excludes them. For comparison, the corresponding fiducial detection significances (with all weights applied) are listed in \cref{tab:Chi2_summary_Table}. We find the detection significances reported here to be consistent, within $1\sigma$, of those in the fiducial analysis.}
    \label{tab:chi2_Sys_Tests_Summary}
\end{table}

In addition to the weight-based systematics tests described above, we perform a dedicated consistency check motivated by an unexpected behavior observed in the patch-based analyses. Specifically, for the auto-correlation measurements we find that, in some cases, the quadrature sum of the detection significances measured in the individual DESI Year-1 footprint patches exceeds the detection significance obtained from the corresponding full Galactic cap (NGC or SGC). Since the full Galactic cap contains a larger effective volume, it should generically yield an equal or higher detection significance than the quadrature combination of its subregions. We attribute this behavior to the fact that our detection significances are normalized using the standard deviation estimated from \textsc{Abacus} mocks, which include box replication; this replication can artificially suppress the mock-to-mock variance on patch scales, leading to inflated patch-level significances when combined in quadrature.

To verify that this effect is not intrinsic to the DESI data or indicative of an unaccounted-for systematic, we repeat the full analysis using \textsc{EZMocks}, which do not employ replication. Treating each of the \num{1000} \textsc{EZMocks} as mock ``data,'' we compute, for each realization, the ratio between the detection significance measured in the full Galactic cap and the quadrature sum of the detection significances measured in the corresponding patches. We then construct the distribution of this ratio across the \num{1000} \textsc{EZMocks} realizations and compare it to the analogous ratio measured from the DESI data.

\begin{figure}
    \centering
    \includegraphics[width=0.5\textwidth]{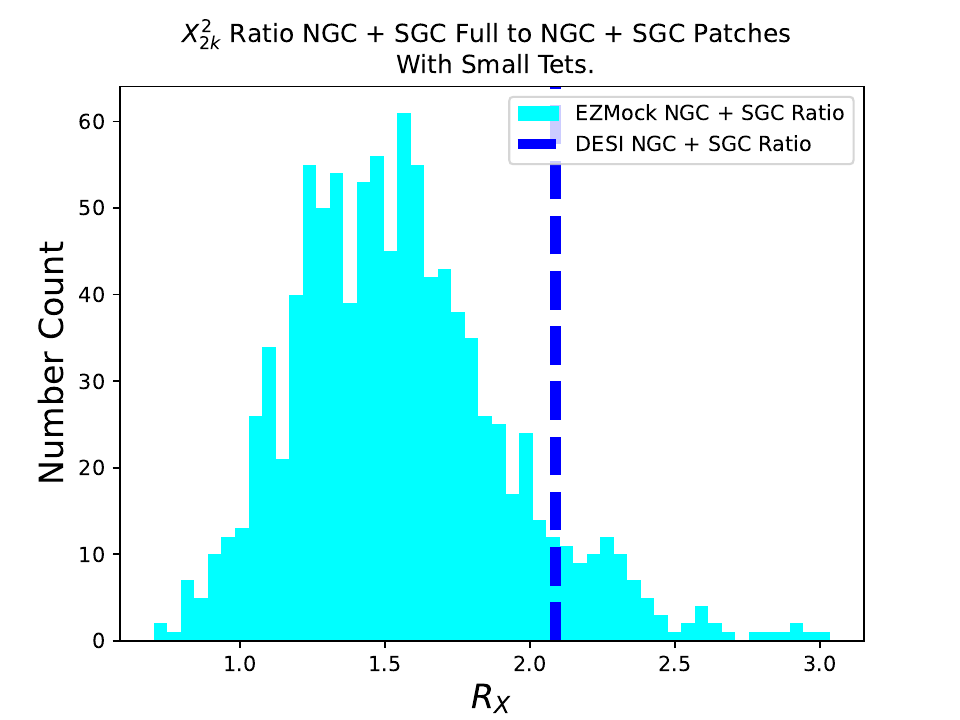}%
    \includegraphics[width=0.5\textwidth]{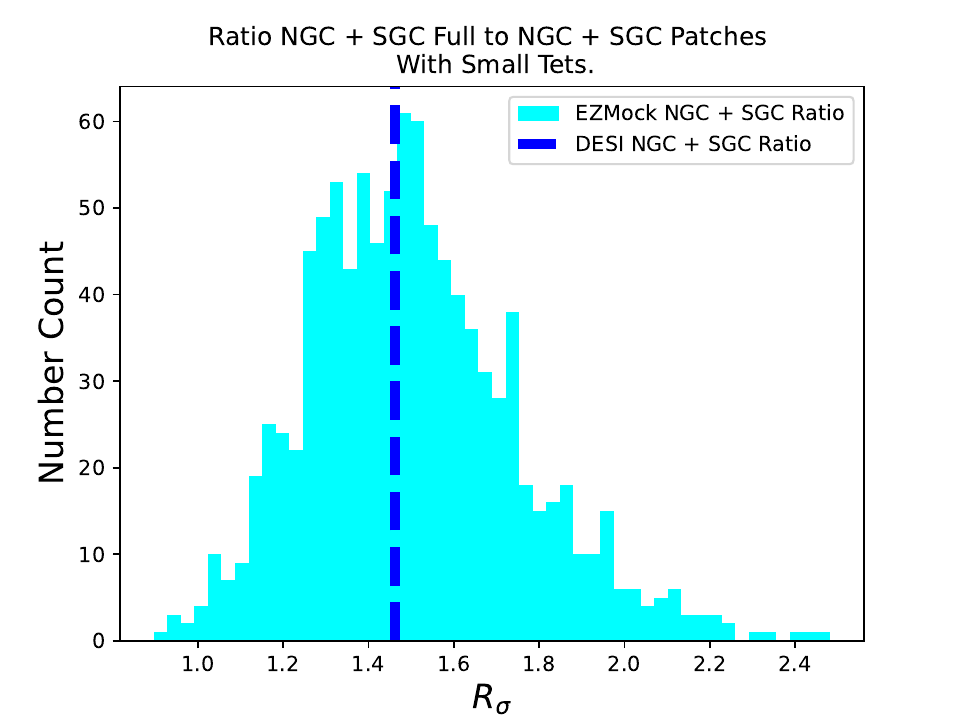}

    \caption{Distribution of the ratio between the detection significance measured in the full NGC~$+$~SGC and the quadrature sum of the patch-level detection significances, obtained from \num{1000} \textsc{EZMocks} realizations (cyan histograms), compared to the DESI measurement (blue dashed line). The left panel shows the analogous ratio constructed after combining patch-level $p$-values using Fisher's method, while the right panel shows the ratio formed directly from the detection significances. In both cases, the DESI result is consistent with the \textsc{EZMocks} distributions, indicating that the observed patch-sum behavior is not driven by any issues with the data.}
    \label{fig:Patch_vs_Cap_Ratio}
\end{figure}

We perform this ratio test in two complementary ways. First, we form the ratio directly using the detection significances themselves. Second, we convert the detection significances to $p$-values and combine regions using Fisher's method \cite{Fisher_Method_Brown},
\begin{equation}
    X^{2}_{2k} = -2 \sum_{i=1}^{k} \ln(p_i)\,,
\end{equation}
where $k$ is the number of regions being combined (with $k=1$ for a full Galactic cap and $k>1$ for the patch-based combinations). After converting the resulting $X^{2}_{2k}$ to an equivalent combined detection significance, we again form the ratio between the full-cap and patch-level results. In the absence of pathological behavior, one expects these ratios to be unity or greater; however, if values below unity arise, consistency between DESI and \textsc{EZMocks} would indicate that the effect is driven by the covariance normalization rather than by the data themselves. We find that the DESI ratios are fully consistent with the \textsc{EZMocks} distributions in both approaches, as shown in \cref{fig:Patch_vs_Cap_Ratio}, confirming that the observed behavior is an expected consequence of the \textsc{Abacus} mock replication and does not impact the robustness of our conclusions.

For clarity, in \cref{fig:Patch_vs_Cap_Ratio} we present these consistency checks in terms of the following ratios:
\begin{equation}
    R_{X} \equiv \frac{X^{2}_{2k}\big|_{\mathrm{GC}}}{X^{2}_{2k}\big|_{\mathrm{P}}}, 
    \qquad
    R_{\sigma} \equiv \frac{\sigma\big|_{\mathrm{GC}}}{\sigma\big|_{\mathrm{P}}}\,,
\end{equation}
where ``GC'' denotes the value obtained from the full Galactic cap under consideration (NGC, SGC, or their combination), and ``P'' denotes the corresponding result obtained by combining the individual footprint patches. Here, $R_X$ corresponds to the ratio formed using the Fisher's-method combination of $p$-values, while $R_\sigma$ corresponds to the ratio obtained by directly comparing the detection significances.

% Final appendix containing all author affiliations.

\section{Author affiliations}
\label{app:author-affiliations}

\begingroup
\small
\begin{list}{}{
  \setlength{\leftmargin}{0.65cm}
  \setlength{\labelsep}{0.08cm}
  \setlength{\itemsep}{2pt}
  \setlength{\parsep}{0pt}
  \setlength{\topsep}{2pt}
}
\item[$^{1}$] Department of Astronomy, University of Florida, 211 Bryant Space Science Center, Gainesville, FL 32611, USA
\item[$^{2}$] Waterloo Centre for Astrophysics, University of Waterloo, 200 University Ave W, Waterloo, ON N2L 3G1, Canada
\item[$^{3}$] Department of Physics and Astronomy, University of Waterloo, 200 University Ave W, Waterloo, ON N2L 3G1, Canada
\item[$^{4}$] California Institute of Technology, 1200~East California Boulevard, Pasadena, CA 91125, USA
\item[$^{*}$] CITA National Fellow
\item[$^{5}$] Fakultät für Physik, Universität Bielefeld, Universitätsstraße~25, 33615 Bielefeld, Germany
\item[$^{6}$] Lawrence Berkeley National Laboratory, 1 Cyclotron Road, Berkeley, CA 94720, USA
\item[$^{7}$] Department of Physics, Boston University, 590 Commonwealth Avenue, Boston, MA 02215 USA
\item[$^{8}$] University of Michigan, 500 S. State Street, Ann Arbor, MI 48109, USA
\item[$^{9}$] Instituto Avanzado de Cosmolog\'{\i}a A.~C., San Marcos 11 - Atenas 202. Magdalena Contreras. Ciudad de M\'{e}xico C.~P.~10720, M\'{e}xico
\item[$^{10}$] Instituto de Ciencias F\'{\i}sicas, Universidad Nacional Aut\'onoma de M\'exico, Av. Universidad s/n, Cuernavaca, Morelos, C.~P.~62210, M\'exico
\item[$^{11}$] Institute for Astronomy, University of Edinburgh, Royal Observatory, Blackford Hill, Edinburgh EH9 3HJ, UK
\item[$^{12}$] Dipartimento di Fisica ``Aldo Pontremoli'', Universit\`a degli Studi di Milano, Via Celoria 16, I-20133 Milano, Italy
\item[$^{13}$] INAF-Osservatorio Astronomico di Brera, Via Brera 28, 20122 Milano, Italy
\item[$^{14}$] Department of Physics \& Astronomy, University College London, Gower Street, London, WC1E 6BT, UK
\item[$^{15}$] Departamento de Astrof\'{\i}sica, Universidad de La Laguna (ULL), E-38206, La Laguna, Tenerife, Spain
\item[$^{16}$] Instituto de Astrof\'{\i}sica de Canarias, C/ V\'{\i}a L\'{a}ctea, s/n, E-38205 La Laguna, Tenerife, Spain
\item[$^{17}$] Instituto de F\'{\i}sica, Universidad Nacional Aut\'{o}noma de M\'{e}xico,  Circuito de la Investigaci\'{o}n Cient\'{\i}fica, Ciudad Universitaria, Cd. de M\'{e}xico  C.~P.~04510,  M\'{e}xico
\item[$^{18}$] Department of Astronomy \& Astrophysics, University of Toronto, Toronto, ON M5S 3H4, Canada
\item[$^{19}$] Department of Physics \& Astronomy and Pittsburgh Particle Physics, Astrophysics, and Cosmology Center (PITT PACC), University of Pittsburgh, 3941 O'Hara Street, Pittsburgh, PA 15260, USA
\item[$^{20}$] University of California, Berkeley, 110 Sproul Hall \#5800 Berkeley, CA 94720, USA
\item[$^{21}$] Departamento de F\'isica, Universidad de los Andes, Cra. 1 No. 18A-10, Edificio Ip, CP 111711, Bogot\'a, Colombia
\item[$^{22}$] Observatorio Astron\'omico, Universidad de los Andes, Cra. 1 No. 18A-10, Edificio H, CP 111711 Bogot\'a, Colombia
\item[$^{23}$] Institut d'Estudis Espacials de Catalunya (IEEC), c/ Esteve Terradas 1, Edifici RDIT, Campus PMT-UPC, 08860 Castelldefels, Spain
\item[$^{24}$] Institute of Cosmology and Gravitation, University of Portsmouth, Dennis Sciama Building, Portsmouth, PO1 3FX, UK
\item[$^{25}$] Institute of Space Sciences, ICE-CSIC, Campus UAB, Carrer de Can Magrans s/n, 08913 Bellaterra, Barcelona, Spain
\item[$^{26}$] University of Virginia, Department of Astronomy, Charlottesville, VA 22904, USA
\item[$^{27}$] Fermi National Accelerator Laboratory, PO Box 500, Batavia, IL 60510, USA
\item[$^{28}$] Center for Cosmology and AstroParticle Physics, The Ohio State University, 191 West Woodruff Avenue, Columbus, OH 43210, USA
\item[$^{29}$] Department of Physics, The Ohio State University, 191 West Woodruff Avenue, Columbus, OH 43210, USA
\item[$^{30}$] The Ohio State University, Columbus, 43210 OH, USA
\item[$^{31}$] Department of Physics, University of Michigan, 450 Church Street, Ann Arbor, MI 48109, USA
\item[$^{32}$] Department of Physics, The University of Texas at Dallas, 800 W. Campbell Rd., Richardson, TX 75080, USA
\item[$^{33}$] NSF NOIRLab, 950 N. Cherry Ave., Tucson, AZ 85719, USA
\item[$^{34}$] Department of Physics, Southern Methodist University, 3215 Daniel Avenue, Dallas, TX 75275, USA
\item[$^{35}$] Department of Physics and Astronomy, University of California, Irvine, 92697, USA
\item[$^{36}$] Sorbonne Universit\'{e}, CNRS/IN2P3, Laboratoire de Physique Nucl\'{e}aire et de Hautes Energies (LPNHE), FR-75005 Paris, France
\item[$^{37}$] Departament de F\'{i}sica, Serra H\'{u}nter, Universitat Aut\`{o}noma de Barcelona, 08193 Bellaterra (Barcelona), Spain
\item[$^{38}$] Institut de F\'{i}sica d’Altes Energies (IFAE), The Barcelona Institute of Science and Technology, Edifici Cn, Campus UAB, 08193, Bellaterra (Barcelona), Spain
\item[$^{39}$] Instituci\'{o} Catalana de Recerca i Estudis Avan\c{c}ats, Passeig de Llu\'{\i}s Companys, 23, 08010 Barcelona, Spain
\item[$^{40}$] Instituto de Estudios Astrof\'isicos, Facultad de Ingenier\'ia y Ciencias, Universidad Diego Portales, Av. Ej\'ercito Libertador 441, Santiago, Chile
\item[$^{41}$] Steward Observatory, University of Arizona, 933 N. Cherry Avenue, Tucson, AZ 85721, USA
\item[$^{42}$] Perimeter Institute for Theoretical Physics, 31 Caroline St. North, Waterloo, ON N2L 2Y5, Canada
\item[$^{43}$] Instituto de Astrof\'{i}sica de Andaluc\'{i}a (CSIC), Glorieta de la Astronom\'{i}a, s/n, E-18008 Granada, Spain
\item[$^{44}$] Departament de F\'isica, EEBE, Universitat Polit\`ecnica de Catalunya, c/Eduard Maristany 10, 08930 Barcelona, Spain
\item[$^{45}$] Universit\'{e} Clermont-Auvergne, CNRS, LPCA, 63000 Clermont-Ferrand, France
\item[$^{46}$] Queensland University of Technology,  School of Chemistry \& Physics, George St, Brisbane 4001, Australia
\item[$^{47}$] Abastumani Astrophysical Observatory, Tbilisi, GE-0179, Georgia
\item[$^{48}$] Department of Physics, Kansas State University, 116 Cardwell Hall, Manhattan, KS 66506, USA
\item[$^{49}$] CIEMAT, Avenida Complutense 40, E-28040 Madrid, Spain
\item[$^{50}$] Max Planck Institute for Extraterrestrial Physics, Gie\ss enbachstra\ss e 1, 85748 Garching, Germany
\end{list}
\endgroup

\FloatBarrier

\bibliographystyle{JHEP}
\bibliography{references}

@article{zenodo_complete_results,
  author       = {Ortolá Leonard, William and
                  Krolewski, Alex and
                  Slepian, Zachary and
                  Greco, Alessandro and
                  May, Simon},
  title        = {Complete results for “{A} Finely-Binned Measurement
                   of the Connected Even-Parity Galaxy 4-Point
                   Correlation Function of {DESI Year 1} {Luminous Red
                   Galaxies}”
                  },
  month        = may,
  year         = 2026,
  publisher    = {Zenodo},
  doi          = {10.5281/zenodo.20185242},
  url          = {https://doi.org/10.5281/zenodo.20185242},
  journal = {Zenodo},
  pages = {doi:10.5281/zenodo.20185242}
}

@ARTICLE{Chen20b,
       author = {{Chen}, Shi-Fan and {Vlah}, Zvonimir and {Castorina}, Emanuele and {White}, Martin},
        title = "{Redshift-space distortions in Lagrangian perturbation theory}",
      journal = {\jcap},
         year = 2021,
        month = mar,
       volume = {2021},
       number = {3},
          eid = {100},
        pages = {100},
          doi = {10.1088/1475-7516/2021/03/100},
archivePrefix = {arXiv},
       eprint = {2012.04636},
 primaryClass = {astro-ph.CO},
       adsurl = {https://ui.adsabs.harvard.edu/abs/2021JCAP...03..100C}
}

@ARTICLE{Chen20a,
       author = {{Chen}, Shi-Fan and {Vlah}, Zvonimir and {White}, Martin},
        title = "{Consistent modeling of velocity statistics and redshift-space distortions in one-loop perturbation theory}",
      journal = {\jcap},
         year = 2020,
        month = jul,
       volume = {2020},
       number = {7},
          eid = {062},
        pages = {062},
          doi = {10.1088/1475-7516/2020/07/062},
archivePrefix = {arXiv},
       eprint = {2005.00523},
 primaryClass = {astro-ph.CO},
       adsurl = {https://ui.adsabs.harvard.edu/abs/2020JCAP...07..062C}
}

@ARTICLE{Pinon24,
       author = {{Pinon}, M. and {de Mattia}, A. and {McDonald}, P. and {Burtin}, E. and {Ruhlmann-Kleider}, V. and {White}, M. and {Bianchi}, D. and {Ross}, A.~J. and {Aguilar}, J. and {Ahlen}, S. and {Brooks}, D. and {Cahn}, R.~N. and {Chaussidon}, E. and {Claybaugh}, T. and {Cole}, S. and {de la Macorra}, A. and {Dey}, B. and {Doel}, P. and {Fanning}, K. and {Forero-Romero}, J.~E. and {Gazta{\~n}aga}, E. and {Gontcho A Gontcho}, S. and {Howlett}, C. and {Kirkby}, D. and {Kisner}, T. and {Kremin}, A. and {Lambert}, A. and {Landriau}, M. and {Lasker}, J. and {Le Guillou}, L. and {Levi}, M.~E. and {Manera}, M. and {Martini}, P. and {Meisner}, A. and {Miquel}, R. and {Moustakas}, J. and {Myers}, A.~D. and {Niz}, G. and {Palanque-Delabrouille}, N. and {Percival}, W.~J. and {Poppett}, C. and {Rossi}, G. and {Sanchez}, E. and {Schlegel}, D. and {Schubnell}, M. and {Seo}, H. and {Sprayberry}, D. and {Tarl{\'e}}, G. and {Vargas-Maga{\~n}a}, M. and {Weaver}, B.~A. and {Zarrouk}, P. and {Zhou}, R. and {Zou}, H.},
        title = "{Mitigation of DESI fiber assignment incompleteness effect on two-point clustering with small angular scale truncated estimators}",
      journal = {\jcap},
         year = 2025,
        month = jan,
       volume = {2025},
       number = {1},
          eid = {131},
        pages = {131},
          doi = {10.1088/1475-7516/2025/01/131},
archivePrefix = {arXiv},
       eprint = {2406.04804},
 primaryClass = {astro-ph.CO},
       adsurl = {https://ui.adsabs.harvard.edu/abs/2025JCAP...01..131P}
}

@ARTICLE{phil_uchuu,
       author = {{Philcox}, O.~H.~E. and {Ereza}, J.},
        title = "{Could sample variance be responsible for the parity-violating signal seen in the Baryon Oscillation Spectroscopic Survey?}",
      journal = {Philosophical Transactions of the Royal Society of London Series A},
         year = 2025,
        month = feb,
       volume = {383},
       number = {2290},
          eid = {20240034},
        pages = {20240034},
          doi = {10.1098/rsta.2024.0034},
archivePrefix = {arXiv},
       eprint = {2401.09523},
 primaryClass = {astro-ph.CO},
       adsurl = {https://ui.adsabs.harvard.edu/abs/2025RSPTA.38340034P}
}

@ARTICLE{uchuu_rep,
       author = {{Ereza}, Julia and {Prada}, Francisco and {Klypin}, Anatoly and {Ishiyama}, Tomoaki and {Smith}, Alex and {Baugh}, Carlton M. and {Li}, Baojiu and {Hern{\'a}ndez-Aguayo}, C{\'e}sar and {Ruedas}, Jos{\'e}},
        title = "{The UCHUU-GLAM BOSS and eBOSS LRG lightcones: exploring clustering and covariance errors}",
      journal = {\mnras},
         year = 2024,
        month = aug,
       volume = {532},
       number = {2},
        pages = {1659-1682},
          doi = {10.1093/mnras/stae1543},
archivePrefix = {arXiv},
       eprint = {2311.14456},
 primaryClass = {astro-ph.CO},
       adsurl = {https://ui.adsabs.harvard.edu/abs/2024MNRAS.532.1659E}
}

@ARTICLE{gouyou,
       author = {{Gouyou Beauchamps}, S. and {Baratta}, P. and {Escoffier}, S. and {Gillard}, W. and {Bel}, J. and {Bautista}, J. and {Carbone}, C.},
        title = "{Cosmological inference including massive neutrinos from the matter power spectrum: Biases induced by uncertainties in the covariance matrix}",
      journal = {\aap},
         year = 2025,
        month = jan,
       volume = {693},
          eid = {A226},
        pages = {A226},
          doi = {10.1051/0004-6361/202347164},
archivePrefix = {arXiv},
       eprint = {2306.05988},
 primaryClass = {astro-ph.CO},
       adsurl = {https://ui.adsabs.harvard.edu/abs/2025A&A...693A.226G}
}

@ARTICLE{shiraishii,
       author = {{Shiraishi}, Maresuke},
        title = "{Parity violation in the CMB trispectrum from the scalar sector}",
      journal = {\prd},
         year = 2016,
        month = oct,
       volume = {94},
       number = {8},
          eid = {083503},
        pages = {083503},
          doi = {10.1103/PhysRevD.94.083503},
archivePrefix = {arXiv},
       eprint = {1608.00368},
 primaryClass = {astro-ph.CO},
       adsurl = {https://ui.adsabs.harvard.edu/abs/2016PhRvD..94h3503S}
}

@ARTICLE{Gil-Marin_2017_RSDMeasurement,
       author = {{Gil-Mar{\'\i}n}, H{\'e}ctor and {Percival}, Will J. and {Verde}, Licia and {Brownstein}, Joel R. and {Chuang}, Chia-Hsun and {Kitaura}, Francisco-Shu and {Rodr{\'\i}guez-Torres}, Sergio A. and {Olmstead}, Matthew D.},
        title = "{The clustering of galaxies in the SDSS-III Baryon Oscillation Spectroscopic Survey: RSD measurement from the power spectrum and bispectrum of the DR12 BOSS galaxies}",
      journal = {\mnras},
         year = 2017,
        month = feb,
       volume = {465},
       number = {2},
        pages = {1757-1788},
          doi = {10.1093/mnras/stw2679},
archivePrefix = {arXiv},
       eprint = {1606.00439},
 primaryClass = {astro-ph.CO},
       adsurl = {https://ui.adsabs.harvard.edu/abs/2017MNRAS.465.1757G}
}

@BOOK{Zwicky_1961_CatalogueGalaxies,
       author = {{Zwicky}, F. and {Herzog}, E. and {Wild}, P. and {Karpowicz}, M. and {Kowal}, C.~T.},
        title = "{Catalogue of galaxies and of clusters of galaxies, Vol. I}",
         year = 1961,
       adsurl = {https://ui.adsabs.harvard.edu/abs/1961cgcg.book.....Z},
    publisher={California Institute of Technology}
}

@ARTICLE{moresco_3pcf_bao,
       author = {{Moresco}, Michele and {Veropalumbo}, Alfonso and {Marulli}, Federico and {Moscardini}, Lauro and {Cimatti}, Andrea},
        title = "{C$^{\text{3}}$: Cluster Clustering Cosmology. II. First Detection of the Baryon Acoustic Oscillations Peak in the Three-point Correlation Function of Galaxy Clusters}",
      journal = {\apj},
         year = 2021,
        month = oct,
       volume = {919},
       number = {2},
          eid = {144},
        pages = {144},
          doi = {10.3847/1538-4357/ac10c9},
archivePrefix = {arXiv},
       eprint = {2011.04665},
 primaryClass = {astro-ph.CO},
       adsurl = {https://ui.adsabs.harvard.edu/abs/2021ApJ...919..144M}
}

@ARTICLE{pearson_bao_bispec,
       author = {{Pearson}, David W. and {Samushia}, Lado},
        title = "{A Detection of the Baryon Acoustic Oscillation features in the SDSS BOSS DR12 Galaxy Bispectrum}",
      journal = {\mnras},
         year = 2018,
        month = aug,
       volume = {478},
       number = {4},
        pages = {4500-4512},
          doi = {10.1093/mnras/sty1266},
archivePrefix = {arXiv},
       eprint = {1712.04970},
 primaryClass = {astro-ph.CO},
       adsurl = {https://ui.adsabs.harvard.edu/abs/2018MNRAS.478.4500P}
}

@ARTICLE{gaztanaga_2009_3pcf,
       author = {{Gazta{\~n}aga}, Enrique and {Cabr{\'e}}, Anna and {Castander}, Francisco and {Crocce}, Martin and {Fosalba}, Pablo},
        title = "{Clustering of luminous red galaxies - III. Baryon acoustic peak in the three-point correlation}",
      journal = {\mnras},
         year = 2009,
        month = oct,
       volume = {399},
       number = {2},
        pages = {801-811},
          doi = {10.1111/j.1365-2966.2009.15313.x},
archivePrefix = {arXiv},
       eprint = {0807.2448},
 primaryClass = {astro-ph},
       adsurl = {https://ui.adsabs.harvard.edu/abs/2009MNRAS.399..801G}
}

@ARTICLE{fry_n_point_theory,
       author = {{Fry}, J.~N.},
        title = "{Galaxy $N$-point correlation functions – Theoretical amplitudes for arbitrary $N$}",
      journal = {\apjl},
         year = 1984,
        month = feb,
       volume = {277},
        pages = {L5-L8},
          doi = {10.1086/184189},
       adsurl = {https://ui.adsabs.harvard.edu/abs/1984ApJ...277L...5F}
}

@ARTICLE{ezmock,
       author = {{Chuang}, Chia-Hsun and {Kitaura}, Francisco-Shu and {Prada}, Francisco and {Zhao}, Cheng and {Yepes}, Gustavo},
        title = "{EZmocks: extending the Zel'dovich approximation to generate mock galaxy catalogues with accurate clustering statistics}",
      journal = {\mnras},
         year = 2015,
        month = jan,
       volume = {446},
       number = {3},
        pages = {2621-2628},
          doi = {10.1093/mnras/stu2301},
archivePrefix = {arXiv},
       eprint = {1409.1124},
 primaryClass = {astro-ph.CO},
       adsurl = {https://ui.adsabs.harvard.edu/abs/2015MNRAS.446.2621C}
}

@ARTICLE{desi_y1_2ps,
       author = {{Adame}, A.~G. and {Aguilar}, J. and {Ahlen}, S. and {Alam}, S. and {Alexander}, D.~M. and {Alvarez}, M. and {Alves}, O. and {Anand}, A. and {Andrade}, U. and {Armengaud}, E. and {Avila}, S. and {Aviles}, A. and {Awan}, H. and {Bailey}, S. and {Baltay}, C. and {Bault}, A. and {Behera}, J. and {BenZvi}, S. and {Beutler}, F. and {Bianchi}, D. and {Blake}, C. and {Blum}, R. and {Brieden}, S. and {Brodzeller}, A. and {Brooks}, D. and {Brown}, Z. and {Buckley-Geer}, E. and {Burtin}, E. and {Calderon}, R. and {Canning}, R. and {Carnero Rosell}, A. and {Cereskaite}, R. and {Cervantes-Cota}, J.~L. and {Chabanier}, S. and {Chaussidon}, E. and {Chaves-Montero}, J. and {Chen}, S. and {Chen}, X. and {Claybaugh}, T. and {Cole}, S. and {Cuceu}, A. and {Davis}, T.~M. and {Dawson}, K. and {de la Macorra}, A. and {de Mattia}, A. and {Deiosso}, N. and {Demina}, R. and {Dey}, A. and {Dey}, B. and {Ding}, Z. and {Doel}, P. and {Edelstein}, J. and {Eftekharzadeh}, S. and {Eisenstein}, D.~J. and {Elliott}, A. and {Fagrelius}, P. and {Fanning}, K. and {Ferraro}, S. and {Ereza}, J. and {Findlay}, N. and {Flaugher}, B. and {Font-Ribera}, A. and {Forero-S{\'a}nchez}, D. and {Forero-Romero}, J.~E. and {Frenk}, C.~S. and {Garcia-Quintero}, C. and {Gazta{\~n}aga}, E. and {Gil-Mar{\'\i}n}, H. and {Gontcho}, S. Gontcho A. and {Gonzalez-Morales}, A.~X. and {Gonzalez-Perez}, V. and {Gordon}, C. and {Green}, D. and {Gruen}, D. and {Gsponer}, R. and {Gutierrez}, G. and {Guy}, J. and {Hadzhiyska}, B. and {Hahn}, C. and {Hanif}, M.~M.~S. and {Herrera-Alcantar}, H.~K. and {Honscheid}, K. and {Hou}, J. and {Howlett}, C. and {Huterer}, D. and {Ir{\v{s}}i{\v{c}}}, V. and {Ishak}, M. and {Juneau}, S. and {Kara{\c{c}}ayl{\i}}, N.~G. and {Kehoe}, R. and {Kent}, S. and {Kirkby}, D. and {Kitaura}, F.-S. and {Kong}, H. and {Kremin}, A. and {Krolewski}, A. and {Lai}, Y. and {Lan}, T.-W. and {Landriau}, M. and {Lang}, D. and {Lasker}, J. and {Le Goff}, J.~M. and {Le Guillou}, L. and {Leauthaud}, A. and {Levi}, M.~E. and {Li}, T.~S. and {Lodha}, K. and {Magneville}, C. and {Manera}, M. and {Margala}, D. and {Martini}, P. and {Maus}, M. and {McDonald}, P. and {Medina-Varela}, L. and {Meisner}, A. and {Mena-Fern{\'a}ndez}, J. and {Miquel}, R. and {Moon}, J. and {Moore}, S. and {Moustakas}, J. and {Mudur}, N. and {Mueller}, E. and {Mu{\~n}oz-Guti{\'e}rrez}, A. and {Myers}, A.~D. and {Nadathur}, S. and {Napolitano}, L. and {Neveux}, R. and {Newman}, J.~A. and {Nguyen}, N.~M. and {Nie}, J. and {Niz}, G. and {Noriega}, H.~E. and {Padmanabhan}, N. and {Paillas}, E. and {Palanque-Delabrouille}, N. and {Pan}, J. and {Penmetsa}, S. and {Percival}, W.~J. and {Pieri}, M.~M. and {Pinon}, M. and {Poppett}, C. and {Porredon}, A. and {Prada}, F. and {P{\'e}rez-Fern{\'a}ndez}, A. and {P{\'e}rez-R{\`a}fols}, I. and {Rabinowitz}, D. and {Raichoor}, A. and {Ram{\'\i}rez-P{\'e}rez}, C. and {Ramirez-Solano}, S. and {Rashkovetskyi}, M. and {Ravoux}, C. and {Rezaie}, M. and {Rich}, J. and {Rocher}, A. and {Rockosi}, C. and {Roe}, N.~A. and {Rosado-Marin}, A. and {Ross}, A.~J. and {Rossi}, G. and {Ruggeri}, R. and {Ruhlmann-Kleider}, V. and {Samushia}, L. and {Sanchez}, E. and {Saulder}, C. and {Schlafly}, E.~F. and {Schlegel}, D. and {Scholte}, D. and {Schubnell}, M. and {Seo}, H. and {Sharples}, R. and {Silber}, J. and {Slosar}, A. and {Smith}, A. and {Sprayberry}, D. and {Tan}, T. and {Tarl{\'e}}, G. and {Trusov}, S. and {Vaisakh}, R. and {Valcin}, D. and {Valdes}, F. and {Vargas-Maga{\~n}a}, M. and {Verde}, L. and {Walther}, M. and {Wang}, B. and {Wang}, M.~S. and {Weaver}, B.~A. and {Weaverdyck}, N. and {Wechsler}, R.~H. and {Weinberg}, D.~H. and {White}, M. and {Wilson}, M.~J. and {Yu}, J. and {Yu}, Y. and {Yuan}, S. and {Y{\`e}che}, C. and {Zaborowski}, E.~A. and {Zarrouk}, P. and {Zhang}, H. and {Zhao}, C. and {Zhao}, R.},
        title = "{DESI 2024 II: sample definitions, characteristics, and two-point clustering statistics}",
      journal = {\jcap},
         year = 2025,
        month = jul,
       volume = {2025},
       number = {7},
          eid = {017},
        pages = {017},
          doi = {10.1088/1475-7516/2025/07/017},
archivePrefix = {arXiv},
       eprint = {2411.12020},
 primaryClass = {astro-ph.CO},
       adsurl = {https://ui.adsabs.harvard.edu/abs/2025JCAP...07..017A}
}

@ARTICLE{DESI_Science_PartI,
   author = {{DESI Collaboration} and {Aghamousa}, A. and {Aguilar}, J. and
	{Ahlen}, S. and {Alam}, S. and {Allen}, L.~E. and {Allende Prieto}, C. and
	{Annis}, J. and {Bailey}, S. and {Balland}, C. and et al.},
    title = "{The DESI Experiment Part I: Science,Targeting, and Survey Design}",
  journal = {ArXiv e-prints},
archivePrefix = "arXiv",
   eprint = {1611.00036},
 primaryClass = "astro-ph.IM",
     year = 2016,
    month = oct,
   adsurl = {http://adsabs.harvard.edu/abs/2016arXiv161100036D}
}

@ARTICLE{DESI_Science_PartII,
       author = {{DESI Collaboration} and {Aghamousa}, Amir and {Aguilar}, Jessica and {Ahlen}, Steve and {Alam}, Shadab and {Allen}, Lori E. and {Allende Prieto}, Carlos and {Annis}, James and {Bailey}, Stephen and {Balland}, Christophe and {Ballester}, Otger and {Baltay}, Charles and {Beaufore}, Lucas and {Bebek}, Chris and {Beers}, Timothy C. and {Bell}, Eric F. and {Bernal}, Jos{\'e} Luis and {Besuner}, Robert and {Beutler}, Florian and {Blake}, Chris and {Bleuler}, Hannes and {Blomqvist}, Michael and {Blum}, Robert and {Bolton}, Adam S. and {Briceno}, Cesar and {Brooks}, David and {Brownstein}, Joel R. and {Buckley-Geer}, Elizabeth and {Burden}, Angela and {Burtin}, Etienne and {Busca}, Nicolas G. and {Cahn}, Robert N. and {Cai}, Yan-Chuan and {Cardiel-Sas}, Laia and {Carlberg}, Raymond G. and {Carton}, Pierre-Henri and {Casas}, Ricard and {Castander}, Francisco J. and {Cervantes-Cota}, Jorge L. and {Claybaugh}, Todd M. and {Close}, Madeline and {Coker}, Carl T. and {Cole}, Shaun and {Comparat}, Johan and {Cooper}, Andrew P. and {Cousinou}, M.-C. and {Crocce}, Martin and {Cuby}, Jean-Gabriel and {Cunningham}, Daniel P. and {Davis}, Tamara M. and {Dawson}, Kyle S. and {de la Macorra}, Axel and {De Vicente}, Juan and {Delubac}, Timoth{\'e}e and {Derwent}, Mark and {Dey}, Arjun and {Dhungana}, Govinda and {Ding}, Zhejie and {Doel}, Peter and {Duan}, Yutong T. and {Ealet}, Anne and {Edelstein}, Jerry and {Eftekharzadeh}, Sarah and {Eisenstein}, Daniel J. and {Elliott}, Ann and {Escoffier}, St{\'e}phanie and {Evatt}, Matthew and {Fagrelius}, Parker and {Fan}, Xiaohui and {Fanning}, Kevin and {Farahi}, Arya and {Farihi}, Jay and {Favole}, Ginevra and {Feng}, Yu and {Fernandez}, Enrique and {Findlay}, Joseph R. and {Finkbeiner}, Douglas P. and {Fitzpatrick}, Michael J. and {Flaugher}, Brenna and {Flender}, Samuel and {Font-Ribera}, Andreu and {Forero-Romero}, Jaime E. and {Fosalba}, Pablo and {Frenk}, Carlos S. and {Fumagalli}, Michele and {Gaensicke}, Boris T. and {Gallo}, Giuseppe and {Garcia-Bellido}, Juan and {Gaztanaga}, Enrique and {Pietro Gentile Fusillo}, Nicola and {Gerard}, Terry and {Gershkovich}, Irena and {Giannantonio}, Tommaso and {Gillet}, Denis and {Gonzalez-de-Rivera}, Guillermo and {Gonzalez-Perez}, Violeta and {Gott}, Shelby and {Graur}, Or and {Gutierrez}, Gaston and {Guy}, Julien and {Habib}, Salman and {Heetderks}, Henry and {Heetderks}, Ian and {Heitmann}, Katrin and {Hellwing}, Wojciech A. and {Herrera}, David A. and {Ho}, Shirley and {Holland}, Stephen and {Honscheid}, Klaus and {Huff}, Eric and {Hutchinson}, Timothy A. and {Huterer}, Dragan and {Hwang}, Ho Seong and {Illa Laguna}, Joseph Maria and {Ishikawa}, Yuzo and {Jacobs}, Dianna and {Jeffrey}, Niall and {Jelinsky}, Patrick and {Jennings}, Elise and {Jiang}, Linhua and {Jimenez}, Jorge and {Johnson}, Jennifer and {Joyce}, Richard and {Jullo}, Eric and {Juneau}, St{\'e}phanie and {Kama}, Sami and {Karcher}, Armin and {Karkar}, Sonia and {Kehoe}, Robert and {Kennamer}, Noble and {Kent}, Stephen and {Kilbinger}, Martin and {Kim}, Alex G. and {Kirkby}, David and {Kisner}, Theodore and {Kitanidis}, Ellie and {Kneib}, Jean-Paul and {Koposov}, Sergey and {Kovacs}, Eve and {Koyama}, Kazuya and {Kremin}, Anthony and {Kron}, Richard and {Kronig}, Luzius and {Kueter-Young}, Andrea and {Lacey}, Cedric G. and {Lafever}, Robin and {Lahav}, Ofer and {Lambert}, Andrew and {Lampton}, Michael and {Landriau}, Martin and {Lang}, Dustin and {Lauer}, Tod R. and {Le Goff}, Jean-Marc and {Le Guillou}, Laurent and {Le Van Suu}, Auguste and {Lee}, Jae Hyeon and {Lee}, Su-Jeong and {Leitner}, Daniela and {Lesser}, Michael and {Levi}, Michael E. and {L'Huillier}, Benjamin and {Li}, Baojiu and {Liang}, Ming and {Lin}, Huan and {Linder}, Eric and {Loebman}, Sarah R. and {Luki{\'c}}, Zarija and {Ma}, Jun and {MacCrann}, Niall and {Magneville}, Christophe and {Makarem}, Laleh and {Manera}, Marc and {Manser}, Christopher J. and {Marshall}, Robert and {Martini}, Paul and {Massey}, Richard and {Matheson}, Thomas and {McCauley}, Jeremy and {McDonald}, Patrick and {McGreer}, Ian D. and {Meisner}, Aaron and {Metcalfe}, Nigel and {Miller}, Timothy N. and {Miquel}, Ramon and {Moustakas}, John and {Myers}, Adam and {Naik}, Milind and {Newman}, Jeffrey A. and {Nichol}, Robert C. and {Nicola}, Andrina and {Nicolati da Costa}, Luiz and {Nie}, Jundan and {Niz}, Gustavo and {Norberg}, Peder and {Nord}, Brian and {Norman}, Dara and {Nugent}, Peter and {O'Brien}, Thomas and {Oh}, Minji and {Olsen}, Knut A.~G.},
        title = "{The DESI Experiment Part II: Instrument Design}",
      journal = {arXiv e-prints},
         year = 2016,
        month = oct,
          eid = {arXiv:1611.00037},
        pages = {arXiv:1611.00037},
          doi = {10.48550/arXiv.1611.00037},
archivePrefix = {arXiv},
       eprint = {1611.00037},
 primaryClass = {astro-ph.IM},
       adsurl = {https://ui.adsabs.harvard.edu/abs/2016arXiv161100037D}
}

@ARTICLE{Survey_Operations_DESI,
       author = {{Schlafly}, Edward F. and {Kirkby}, David and {Schlegel}, David J. and {Myers}, Adam D. and {Raichoor}, Anand and {Dawson}, Kyle and {Aguilar}, Jessica and {Allende Prieto}, Carlos and {Bailey}, Stephen and {BenZvi}, Segev and {Bermejo-Climent}, Jose and {Brooks}, David and {de la Macorra}, Axel and {Dey}, Arjun and {Doel}, Peter and {Fanning}, Kevin and {Font-Ribera}, Andreu and {Forero-Romero}, Jaime E. and {Garc{\'\i}a-Bellido}, Juan and {Gontcho A Gontcho}, Satya and {Guy}, Julien and {Hahn}, ChangHoon and {Honscheid}, Klaus and {Ishak}, Mustapha and {Juneau}, St{\'e}phanie and {Kehoe}, Robert and {Kisner}, Theodore and {Kremin}, Anthony and {Landriau}, Martin and {Lang}, Dustin A. and {Lasker}, James and {Levi}, Michael E. and {Magneville}, Christophe and {Manser}, Christopher J. and {Martini}, Paul and {Meisner}, Aaron M. and {Miquel}, Ramon and {Moustakas}, John and {Newman}, Jeffrey A. and {Nie}, Jundan and {Palanque-Delabrouille}, Nathalie. and {Percival}, Will J. and {Poppett}, Claire and {Rockosi}, Constance and {Ross}, Ashley J. and {Rossi}, Graziano and {Tarl{\'e}}, Gregory and {Weaver}, Benjamin A. and {Y{\`e}che}, Christophe and {Zhou}, Rongpu and {DESI Collaboration}},
        title = "{Survey Operations for the Dark Energy Spectroscopic Instrument}",
      journal = {\aj},
         year = 2023,
        month = dec,
       volume = {166},
       number = {6},
          eid = {259},
        pages = {259},
          doi = {10.3847/1538-3881/ad0832},
archivePrefix = {arXiv},
       eprint = {2306.06309},
 primaryClass = {astro-ph.CO},
       adsurl = {https://ui.adsabs.harvard.edu/abs/2023AJ....166..259S}
}

@ARTICLE{Data_Processing_DESI,
       author = {{Guy}, J. and {Bailey}, S. and {Kremin}, A. and {Alam}, Shadab and {Alexander}, D.~M. and {Allende Prieto}, C. and {BenZvi}, S. and {Bolton}, A.~S. and {Brooks}, D. and {Chaussidon}, E. and {Cooper}, A.~P. and {Dawson}, K. and {de la Macorra}, A. and {Dey}, A. and {Dey}, Biprateep and {Dhungana}, G. and {Eisenstein}, D.~J. and {Font-Ribera}, A. and {Forero-Romero}, J.~E. and {Gazta{\~n}aga}, E. and {Gontcho A Gontcho}, S. and {Green}, D. and {Honscheid}, K. and {Ishak}, M. and {Kehoe}, R. and {Kirkby}, D. and {Kisner}, T. and {Koposov}, Sergey E. and {Lan}, Ting-Wen and {Landriau}, M. and {Le Guillou}, L. and {Levi}, Michael E. and {Magneville}, C. and {Manser}, Christopher J. and {Martini}, P. and {Meisner}, Aaron M. and {Miquel}, R. and {Moustakas}, J. and {Myers}, Adam D. and {Newman}, Jeffrey A. and {Nie}, Jundan and {Palanque-Delabrouille}, N. and {Percival}, W.~J. and {Poppett}, C. and {Prada}, F. and {Raichoor}, A. and {Ravoux}, C. and {Ross}, A.~J. and {Schlafly}, E.~F. and {Schlegel}, D. and {Schubnell}, M. and {Sharples}, Ray M. and {Tarl{\'e}}, Gregory and {Weaver}, B.~A. and {Y{\'e}che}, Christophe and {Zhou}, Rongpu and {Zhou}, Zhimin and {Zou}, H.},
        title = "{The Spectroscopic Data Processing Pipeline for the Dark Energy Spectroscopic Instrument}",
      journal = {\aj},
         year = 2023,
        month = apr,
       volume = {165},
       number = {4},
          eid = {144},
        pages = {144},
          doi = {10.3847/1538-3881/acb212},
archivePrefix = {arXiv},
       eprint = {2209.14482},
 primaryClass = {astro-ph.IM},
       adsurl = {https://ui.adsabs.harvard.edu/abs/2023AJ....165..144G}
}

@ARTICLE{Fiber_system_DESI,
       author = {{Poppett}, Claire and {Tyas}, Luke and {Aguilar}, J. and {Bebek}, Christopher and {Bramall}, D. and {Claybaugh}, T. and {Edelstein}, J. and {Fagrelius}, P. and {Heetderks}, H. and {Jelinsky}, P. and {Jelinsky}, S. and {Lafever}, Robin and {Lambert}, A. and {Lampton}, M. and {Levi}, Michael E. and {Martini}, P. and {Rockosi}, C. and {Schmoll}, J. and {Sharples}, Ray M. and {Sirk}, Martin and {Wishnow}, Edward and {Yu}, Jiaxi and {Ahlen}, S. and {Bault}, A. and {BenZvi}, S. and {Brooks}, D. and {Cole}, S. and {de la Macorra}, A. and {Dey}, Arjun and {Doel}, P. and {Fanning}, K. and {Font-Ribera}, A. and {Forero-Romero}, J.~E. and {Gazta{\~n}aga}, E. and {Gontcho A Gontcho}, S. and {Gonzalez-Morales}, A.~X. and {Hahn}, C. and {Honscheid}, K. and {Jimenez}, J. and {Juneau}, S. and {Kirkby}, D. and {Kremin}, A. and {Landriau}, M. and {Le Guillou}, L. and {Manera}, M. and {Meisner}, A. and {Miquel}, R. and {Moustakas}, J. and {Mueller}, E. and {Mu{\~n}oz-Guti{\'e}rrez}, A. and {Myers}, A.~D. and {Nie}, J. and {Niz}, G. and {Palanque-Delabrouille}, N. and {Percival}, W.~J. and {Prada}, F. and {Rabinowitz}, D. and {Rezaie}, M. and {Rossi}, G. and {Sanchez}, E. and {Schlafly}, Edward F. and {Schlegel}, D. and {Schubnell}, M. and {Seo}, H. and {Sprayberry}, D. and {Tarl{\'e}}, G. and {Vargas-Maga{\~n}a}, M. and {Weaver}, B.~A. and {Zhou}, R.},
        title = "{Overview of the Fiber System for the Dark Energy Spectroscopic Instrument}",
      journal = {\aj},
         year = 2024,
        month = dec,
       volume = {168},
       number = {6},
          eid = {245},
        pages = {245},
          doi = {10.3847/1538-3881/ad76a4},
       adsurl = {https://ui.adsabs.harvard.edu/abs/2024AJ....168..245P}
}

@ARTICLE{Optical_corrector_DESI,
       author = {{Miller}, Timothy N. and {Doel}, Peter and {Gutierrez}, Gaston and {Besuner}, Robert and {Brooks}, David and {Gallo}, Giuseppe and {Heetderks}, Henry and {Jelinsky}, Patrick and {Kent}, Stephen M. and {Lampton}, Michael and {Levi}, Michael E. and {Liang}, Ming and {Meisner}, Aaron and {Sholl}, Michael J. and {Silber}, Joseph Harry and {Sprayberry}, David and {Aguilar}, Jessica Nicole and {de la Macorra}, Axel and {Eisenstein}, Daniel and {Fanning}, Kevin and {Font-Ribera}, Andreu and {Gazta{\~n}aga}, Enrique and {Gontcho A Gontcho}, Satya and {Honscheid}, Klaus and {Jimenez}, Jorge and {Joyce}, Dick and {Kehoe}, Robert and {Kisner}, Theodore and {Kremin}, Anthony and {Landriau}, Martin and {Le Guillou}, Laurent and {Magneville}, Christophe and {Martini}, Paul and {Miquel}, Ramon and {Moustakas}, John and {Nie}, Jundan and {Percival}, Will and {Poppett}, Claire and {Prada}, Francisco and {Rossi}, Graziano and {Schlegel}, David and {Schubnell}, Michael and {Seo}, Hee-Jong and {Sharples}, Ray and {Tarl{\'e}}, Gregory and {Vargas-Maga{\~n}a}, Mariana and {Zhou}, Zhimin and {the DESI Collaboration}},
        title = "{The Optical Corrector for the Dark Energy Spectroscopic Instrument}",
      journal = {\aj},
         year = 2024,
        month = aug,
       volume = {168},
       number = {2},
          eid = {95},
        pages = {95},
          doi = {10.3847/1538-3881/ad45fe},
archivePrefix = {arXiv},
       eprint = {2306.06310},
 primaryClass = {astro-ph.IM},
       adsurl = {https://ui.adsabs.harvard.edu/abs/2024AJ....168...95M}
}

@ARTICLE{Instrument_Overview_DESI,
       author = {{DESI Collaboration} and {Abareshi}, B. and {Aguilar}, J. and {Ahlen}, S. and {Alam}, Shadab and {Alexander}, David M. and {Alfarsy}, R. and {Allen}, L. and {Allende Prieto}, C. and {Alves}, O. and {Ameel}, J. and {Armengaud}, E. and {Asorey}, J. and {Aviles}, Alejandro and {Bailey}, S. and {Balaguera-Antol{\'\i}nez}, A. and {Ballester}, O. and {Baltay}, C. and {Bault}, A. and {Beltran}, S.~F. and {Benavides}, B. and {BenZvi}, S. and {Berti}, A. and {Besuner}, R. and {Beutler}, Florian and {Bianchi}, D. and {Blake}, C. and {Blanc}, P. and {Blum}, R. and {Bolton}, A. and {Bose}, S. and {Bramall}, D. and {Brieden}, S. and {Brodzeller}, A. and {Brooks}, D. and {Brownewell}, C. and {Buckley-Geer}, E. and {Cahn}, R.~N. and {Cai}, Z. and {Canning}, R. and {Capasso}, R. and {Carnero Rosell}, A. and {Carton}, P. and {Casas}, R. and {Castander}, F.~J. and {Cervantes-Cota}, J.~L. and {Chabanier}, S. and {Chaussidon}, E. and {Chuang}, C. and {Circosta}, C. and {Cole}, S. and {Cooper}, A.~P. and {da Costa}, L. and {Cousinou}, M.-C. and {Cuceu}, A. and {Davis}, T.~M. and {Dawson}, K. and {de la Cruz-Noriega}, R. and {de la Macorra}, A. and {de Mattia}, A. and {Della Costa}, J. and {Demmer}, P. and {Derwent}, M. and {Dey}, A. and {Dey}, B. and {Dhungana}, G. and {Ding}, Z. and {Dobson}, C. and {Doel}, P. and {Donald-McCann}, J. and {Donaldson}, J. and {Douglass}, K. and {Duan}, Y. and {Dunlop}, P. and {Edelstein}, J. and {Eftekharzadeh}, S. and {Eisenstein}, D.~J. and {Enriquez-Vargas}, M. and {Escoffier}, S. and {Evatt}, M. and {Fagrelius}, P. and {Fan}, X. and {Fanning}, K. and {Fawcett}, V.~A. and {Ferraro}, S. and {Ereza}, J. and {Flaugher}, B. and {Font-Ribera}, A. and {Forero-Romero}, J.~E. and {Frenk}, C.~S. and {Fromenteau}, S. and {G{\"a}nsicke}, B.~T. and {Garcia-Quintero}, C. and {Garrison}, L. and {Gazta{\~n}aga}, E. and {Gerardi}, F. and {Gil-Mar{\'\i}n}, H. and {Gontcho A Gontcho}, S. and {Gonzalez-Morales}, Alma X. and {Gonzalez-de-Rivera}, G. and {Gonzalez-Perez}, V. and {Gordon}, C. and {Graur}, O. and {Green}, D. and {Grove}, C. and {Gruen}, D. and {Gutierrez}, G. and {Guy}, J. and {Hahn}, C. and {Harris}, S. and {Herrera}, D. and {Herrera-Alcantar}, Hiram K. and {Honscheid}, K. and {Howlett}, C. and {Huterer}, D. and {Ir{\v{s}}i{\v{c}}}, V. and {Ishak}, M. and {Jelinsky}, P. and {Jiang}, L. and {Jimenez}, J. and {Jing}, Y.~P. and {Joyce}, R. and {Jullo}, E. and {Juneau}, S. and {Kara{\c{c}}ayl{\i}}, N.~G. and {Karamanis}, M. and {Karcher}, A. and {Karim}, T. and {Kehoe}, R. and {Kent}, S. and {Kirkby}, D. and {Kisner}, T. and {Kitaura}, F. and {Koposov}, S.~E. and {Kov{\'a}cs}, A. and {Kremin}, A. and {Krolewski}, Alex and {L'Huillier}, B. and {Lahav}, O. and {Lambert}, A. and {Lamman}, C. and {Lan}, Ting-Wen and {Landriau}, M. and {Lane}, S. and {Lang}, D. and {Lange}, J.~U. and {Lasker}, J. and {Le Guillou}, L. and {Leauthaud}, A. and {Le Van Suu}, A. and {Levi}, Michael E. and {Li}, T.~S. and {Magneville}, C. and {Manera}, M. and {Manser}, Christopher J. and {Marshall}, B. and {Martini}, Paul and {McCollam}, W. and {McDonald}, P. and {Meisner}, Aaron M. and {Mena-Fern{\'a}ndez}, J. and {Meneses-Rizo}, J. and {Mezcua}, M. and {Miller}, T. and {Miquel}, R. and {Montero-Camacho}, P. and {Moon}, J. and {Moustakas}, J. and {Mueller}, E. and {Mu{\~n}oz-Guti{\'e}rrez}, Andrea and {Myers}, Adam D. and {Nadathur}, S. and {Najita}, J. and {Napolitano}, L. and {Neilsen}, E. and {Newman}, Jeffrey A. and {Nie}, J.~D. and {Ning}, Y. and {Niz}, G. and {Norberg}, P. and {Noriega}, Hern{\'a}n E. and {O'Brien}, T. and {Obuljen}, A. and {Palanque-Delabrouille}, N. and {Palmese}, A. and {Zhiwei}, P. and {Pappalardo}, D. and {PENG}, X. and {Percival}, W.~J. and {Perruchot}, S. and {Pogge}, R. and {Poppett}, C. and {Porredon}, A. and {Prada}, F. and {Prochaska}, J. and {Pucha}, R. and {P{\'e}rez-Fern{\'a}ndez}, A. and {P{\'e}rez-R{\`a}fols}, I. and {Rabinowitz}, D. and {Raichoor}, A.},
        title = "{Overview of the Instrumentation for the Dark Energy Spectroscopic Instrument}",
      journal = {\aj},
         year = 2022,
        month = nov,
       volume = {164},
       number = {5},
          eid = {207},
        pages = {207},
          doi = {10.3847/1538-3881/ac882b},
archivePrefix = {arXiv},
       eprint = {2205.10939},
 primaryClass = {astro-ph.IM},
       adsurl = {https://ui.adsabs.harvard.edu/abs/2022AJ....164..207D}
}

@article{Ivanov_2022,
   title={Precision analysis of the redshift-space galaxy bispectrum},
   volume={105},
   ISSN={2470-0029},
   url={http://dx.doi.org/10.1103/PhysRevD.105.063512},
   DOI={10.1103/physrevd.105.063512},
   number={6},
   journal={Physical Review D},
   publisher={American Physical Society (APS)},
   author={Ivanov, Mikhail M. and Philcox, Oliver H. E. and Nishimichi, Takahiro and Simonović, Marko and Takada, Masahiro and Zaldarriaga, Matias},
   year={2022},
   month=mar }

@ARTICLE{Scoccimarro_Theory_To_observation_Bisp,
       author = {{Scoccimarro}, Rom{\'a}n},
        title = "{The Bispectrum: From Theory to Observations}",
      journal = {\apj},
         year = 2000,
        month = dec,
       volume = {544},
       number = {2},
        pages = {597-615},
          doi = {10.1086/317248},
archivePrefix = {arXiv},
       eprint = {astro-ph/0004086},
 primaryClass = {astro-ph},
       adsurl = {https://ui.adsabs.harvard.edu/abs/2000ApJ...544..597S}
}

@article{Desjacques_2018,
   title={Large-scale galaxy bias},
   volume={733},
   ISSN={0370-1573},
   url={http://dx.doi.org/10.1016/j.physrep.2017.12.002},
   DOI={10.1016/j.physrep.2017.12.002},
   journal={\physrep},
   publisher={Elsevier BV},
   author={Desjacques, Vincent and Jeong, Donghui and Schmidt, Fabian},
   year={2018},
   month=feb, pages={1–193} }

@ARTICLE{hartlap,
    author = "Hartlap, J. and Simon, Patrick and Schneider, P.",
    title = "{Why your model parameter confidences might be too optimistic: Unbiased estimation of the inverse covariance matrix}",
    eprint = "astro-ph/0608064",
    archivePrefix = "arXiv",
    doi = "10.1051/0004-6361:20066170",
    journal = "Astron. Astrophys.",
    volume = "464",
    pages = "399",
    year = "2007"
}

@ARTICLE{sellentin,
       author = {{Sellentin}, Elena and {Heavens}, Alan F.},
        title = "{Parameter inference with estimated covariance matrices}",
      journal = {\mnras},
         year = 2016,
        month = feb,
       volume = {456},
       number = {1},
        pages = {L132-L136},
          doi = {10.1093/mnrasl/slv190},
archivePrefix = {arXiv},
       eprint = {1511.05969},
 primaryClass = {astro-ph.CO},
       adsurl = {https://ui.adsabs.harvard.edu/abs/2016MNRAS.456L.132S}
}

@article{Bernardeau_2002,
   title={Large-scale structure of the Universe and cosmological perturbation theory},
   volume={367},
   ISSN={0370-1573},
   url={http://dx.doi.org/10.1016/S0370-1573(02)00135-7},
   DOI={10.1016/s0370-1573(02)00135-7},
   number={1–3},
   journal={\physrep},
   publisher={Elsevier BV},
   author={Bernardeau, F. and Colombi, S. and Gaztañaga, E. and Scoccimarro, R.},
   year={2002},
   month=sep, pages={1–248} }

@article{adame2024desi,
  title={DESI 2024 VII: Cosmological Constraints from the Full-Shape Modeling of Clustering Measurements},
  author={Adame, AG and Aguilar, J and Ahlen, S and Alam, S and Alexander, DM and Prieto, C Allende and Alvarez, M and Alves, O and Anand, A and Andrade, U and others},
  journal={arXiv preprint arXiv:2411.12022},
  year={2024}
}

@article{philcox2022boss,
       author = {{Philcox}, Oliver H.~E. and {Ivanov}, Mikhail M.},
        title = "{BOSS DR12 full-shape cosmology: {\ensuremath{\Lambda}}CDM constraints from the large-scale galaxy power spectrum and bispectrum monopole}",
      journal = {\prd},
         year = 2022,
        month = feb,
       volume = {105},
       number = {4},
          eid = {043517},
        pages = {043517},
          doi = {10.1103/PhysRevD.105.043517},
archivePrefix = {arXiv},
       eprint = {2112.04515},
 primaryClass = {astro-ph.CO},
       adsurl = {https://ui.adsabs.harvard.edu/abs/2022PhRvD.105d3517P}
}

@ARTICLE{inomata,
       author = {{Inomata}, Keisuke and {Jenks}, Leah and {Kamionkowski}, Marc},
        title = "{Parity-breaking galaxy 4-point function from lensing by chiral gravitational waves}",
      journal = {\prd},
         year = 2025,
        month = feb,
       volume = {111},
       number = {4},
          eid = {043504},
        pages = {043504},
          doi = {10.1103/PhysRevD.111.043504},
archivePrefix = {arXiv},
       eprint = {2408.03994},
 primaryClass = {astro-ph.CO},
       adsurl = {https://ui.adsabs.harvard.edu/abs/2025PhRvD.111d3504I}
}

@ARTICLE{hou_covar,
       author = {{Hou}, Jiamin and {Cahn}, Robert N. and {Philcox}, Oliver H.~E. and {Slepian}, Zachary},
        title = "{Analytic Gaussian covariance matrices for galaxy N -point correlation functions}",
      journal = {\prd},
         year = 2022,
        month = aug,
       volume = {106},
       number = {4},
          eid = {043515},
        pages = {043515},
          doi = {10.1103/PhysRevD.106.043515},
archivePrefix = {arXiv},
       eprint = {2108.01714},
 primaryClass = {astro-ph.CO},
       adsurl = {https://ui.adsabs.harvard.edu/abs/2022PhRvD.106d3515H}
}

@ARTICLE{phil_parity,
       author = {{Philcox}, Oliver H.~E.},
        title = "{Probing parity violation with the four-point correlation function of BOSS galaxies}",
      journal = {\prd},
         year = 2022,
        month = sep,
       volume = {106},
       number = {6},
          eid = {063501},
        pages = {063501},
          doi = {10.1103/PhysRevD.106.063501},
archivePrefix = {arXiv},
       eprint = {2206.04227},
 primaryClass = {astro-ph.CO},
       adsurl = {https://ui.adsabs.harvard.edu/abs/2022PhRvD.106f3501P}
}

@ARTICLE{jeong_fossil,
    author = "Jeong, Donghui and Kamionkowski, Marc",
    title = "{Clustering Fossils from the Early Universe}",
    eprint = "1203.0302",
    archivePrefix = "arXiv",
    primaryClass = "astro-ph.CO",
    doi = "10.1103/PhysRevLett.108.251301",
    journal = "\prl",
    volume = "108",
    pages = "251301",
    year = "2012"
}

@ARTICLE{bertolini,
       author = {{Bertolini}, Daniele and {Schutz}, Katelin and {Solon}, Mikhail P. and {Zurek}, Kathryn M.},
        title = "{The trispectrum in the Effective Field Theory of Large Scale Structure}",
      journal = {\jcap},
         year = 2016,
        month = jun,
       volume = {2016},
       number = {6},
          eid = {052},
        pages = {052},
          doi = {10.1088/1475-7516/2016/06/052},
archivePrefix = {arXiv},
       eprint = {1604.01770},
 primaryClass = {astro-ph.CO},
       adsurl = {https://ui.adsabs.harvard.edu/abs/2016JCAP...06..052B}
}

@ARTICLE{gualdi_joint_bi_tri,
       author = {{Gualdi}, Davide and {Gil-Mar{\'\i}n}, H{\'e}ctor and {Verde}, Licia},
        title = "{Joint analysis of anisotropic power spectrum, bispectrum and trispectrum: application to N-body simulations}",
      journal = {\jcap},
         year = 2021,
        month = jul,
       volume = {2021},
       number = {7},
          eid = {008},
        pages = {008},
          doi = {10.1088/1475-7516/2021/07/008},
archivePrefix = {arXiv},
       eprint = {2104.03976},
 primaryClass = {astro-ph.CO},
       adsurl = {https://ui.adsabs.harvard.edu/abs/2021JCAP...07..008G}
}

@ARTICLE{gualdi_2022_data_trispec,
       author = {{Gualdi}, Davide and {Verde}, Licia},
        title = "{Integrated trispectrum detection from BOSS DR12 NGC CMASS}",
      journal = {\jcap},
         year = 2022,
        month = sep,
       volume = {2022},
       number = {9},
          eid = {050},
        pages = {050},
          doi = {10.1088/1475-7516/2022/09/050},
archivePrefix = {arXiv},
       eprint = {2201.06932},
 primaryClass = {astro-ph.CO},
       adsurl = {https://ui.adsabs.harvard.edu/abs/2022JCAP...09..050G}
}

@ARTICLE{gualdi_2021_trispec,
       author = {{Gualdi}, Davide and {Novell}, Sergi and {Gil-Mar{\'\i}n}, H{\'e}ctor and {Verde}, Licia},
        title = "{Matter trispectrum: theoretical modelling and comparison to N-body simulations}",
      journal = {\jcap},
         year = 2021,
        month = jan,
       volume = {2021},
       number = {1},
          eid = {015},
        pages = {015},
          doi = {10.1088/1475-7516/2021/01/015},
archivePrefix = {arXiv},
       eprint = {2009.02290},
 primaryClass = {astro-ph.CO},
       adsurl = {https://ui.adsabs.harvard.edu/abs/2021JCAP...01..015G}
}

@ARTICLE{fry_bbgky,
       author = {{Fry}, J.~N.},
        title = "{The four-point function in the BBGKY hierarchy}",
      journal = {\apj},
         year = 1982,
        month = nov,
       volume = {262},
        pages = {424-431},
          doi = {10.1086/160437},
       adsurl = {https://ui.adsabs.harvard.edu/abs/1982ApJ...262..424F}
}

@ARTICLE{fry_peebles_4pcf,
       author = {{Fry}, J.~N. and {Peebles}, P.~J.~E.},
        title = "{Statistical analysis of catalogs of extragalactic objects. IX. The four-point galaxy correlation function.}",
      journal = {\apj},
         year = 1978,
        month = apr,
       volume = {221},
        pages = {19-33},
          doi = {10.1086/156001},
       adsurl = {https://ui.adsabs.harvard.edu/abs/1978ApJ...221...19F}
}

@ARTICLE{reinhard_axion,
       author = {{Reinhard}, Matthew A. and {Slepian}, Zachary and {Hou}, Jiamin and {Greco}, Alessandro},
        title = "{Full parity-violating trispectrum in axion inflation: Reduction to low-D integrals}",
      journal = {\jcap},
         year = 2026,
        month = may,
       volume = {2026},
       number = {5},
          eid = {099},
        pages = {099},
          doi = {10.1088/1475-7516/2026/05/099},
archivePrefix = {arXiv},
       eprint = {2412.16037},
 primaryClass = {astro-ph.CO},
       adsurl = {https://ui.adsabs.harvard.edu/abs/2026JCAP...05..099R}
}

@ARTICLE{jamieson_pops,
       author = {{Jamieson}, Drew and {Caravano}, Angelo and {Hou}, Jiamin and {Slepian}, Zachary and {Komatsu}, Eiichiro},
        title = "{Parity-odd power spectra: concise statistics for cosmological parity violation}",
      journal = {\mnras},
         year = 2024,
        month = sep,
       volume = {533},
       number = {3},
        pages = {2582-2598},
          doi = {10.1093/mnras/stae1924},
archivePrefix = {arXiv},
       eprint = {2406.15683},
 primaryClass = {astro-ph.CO},
       adsurl = {https://ui.adsabs.harvard.edu/abs/2024MNRAS.533.2582J}
}

@ARTICLE{iso_gen,
       author = {{Slepian}, Zachary and {Chellino}, Jessica and {Hou}, Jiamin and {Greco}, Alessandro},
        title = "{On a generating function for the isotropic basis functions and other connected results}",
      journal = {J. Phys. A: Math. Gen.},
         year = 2024,
        month = dec,
       volume = {57},
       number = {50},
          eid = {505203},
        pages = {505203},
          doi = {10.1088/1751-8121/ad955c},
archivePrefix = {arXiv},
       eprint = {2406.15385},
 primaryClass = {astro-ph.IM},
       adsurl = {https://ui.adsabs.harvard.edu/abs/2024JPhA...57X5203S}
}

@ARTICLE{DESI_Altmtl_Mock_Production,
       author = {{Lasker}, J. and {Carnero Rosell}, A. and {Myers}, A.~D. and {Ross}, A.~J. and {Bianchi}, D. and {Hanif}, M.~M.~S. and {Kehoe}, R. and {de Mattia}, A. and {Napolitano}, L. and {Percival}, W.~J. and {Staten}, R. and {Aguilar}, J. and {Ahlen}, S. and {Bigwood}, L. and {Brooks}, D. and {Claybaugh}, T. and {Cole}, S. and {de la Macorra}, A. and {Ding}, Z. and {Doel}, P. and {Fanning}, K. and {Forero-Romero}, J.~E. and {Gazta{\~n}aga}, E. and {Gontcho A Gontcho}, S. and {Gutierrez}, G. and {Honscheid}, K. and {Howlett}, C. and {Juneau}, S. and {Kremin}, A. and {Landriau}, M. and {Le Guillou}, L. and {Levi}, M.~E. and {Manera}, M. and {Meisner}, A. and {Miquel}, R. and {Moustakas}, J. and {Mueller}, E. and {Nie}, J. and {Niz}, G. and {Oh}, M. and {Palanque-Delabrouille}, N. and {Poppett}, C. and {Prada}, F. and {Rezaie}, M. and {Rossi}, G. and {Sanchez}, E. and {Schlegel}, D. and {Schubnell}, M. and {Seo}, H. and {Sprayberry}, D. and {Tarl{\'e}}, G. and {Vargas-Maga{\~n}a}, M. and {Weaver}, B.~A. and {Wilson}, Michael J. and {Zheng}, Y. and {DESI Collaboration}},
        title = "{Production of alternate realizations of DESI fiber assignment for unbiased clustering measurement in data and simulations}",
      journal = {\jcap},
         year = 2025,
        month = jan,
       volume = {2025},
       number = {1},
          eid = {127},
        pages = {127},
          doi = {10.1088/1475-7516/2025/01/127},
archivePrefix = {arXiv},
       eprint = {2404.03006},
 primaryClass = {astro-ph.CO},
       adsurl = {https://ui.adsabs.harvard.edu/abs/2025JCAP...01..127L}
}

@ARTICLE{cahn_parity,
    author = "Cahn, Robert N. and Slepian, Zachary and Hou, Jiamin",
    title = "{Test for Cosmological Parity Violation Using the 3D Distribution of Galaxies}",
    eprint = "2110.12004",
    archivePrefix = "arXiv",
    primaryClass = "astro-ph.CO",
    doi = "10.1103/PhysRevLett.130.201002",
    journal = "\prl",
    volume = "130",
    number = "20",
    pages = "201002",
    year = "2023"
}

@ARTICLE{se_rv_boss,
       author = {{Slepian}, Zachary and {Eisenstein}, Daniel J. and {Blazek}, Jonathan A. and {Brownstein}, Joel R. and {Chuang}, Chia-Hsun and {Gil-Mar{\'\i}n}, H{\'e}ctor and {Ho}, Shirley and {Kitaura}, Francisco-Shu and {McEwen}, Joseph E. and {Percival}, Will J. and {Ross}, Ashley J. and {Rossi}, Graziano and {Seo}, Hee-Jong and {Slosar}, An{\v{z}}e and {Vargas-Maga{\~n}a}, Mariana},
        title = "{Constraining the baryon-dark matter relative velocity with the large-scale three-point correlation function of the SDSS BOSS DR12 CMASS galaxies}",
      journal = {\mnras},
         year = 2018,
        month = feb,
       volume = {474},
       number = {2},
        pages = {2109-2115},
          doi = {10.1093/mnras/stx2723},
archivePrefix = {arXiv},
       eprint = {1607.06098},
 primaryClass = {astro-ph.CO},
       adsurl = {https://ui.adsabs.harvard.edu/abs/2018MNRAS.474.2109S}
}

@ARTICLE{phil_4pcf,
       author = {{Philcox}, Oliver H.~E. and {Hou}, Jiamin and {Slepian}, Zachary},
        title = "{A First Detection of the Connected 4-Point Correlation Function of Galaxies Using the BOSS CMASS Sample}",
      journal = {arXiv e-prints},
         year = 2021,
        month = aug,
          eid = {arXiv:2108.01670},
        pages = {arXiv:2108.01670},
          doi = {10.48550/arXiv.2108.01670},
archivePrefix = {arXiv},
       eprint = {2108.01670},
 primaryClass = {astro-ph.CO},
       adsurl = {https://ui.adsabs.harvard.edu/abs/2021arXiv210801670P}
}

@ARTICLE{se_rsd_3pcf,
       author = {{Slepian}, Zachary and {Eisenstein}, Daniel J.},
        title = "{Modelling the large-scale redshift-space 3-point correlation function of galaxies}",
      journal = {\mnras},
         year = 2017,
        month = aug,
       volume = {469},
       number = {2},
        pages = {2059-2076},
          doi = {10.1093/mnras/stx490},
archivePrefix = {arXiv},
       eprint = {1607.03109},
 primaryClass = {astro-ph.CO},
       adsurl = {https://ui.adsabs.harvard.edu/abs/2017MNRAS.469.2059S}
}

@ARTICLE{se_rv,
       author = {{Slepian}, Zachary and {Eisenstein}, Daniel J.},
        title = "{On the signature of the baryon-dark matter relative velocity in the two- and three-point galaxy correlation functions}",
      journal = {\mnras},
         year = 2015,
        month = mar,
       volume = {448},
       number = {1},
        pages = {9-26},
          doi = {10.1093/mnras/stu2627},
archivePrefix = {arXiv},
       eprint = {1411.4052},
 primaryClass = {astro-ph.CO},
       adsurl = {https://ui.adsabs.harvard.edu/abs/2015MNRAS.448....9S}
}

@ARTICLE{hou_parity,
       author = {{Hou}, Jiamin and {Slepian}, Zachary and {Cahn}, Robert N.},
        title = "{Measurement of parity-odd modes in the large-scale 4-point correlation function of Sloan Digital Sky Survey Baryon Oscillation Spectroscopic Survey twelfth data release CMASS and LOWZ galaxies}",
      journal = {\mnras},
         year = 2023,
        month = may,
       volume = {522},
       number = {4},
        pages = {5701-5739},
          doi = {10.1093/mnras/stad1062},
archivePrefix = {arXiv},
       eprint = {2206.03625},
 primaryClass = {astro-ph.CO},
       adsurl = {https://ui.adsabs.harvard.edu/abs/2023MNRAS.522.5701H}
}

@ARTICLE{se_boss_3pcf,
       author = {{Slepian}, Zachary and {Eisenstein}, Daniel J. and {Beutler}, Florian and {Chuang}, Chia-Hsun and {Cuesta}, Antonio J. and {Ge}, Jian and {Gil-Mar{\'\i}n}, H{\'e}ctor and {Ho}, Shirley and {Kitaura}, Francisco-Shu and {McBride}, Cameron K. and {Nichol}, Robert C. and {Percival}, Will J. and {Rodr{\'\i}guez-Torres}, Sergio and {Ross}, Ashley J. and {Scoccimarro}, Rom{\'a}n and {Seo}, Hee-Jong and {Tinker}, Jeremy and {Tojeiro}, Rita and {Vargas-Maga{\~n}a}, Mariana},
        title = "{The large-scale three-point correlation function of the SDSS BOSS DR12 CMASS galaxies}",
      journal = {\mnras},
         year = 2017,
        month = jun,
       volume = {468},
       number = {1},
        pages = {1070-1083},
          doi = {10.1093/mnras/stw3234},
archivePrefix = {arXiv},
       eprint = {1512.02231},
 primaryClass = {astro-ph.CO},
       adsurl = {https://ui.adsabs.harvard.edu/abs/2017MNRAS.468.1070S}
}

@ARTICLE{se_3pcf_bao,
       author = {{Slepian}, Zachary and {Eisenstein}, Daniel J. and {Brownstein}, Joel R. and {Chuang}, Chia-Hsun and {Gil-Mar{\'\i}n}, H{\'e}ctor and {Ho}, Shirley and {Kitaura}, Francisco-Shu and {Percival}, Will J. and {Ross}, Ashley J. and {Rossi}, Graziano and {Seo}, Hee-Jong and {Slosar}, An{\v{z}}e and {Vargas-Maga{\~n}a}, Mariana},
        title = "{Detection of baryon acoustic oscillation features in the large-scale three-point correlation function of SDSS BOSS DR12 CMASS galaxies}",
      journal = {\mnras},
         year = 2017,
        month = aug,
       volume = {469},
       number = {2},
        pages = {1738-1751},
          doi = {10.1093/mnras/stx488},
archivePrefix = {arXiv},
       eprint = {1607.06097},
 primaryClass = {astro-ph.CO},
       adsurl = {https://ui.adsabs.harvard.edu/abs/2017MNRAS.469.1738S}
}

@ARTICLE{xu_2012,
       author = {{Xu}, Xiaoying and {Padmanabhan}, Nikhil and {Eisenstein}, Daniel J. and {Mehta}, Kushal T. and {Cuesta}, Antonio J.},
        title = "{A 2 per cent distance to z = 0.35 by reconstructing baryon acoustic oscillations - II. Fitting techniques}",
      journal = {\mnras},
         year = 2012,
        month = dec,
       volume = {427},
       number = {3},
        pages = {2146-2167},
          doi = {10.1111/j.1365-2966.2012.21573.x},
archivePrefix = {arXiv},
       eprint = {1202.0091},
 primaryClass = {astro-ph.CO},
       adsurl = {https://ui.adsabs.harvard.edu/abs/2012MNRAS.427.2146X}
}

@ARTICLE{Galaxy_Halo_DESI_DR2,
       author = {{Wang}, Hanyue and {Eisenstein}, Daniel J. and {Aguilar}, Jessica Nicole and {Ahlen}, Steven and {Bianchi}, Davide and {Brooks}, David and {Claybaugh}, Todd and {de la Macorra}, Axel and {Dey}, Arjun and {Dey}, Biprateep and {Doel}, Peter and {Ferraro}, Simone and {Font-Ribera}, Andreu and {Forero-Romero}, Jaime E. and {Gazta{\~n}aga}, Enrique and {Gutierrez}, Gaston and {Honscheid}, Klaus and {Ishak}, Mustapha and {Joyce}, Richard and {Juneau}, Stephanie and {Kirkby}, David and {Kisner}, Theodore and {Kremin}, Anthony and {Lahav}, Ofer and {Lamman}, Claire and {Landriau}, Martin and {Manera}, Marc and {Meisner}, Aaron and {Miquel}, Ramon and {Mueller}, Eva-Maria and {Nadathur}, Seshadri and {Niz}, Gustavo and {Palanque-Delabrouille}, Nathalie and {Percival}, Will J. and {Prada}, Francisco and {P{\'e}rez-R{\`a}fols}, Ignasi and {Ross}, Ashley J. and {Rossi}, Graziano and {Sanchez}, Eusebio and {Schlegel}, David and {Schubnell}, Michael and {Silber}, Joseph Harry and {Sprayberry}, David and {Tarl{\'e}}, Gregory and {Weaver}, Benjamin Alan and {Zhou}, Rongpu and {Zou}, Hu},
        title = "{Galaxy-multiplet clustering from DESI DR2}",
      journal = {\mnras},
         year = 2026,
        month = feb,
       volume = {545},
       number = {4},
          eid = {staf2069},
        pages = {staf2069},
          doi = {10.1093/mnras/staf2069},
archivePrefix = {arXiv},
       eprint = {2511.15354},
 primaryClass = {astro-ph.CO},
       adsurl = {https://ui.adsabs.harvard.edu/abs/2026MNRAS.545f2069W}
}

@ARTICLE{Cov_Accuracy_Taylor,
       author = {{Taylor}, Andy and {Joachimi}, Benjamin and {Kitching}, Thomas},
        title = "{Putting the precision in precision cosmology: How accurate should your data covariance matrix be?}",
      journal = {\mnras},
         year = 2013,
        month = jul,
       volume = {432},
       number = {3},
        pages = {1928-1946},
          doi = {10.1093/mnras/stt270},
archivePrefix = {arXiv},
       eprint = {1212.4359},
 primaryClass = {astro-ph.CO},
       adsurl = {https://ui.adsabs.harvard.edu/abs/2013MNRAS.432.1928T}
}

@ARTICLE{guth_2007,
       author = {{Guth}, Alan H.},
        title = "{Eternal inflation and its implications}",
      journal = {J. Phys. A: Math. Gen.},
         year = 2007,
        month = jun,
       volume = {40},
       number = {25},
        pages = {6811-6826},
          doi = {10.1088/1751-8113/40/25/S25},
archivePrefix = {arXiv},
       eprint = {hep-th/0702178},
 primaryClass = {hep-th},
       adsurl = {https://ui.adsabs.harvard.edu/abs/2007JPhA...40.6811G}
}

@ARTICLE{esw_07,
       author = {{Eisenstein}, Daniel J. and {Seo}, Hee-Jong and {White}, Martin},
        title = "{On the Robustness of the Acoustic Scale in the Low-Redshift Clustering of Matter}",
      journal = {\apj},
         year = 2007,
        month = aug,
       volume = {664},
       number = {2},
        pages = {660-674},
          doi = {10.1086/518755},
archivePrefix = {arXiv},
       eprint = {astro-ph/0604361},
 primaryClass = {astro-ph},
       adsurl = {https://ui.adsabs.harvard.edu/abs/2007ApJ...664..660E}
}

@ARTICLE{Eisenstein_05,
       author = {{Eisenstein}, Daniel J. and {Zehavi}, Idit and {Hogg}, David W. and {Scoccimarro}, Roman and {Blanton}, Michael R. and {Nichol}, Robert C. and {Scranton}, Ryan and {Seo}, Hee-Jong and {Tegmark}, Max and {Zheng}, Zheng and {Anderson}, Scott F. and {Annis}, Jim and {Bahcall}, Neta and {Brinkmann}, Jon and {Burles}, Scott and {Castander}, Francisco J. and {Connolly}, Andrew and {Csabai}, Istvan and {Doi}, Mamoru and {Fukugita}, Masataka and {Frieman}, Joshua A. and {Glazebrook}, Karl and {Gunn}, James E. and {Hendry}, John S. and {Hennessy}, Gregory and {Ivezi{\'c}}, Zeljko and {Kent}, Stephen and {Knapp}, Gillian R. and {Lin}, Huan and {Loh}, Yeong-Shang and {Lupton}, Robert H. and {Margon}, Bruce and {McKay}, Timothy A. and {Meiksin}, Avery and {Munn}, Jeffery A. and {Pope}, Adrian and {Richmond}, Michael W. and {Schlegel}, David and {Schneider}, Donald P. and {Shimasaku}, Kazuhiro and {Stoughton}, Christopher and {Strauss}, Michael A. and {SubbaRao}, Mark and {Szalay}, Alexander S. and {Szapudi}, Istv{\'a}n and {Tucker}, Douglas L. and {Yanny}, Brian and {York}, Donald G.},
        title = "{Detection of the Baryon Acoustic Peak in the Large-Scale Correlation Function of SDSS Luminous Red Galaxies}",
      journal = {\apj},
         year = 2005,
        month = nov,
       volume = {633},
       number = {2},
        pages = {560-574},
          doi = {10.1086/466512},
archivePrefix = {arXiv},
       eprint = {astro-ph/0501171},
 primaryClass = {astro-ph},
       adsurl = {https://ui.adsabs.harvard.edu/abs/2005ApJ...633..560E}
}

@ARTICLE{bert_jain,
       author = {{Jain}, Bhuvnesh and {Bertschinger}, Edmund},
        title = "{Second-Order Power Spectrum and Nonlinear Evolution at High Redshift}",
      journal = {\apj},
         year = 1994,
        month = aug,
       volume = {431},
        pages = {495},
          doi = {10.1086/174502},
archivePrefix = {arXiv},
       eprint = {astro-ph/9311070},
 primaryClass = {astro-ph},
       adsurl = {https://ui.adsabs.harvard.edu/abs/1994ApJ...431..495J}
}

@ARTICLE{goroff,
       author = {{Goroff}, M.~H. and {Grinstein}, B. and {Rey}, S. -J. and {Wise}, M.~B.},
        title = "{Coupling of modes of cosmological mass density fluctuations}",
      journal = {\apj},
         year = 1986,
        month = dec,
       volume = {311},
        pages = {6-14},
          doi = {10.1086/164749},
       adsurl = {https://ui.adsabs.harvard.edu/abs/1986ApJ...311....6G}
}

@ARTICLE{bartolo,
    author = "Bartolo, N. and Komatsu, E. and Matarrese, Sabino and Riotto, A.",
    title = "{Non-Gaussianity from inflation: Theory and observations}",
    eprint = "astro-ph/0406398",
    archivePrefix = "arXiv",
    reportNumber = "DFPD-04-A-12",
    doi = "10.1016/j.physrep.2004.08.022",
    journal = "\physrep",
    volume = "402",
    pages = "103--266",
    year = "2004"
}

@ARTICLE{SE_BAO_2016,
       author = {{Slepian}, Zachary and {Eisenstein}, Daniel J.},
        title = "{A simple analytic treatment of linear growth of structure with baryon acoustic oscillations}",
      journal = {\mnras},
         year = 2016,
        month = mar,
       volume = {457},
       number = {1},
        pages = {24-37},
          doi = {10.1093/mnras/stv2889},
archivePrefix = {arXiv},
       eprint = {1509.08199},
 primaryClass = {astro-ph.CO},
       adsurl = {https://ui.adsabs.harvard.edu/abs/2016MNRAS.457...24S}
}

@ARTICLE{encore,
       author = {{Philcox}, Oliver H.~E. and {Slepian}, Zachary and {Hou}, Jiamin and {Warner}, Craig and {Cahn}, Robert N. and {Eisenstein}, Daniel J.},
        title = "{ENCORE: an $O(N_{\mathrm{g}}^{2})$ estimator for galaxy $N$-point correlation functions}",
      journal = {\mnras},
         year = 2022,
        month = jan,
       volume = {509},
       number = {2},
        pages = {2457-2481},
          doi = {10.1093/mnras/stab3025},
archivePrefix = {arXiv},
       eprint = {2105.08722},
 primaryClass = {astro-ph.IM},
       adsurl = {https://ui.adsabs.harvard.edu/abs/2022MNRAS.509.2457P}
}

@article{cahn_iso,
doi = {10.1088/1751-8121/acdfc4},
url = {https://dx.doi.org/10.1088/1751-8121/acdfc4},
year = {2023},
month = {jul},
publisher = {IOP Publishing},
volume = {56},
number = {32},
pages = {325204},
author = {Robert N Cahn and Zachary Slepian},
title = {Isotropic $N$-point basis functions and their properties},
journal = {Journal of Physics A: Mathematical and Theoretical}
}

@ARTICLE{se_3pt_alg,
       author = {{Slepian}, Zachary and {Eisenstein}, Daniel J.},
        title = "{Computing the three-point correlation function of galaxies in $O(N^2)$ time}",
      journal = {\mnras},
         year = 2015,
        month = dec,
       volume = {454},
       number = {4},
        pages = {4142-4158},
          doi = {10.1093/mnras/stv2119},
archivePrefix = {arXiv},
       eprint = {1506.02040},
 primaryClass = {astro-ph.CO},
       adsurl = {https://ui.adsabs.harvard.edu/abs/2015MNRAS.454.4142S}
}

@ARTICLE{landy_szalay,
       author = {{Landy}, Stephen D. and {Szalay}, Alexander S.},
        title = "{Bias and Variance of Angular Correlation Functions}",
      journal = {\apj},
         year = 1993,
        month = jul,
       volume = {412},
        pages = {64},
          doi = {10.1086/172900},
       adsurl = {https://ui.adsabs.harvard.edu/abs/1993ApJ...412...64L}
}

@article{ivanov2023cosmology,
       author = {{Ivanov}, Mikhail M. and {Philcox}, Oliver H.~E. and {Cabass}, Giovanni and {Nishimichi}, Takahiro and {Simonovi{\'c}}, Marko and {Zaldarriaga}, Matias},
        title = "{Cosmology with the galaxy bispectrum multipoles: Optimal estimation and application to BOSS data}",
      journal = {\prd},
         year = 2023,
        month = apr,
       volume = {107},
       number = {8},
          eid = {083515},
        pages = {083515},
          doi = {10.1103/PhysRevD.107.083515},
archivePrefix = {arXiv},
       eprint = {2302.04414},
 primaryClass = {astro-ph.CO},
       adsurl = {https://ui.adsabs.harvard.edu/abs/2023PhRvD.107h3515I}
}

@article{gil2015power,
       author = {{Gil-Mar{\'\i}n}, H{\'e}ctor and {Nore{\~n}a}, Jorge and {Verde}, Licia and {Percival}, Will J. and {Wagner}, Christian and {Manera}, Marc and {Schneider}, Donald P.},
        title = "{The power spectrum and bispectrum of SDSS DR11 BOSS galaxies – I. Bias and gravity}",
      journal = {\mnras},
         year = 2015,
        month = jul,
       volume = {451},
       number = {1},
        pages = {539-580},
          doi = {10.1093/mnras/stv961},
archivePrefix = {arXiv},
       eprint = {1407.5668},
 primaryClass = {astro-ph.CO},
       adsurl = {https://ui.adsabs.harvard.edu/abs/2015MNRAS.451..539G}
}

@ARTICLE{Ortola_4PCF,
       author = {{Ortol{\'a} Leonard}, William and {Slepian}, Zachary and {Hou}, Jiamin},
        title = "{A model for the redshift-space galaxy 4-point correlation function}",
      journal = {\jcap},
         year = 2025,
        month = jan,
       volume = {2025},
       number = {1},
          eid = {090},
        pages = {090},
          doi = {10.1088/1475-7516/2025/01/090},
archivePrefix = {arXiv},
       eprint = {2402.15510},
 primaryClass = {astro-ph.CO},
       adsurl = {https://ui.adsabs.harvard.edu/abs/2025JCAP...01..090O}
}

@ARTICLE{Zhou_LRG_Selection,
       author = {{Zhou}, Rongpu and {Dey}, Biprateep and {Newman}, Jeffrey A. and {Eisenstein}, Daniel J. and {Dawson}, K. and {Bailey}, S. and {Berti}, A. and {Guy}, J. and {Lan}, Ting-Wen and {Zou}, H. and {Aguilar}, J. and {Ahlen}, S. and {Alam}, Shadab and {Brooks}, D. and {de la Macorra}, A. and {Dey}, A. and {Dhungana}, G. and {Fanning}, K. and {Font-Ribera}, A. and {Gontcho}, S. Gontcho A. and {Honscheid}, K. and {Ishak}, Mustapha and {Kisner}, T. and {Kov{\'a}cs}, A. and {Kremin}, A. and {Landriau}, M. and {Levi}, Michael E. and {Magneville}, C. and {Manera}, Marc and {Martini}, P. and {Meisner}, Aaron M. and {Miquel}, R. and {Moustakas}, J. and {Myers}, Adam D. and {Nie}, Jundan and {Palanque-Delabrouille}, N. and {Percival}, W.~J. and {Poppett}, C. and {Prada}, F. and {Raichoor}, A. and {Ross}, A.~J. and {Schlafly}, E. and {Schlegel}, D. and {Schubnell}, M. and {Tarl{\'e}}, Gregory and {Weaver}, B.~A. and {Wechsler}, R.~H. and {Y{\'e}che}, Christophe and {Zhou}, Zhimin},
        title = "{Target Selection and Validation of DESI Luminous Red Galaxies}",
      journal = {\aj},
         year = 2023,
        month = feb,
       volume = {165},
       number = {2},
          eid = {58},
        pages = {58},
          doi = {10.3847/1538-3881/aca5fb},
archivePrefix = {arXiv},
       eprint = {2208.08515},
 primaryClass = {astro-ph.CO},
       adsurl = {https://ui.adsabs.harvard.edu/abs/2023AJ....165...58Z}
}

@ARTICLE{Beutler_rv,
       author = {{Beutler}, Florian and {Seljak}, Uro{\v{s}} and {Vlah}, Zvonimir},
        title = "{Constraining the relative velocity effect using the Baryon Oscillation Spectroscopic Survey}",
      journal = {\mnras},
         year = 2017,
        month = sep,
       volume = {470},
       number = {3},
        pages = {2723-2735},
          doi = {10.1093/mnras/stx1196},
archivePrefix = {arXiv},
       eprint = {1612.04720},
 primaryClass = {astro-ph.CO},
       adsurl = {https://ui.adsabs.harvard.edu/abs/2017MNRAS.470.2723B}
}

@ARTICLE{Krowleski_No_PV_det,
       author = {{Krolewski}, Alex and {May}, Simon and {Smith}, Kendrick and {Hopkins}, Hans},
        title = "{No evidence for parity violation in BOSS}",
      journal = {\jcap},
         year = 2024,
        month = aug,
       volume = {2024},
       number = {8},
          eid = {044},
        pages = {044},
          doi = {10.1088/1475-7516/2024/08/044},
archivePrefix = {arXiv},
       eprint = {2407.03397},
 primaryClass = {astro-ph.CO},
       adsurl = {https://ui.adsabs.harvard.edu/abs/2024JCAP...08..044K}
}

@ARTICLE{niu_axion_trispectrum,
       author = {{Niu}, Xuce and {Rahat}, Moinul Hossain and {Srinivasan}, Karthik and {Xue}, Wei},
        title = "{Parity-odd and even trispectrum from axion inflation}",
      journal = {\jcap},
         year = 2023,
        month = may,
       volume = {2023},
       number = {5},
          eid = {018},
        pages = {018},
          doi = {10.1088/1475-7516/2023/05/018},
archivePrefix = {arXiv},
       eprint = {2211.14324},
 primaryClass = {hep-ph},
       adsurl = {https://ui.adsabs.harvard.edu/abs/2023JCAP...05..018N}
}

@ARTICLE{ortola_cov_I,
       author = {{Ortol{\'a} Leonard}, William and {Slepian}, Zachary},
        title = "{Analytical Template for the 4-Point Correlation Function Covariance Beyond the Gaussian Random Field I: 1-Loop Corrections involving Second-Order Densities}",
      journal = {arXiv e-prints},
         year = 2025,
        month = sep,
          eid = {arXiv:2509.05419},
        pages = {arXiv:2509.05419},
          doi = {10.48550/arXiv.2509.05419},
archivePrefix = {arXiv},
       eprint = {2509.05419},
 primaryClass = {astro-ph.CO},
       adsurl = {https://ui.adsabs.harvard.edu/abs/2025arXiv250905419O}
}

@ARTICLE{ortola_cov_II,
       author = {{Ortol{\'a} Leonard}, William and {Slepian}, Zachary},
        title = "{Analytical Template for the 4-Point Correlation Function Covariance Beyond the Gaussian Random Field II: 1-Loop Corrections with Third-Order Densities}",
      journal = {arXiv e-prints},
         year = 2025,
        month = sep,
          eid = {arXiv:2509.05422},
        pages = {arXiv:2509.05422},
          doi = {10.48550/arXiv.2509.05422},
archivePrefix = {arXiv},
       eprint = {2509.05422},
 primaryClass = {astro-ph.CO},
       adsurl = {https://ui.adsabs.harvard.edu/abs/2025arXiv250905422O}
}

@ARTICLE{Williamson_4PCF_MHD,
       author = {{Williamson}, Victoria and {Sunseri}, James and {Slepian}, Zachary and {Hou}, Jiamin and {Greco}, Alessandro},
        title = "{First Measurements of the 4-Point Correlation Function of Magnetohydrodynamic Turbulence as a Novel Probe of the Interstellar Medium}",
      journal = {arXiv e-prints},
         year = 2024,
        month = dec,
          eid = {arXiv:2412.03967},
        pages = {arXiv:2412.03967},
          doi = {10.48550/arXiv.2412.03967},
archivePrefix = {arXiv},
       eprint = {2412.03967},
 primaryClass = {astro-ph.GA},
       adsurl = {https://ui.adsabs.harvard.edu/abs/2024arXiv241203967W}
}

@ARTICLE{Slepian_Parity_Meas_DESI,
       author = {{Slepian}, Zachary and {Krolewski}, Alex and {Greco}, Alessandro and {May}, Simon and {Ortolá Leonard}, William and {Kamalinejad}, Farshad and {Chellino}, Jessica and {Reinhard}, Matthew and {Fernandez}, Elena and {Prada}, Francisco and {Ahlen}, Steven and {Bianchi}, Davide and {Brooks}, David and {Claybaugh}, Todd and {de la Macorra}, Axel and {de Mattia}, Arnaud and {Dey}, Biprateep and {Doel}, Peter and {Gaztanaga}, Enrique and {Gutierrez}, Gaston and {Honscheid}, Klaus and {Huterer}, Dragan and {Joyce}, Dick and {Kehoe}, Robert and {Kirkby}, David and {Kisner}, Theodore and {Landriau}, Martin and {Le Guillou}, Laurent and {Manera}, Marc and {Meisner}, Aaron and {Miquel}, Ramon and {Nadathur}, Seshadri and {Percival}, Will and {Ross}, Ashley and {Sanchez}, Eusebio and {Schlegel}, David and {Schubnell}, Michael and {Seo}, Hee-Jong and {Silber}, Joseph and {Sprayberry}, David and {Tarle}, Gregory},
        title = "{Measurement of Parity-Violating Modes of the Dark Energy Spectroscopic Instrument (DESI) Year 1 Luminous Red Galaxies' 4-Point Correlation Function}",
      journal = {arXiv e-prints},
         year = 2025,
        month = aug,
          eid = {arXiv:2508.09133},
        pages = {arXiv:2508.09133},
          doi = {10.48550/arXiv.2508.09133},
archivePrefix = {arXiv},
       eprint = {2508.09133},
 primaryClass = {astro-ph.CO},
       adsurl = {https://ui.adsabs.harvard.edu/abs/2025arXiv250809133S}
}

@ARTICLE{DESI_Catalog_Construction,
       author = {{Ross}, A.~J. and {Aguilar}, J. and {Ahlen}, S. and {Alam}, S. and {Anand}, A. and {Bailey}, S. and {Bianchi}, D. and {Brieden}, S. and {Brooks}, D. and {Burtin}, E. and {Carnero Rosell}, A. and {Chaussidon}, E. and {Claybaugh}, T. and {Cole}, S. and {Dawson}, K. and {de la Macorra}, A. and {de Mattia}, A. and {Dey}, A. and {Dey}, B. and {Doel}, P. and {Fanning}, K. and {Ferraro}, S. and {Ereza}, J. and {Font-Ribera}, A. and {Forero-Romero}, J.~E. and {Gazta{\~n}aga}, E. and {Gil-Mar{\'\i}n}, H. and {Gontcho A Gontcho}, S. and {Gonzalez-Morales}, A.~X. and {Guy}, J. and {Hahn}, C. and {Heydenreich}, S. and {Honscheid}, K. and {Howlett}, C. and {Ishak}, M. and {Karim}, T. and {Kirkby}, D. and {Kisner}, T. and {Kong}, H. and {Kremin}, A. and {Krolewski}, A. and {Lambert}, A. and {Landriau}, M. and {Lasker}, J. and {Guillou}, L.~L. and {Levi}, M.~E. and {Manera}, M. and {Martini}, P. and {McDonald}, P. and {Meisner}, A. and {Miquel}, R. and {Moon}, J. and {Moustakas}, J. and {Mu{\~n}oz-Guti{\'e}rrez}, A. and {Myers}, A.~D. and {Nadathur}, S. and {Napolitano}, L. and {Newman}, J.~A. and {Nie}, J. and {Niz}, G. and {Palanque-Delabrouille}, N. and {Percival}, W.~J. and {Poppett}, C. and {Prada}, F. and {Raichoor}, A. and {Ravoux}, C. and {Rezaie}, M. and {Rosado-Marin}, A. and {Rossi}, G. and {Samushia}, L. and {Sanchez}, E. and {Schlafly}, E.~F. and {Schlegel}, D. and {Seo}, H. and {Smith}, A. and {Sprayberry}, D. and {Tarl{\'e}}, G. and {Valcin}, D. and {Vargas-Maga{\~n}a}, M. and {Weaver}, B.~A. and {Wilson}, M.~J. and {Yu}, J. and {Zarrouk}, P. and {Zhao}, C. and {Zhou}, R. and {Zou}, H.},
        title = "{The construction of large-scale structure catalogs for the Dark Energy Spectroscopic Instrument}",
      journal = {\jcap},
         year = 2025,
        month = jan,
       volume = {2025},
       number = {1},
          eid = {125},
        pages = {125},
          doi = {10.1088/1475-7516/2025/01/125},
archivePrefix = {arXiv},
       eprint = {2405.16593},
 primaryClass = {astro-ph.CO},
       adsurl = {https://ui.adsabs.harvard.edu/abs/2025JCAP...01..125R}
}

@article{Zenodo_repo_with_Data,
  author       = {Ortolá Leonard, William and
                  Krolewski, Alex and
                  Slepian, Zachary and
                  Greco, Alessandro and
                  May, Simon},
  title        = {Supplementary data for: A Finely-Binned
                   Measurement of the Connected Even-Parity Galaxy
                   4-Point Correlation Function of DESI Year 1
                   Luminous Red Galaxies
                  },
  month        = sep,
  year         = 2026,
  publisher    = {Zenodo},
  doi          = {10.5281/zenodo.22149471},
  url          = {https://doi.org/10.5281/zenodo.22149471},
  journal = {Zenodo},
  pages   = {doi:10.5281/zenodo.22149471}
}

@ARTICLE{Kamalinejad_3PCF_BAO,
       author = {{Kamalinejad}, Farshad and {Slepian}, Zachary and {Krolewski}, Alex and {Greco}, Alessandro and {Ortol{\'a} Leonard}, William and {Chellino}, Jessica and {Reinhard}, Matthew and {Fern{\'a}ndez-Garc{\'\i}a}, Elena and {Prada}, Francisco and {Aguilar}, J. and {Ahlen}, S. and {Anand}, A. and {Bebek}, C. and {Bianchi}, D. and {Brooks}, D. and {Claybaugh}, T. and {Cuceu}, A. and {Dawson}, K.~S. and {de la Macorra}, A. and {Demina}, R. and {Doel}, P. and {Edelstein}, J. and {Forero-Romero}, J.~E. and {Gazta{\~n}aga}, E. and {Gontcho}, S. Gontcho A and {Gutierrez}, G. and {Herrera-Alcantar}, H.~K. and {Honscheid}, K. and {Howlett}, C. and {Huterer}, D. and {Ishak}, M. and {Joyce}, R. and {Juneau}, S. and {Kirkby}, D. and {Kisner}, T. and {Kremin}, A. and {Lahav}, O. and {Lamman}, C. and {Landriau}, M. and {Le Guillou}, L. and {Manera}, M. and {Meisner}, A. and {Miquel}, R. and {Newman}, J.~A. and {Percival}, W.~J. and {Poppett}, C. and {P{\'e}rez-R{\`a}fols}, I. and {Samushia}, L. and {Sanchez}, E. and {Schlegel}, D. and {Schubnell}, M. and {Seo}, H. and {Silber}, J. and {Sprayberry}, D. and {Tarl{\'e}}, G. and {Weaver}, B.~A. and {Zhao}, C. and {Zou}, H.},
        title = "{First Detection of the Baryon Acoustic Oscillation (BAO) Feature in the 3-Point Correlation Function of DESI DR1 Luminous Red Galaxies}",
      journal = {arXiv e-prints},
         year = 2026,
        month = feb,
          eid = {arXiv:2602.16134},
        pages = {arXiv:2602.16134},
          doi = {10.48550/arXiv.2602.16134},
archivePrefix = {arXiv},
       eprint = {2602.16134},
 primaryClass = {astro-ph.CO},
       adsurl = {https://ui.adsabs.harvard.edu/abs/2026arXiv260216134K}
}

@ARTICLE{DESI_VI_Cosmological_Constraints_BAO,
       author = {{Adame}, A.~G. and {Aguilar}, J. and {Ahlen}, S. and {Alam}, S. and {Alexander}, D.~M. and {Alvarez}, M. and {Alves}, O. and {Anand}, A. and {Andrade}, U. and {Armengaud}, E. and {Avila}, S. and {Aviles}, A. and {Awan}, H. and {Bahr-Kalus}, B. and {Bailey}, S. and {Baltay}, C. and {Bault}, A. and {Behera}, J. and {BenZvi}, S. and {Bera}, A. and {Beutler}, F. and {Bianchi}, D. and {Blake}, C. and {Blum}, R. and {Brieden}, S. and {Brodzeller}, A. and {Brooks}, D. and {Buckley-Geer}, E. and {Burtin}, E. and {Calderon}, R. and {Canning}, R. and {Carnero Rosell}, A. and {Cereskaite}, R. and {Cervantes-Cota}, J.~L. and {Chabanier}, S. and {Chaussidon}, E. and {Chaves-Montero}, J. and {Chen}, S. and {Chen}, X. and {Claybaugh}, T. and {Cole}, S. and {Cuceu}, A. and {Davis}, T.~M. and {Dawson}, K. and {de la Macorra}, A. and {de Mattia}, A. and {Deiosso}, N. and {Dey}, A. and {Dey}, B. and {Ding}, Z. and {Doel}, P. and {Edelstein}, J. and {Eftekharzadeh}, S. and {Eisenstein}, D.~J. and {Elliott}, A. and {Fagrelius}, P. and {Fanning}, K. and {Ferraro}, S. and {Ereza}, J. and {Findlay}, N. and {Flaugher}, B. and {Font-Ribera}, A. and {Forero-S{\'a}nchez}, D. and {Forero-Romero}, J.~E. and {Frenk}, C.~S. and {Garcia-Quintero}, C. and {Gazta{\~n}aga}, E. and {Gil-Mar{\'\i}n}, H. and {Gontcho a Gontcho}, S. and {Gonzalez-Morales}, A.~X. and {Gonzalez-Perez}, V. and {Gordon}, C. and {Green}, D. and {Gruen}, D. and {Gsponer}, R. and {Gutierrez}, G. and {Guy}, J. and {Hadzhiyska}, B. and {Hahn}, C. and {Hanif}, M.~M.~S. and {Herrera-Alcantar}, H.~K. and {Honscheid}, K. and {Howlett}, C. and {Huterer}, D. and {Ir{\v{s}}i{\v{c}}}, V. and {Ishak}, M. and {Juneau}, S. and {Kara{\c{c}}ayl{\i}}, N.~G. and {Kehoe}, R. and {Kent}, S. and {Kirkby}, D. and {Kremin}, A. and {Krolewski}, A. and {Lai}, Y. and {Lan}, T.-W. and {Landriau}, M. and {Lang}, D. and {Lasker}, J. and {Le Goff}, J.~M. and {Le Guillou}, L. and {Leauthaud}, A. and {Levi}, M.~E. and {Li}, T.~S. and {Linder}, E. and {Lodha}, K. and {Magneville}, C. and {Manera}, M. and {Margala}, D. and {Martini}, P. and {Maus}, M. and {McDonald}, P. and {Medina-Varela}, L. and {Meisner}, A. and {Mena-Fern{\'a}ndez}, J. and {Miquel}, R. and {Moon}, J. and {Moore}, S. and {Moustakas}, J. and {Mueller}, E. and {Mu{\~n}oz-Guti{\'e}rrez}, A. and {Myers}, A.~D. and {Nadathur}, S. and {Napolitano}, L. and {Neveux}, R. and {Newman}, J.~A. and {Nguyen}, N.~M. and {Nie}, J. and {Niz}, G. and {Noriega}, H.~E. and {Padmanabhan}, N. and {Paillas}, E. and {Palanque-Delabrouille}, N. and {Pan}, J. and {Penmetsa}, S. and {Percival}, W.~J. and {Pieri}, M.~M. and {Pinon}, M. and {Poppett}, C. and {Porredon}, A. and {Prada}, F. and {P{\'e}rez-Fern{\'a}ndez}, A. and {P{\'e}rez-R{\`a}fols}, I. and {Rabinowitz}, D. and {Raichoor}, A. and {Ram{\'\i}rez-P{\'e}rez}, C. and {Ramirez-Solano}, S. and {Rashkovetskyi}, M. and {Ravoux}, C. and {Rezaie}, M. and {Rich}, J. and {Rocher}, A. and {Rockosi}, C. and {Roe}, N.~A. and {Rosado-Marin}, A. and {Ross}, A.~J. and {Rossi}, G. and {Ruggeri}, R. and {Ruhlmann-Kleider}, V. and {Samushia}, L. and {Sanchez}, E. and {Saulder}, C. and {Schlafly}, E.~F. and {Schlegel}, D. and {Schubnell}, M. and {Seo}, H. and {Shafieloo}, A. and {Sharples}, R. and {Silber}, J. and {Slosar}, A. and {Smith}, A. and {Sprayberry}, D. and {Tan}, T. and {Tarl{\'e}}, G. and {Taylor}, P. and {Trusov}, S. and {Ure{\~n}a-L{\'o}pez}, L.~A. and {Vaisakh}, R. and {Valcin}, D. and {Valdes}, F. and {Vargas-Maga{\~n}a}, M. and {Verde}, L. and {Walther}, M. and {Wang}, B. and {Wang}, M.~S. and {Weaver}, B.~A. and {Weaverdyck}, N. and {Wechsler}, R.~H. and {Weinberg}, D.~H. and {White}, M. and {Yu}, J. and {Yu}, Y. and {Yuan}, S. and {Y{\`e}che}, C. and {Zaborowski}, E.~A. and {Zarrouk}, P. and {Zhang}, H. and {Zhao}, C. and {Zhao}, R. and {Zhou}, R. and {Zhuang}, T.},
        title = "{DESI 2024 VI: cosmological constraints from the measurements of baryon acoustic oscillations}",
      journal = {\jcap},
         year = 2025,
        month = feb,
       volume = {2025},
       number = {2},
          eid = {021},
        pages = {021},
          doi = {10.1088/1475-7516/2025/02/021},
archivePrefix = {arXiv},
       eprint = {2404.03002},
 primaryClass = {astro-ph.CO},
       adsurl = {https://ui.adsabs.harvard.edu/abs/2025JCAP...02..021A}
}

@ARTICLE{DESI_III_BAO_from_Galaxies,
       author = {{Adame}, A.~G. and {Aguilar}, J. and {Ahlen}, S. and {Alam}, S. and {Alexander}, D.~M. and {Alvarez}, M. and {Alves}, O. and {Anand}, A. and {Andrade}, U. and {Armengaud}, E. and {Avila}, S. and {Aviles}, A. and {Awan}, H. and {Bailey}, S. and {Baltay}, C. and {Bault}, A. and {Behera}, J. and {BenZvi}, S. and {Beutler}, F. and {Bianchi}, D. and {Blake}, C. and {Blum}, R. and {Brieden}, S. and {Brodzeller}, A. and {Brooks}, D. and {Buckley-Geer}, E. and {Burtin}, E. and {Calderon}, R. and {Canning}, R. and {Carnero Rosell}, A. and {Cereskaite}, R. and {Cervantes-Cota}, J.~L. and {Chabanier}, S. and {Chaussidon}, E. and {Chaves-Montero}, J. and {Chen}, S. and {Chen}, X. and {Claybaugh}, T. and {Cole}, S. and {Cuceu}, A. and {Davis}, T.~M. and {Dawson}, K. and {de la Macorra}, A. and {de Mattia}, A. and {Deiosso}, N. and {Dey}, A. and {Dey}, B. and {Ding}, Z. and {Doel}, P. and {Edelstein}, J. and {Eftekharzadeh}, S. and {Eisenstein}, D.~J. and {Elliott}, A. and {Fagrelius}, P. and {Fanning}, K. and {Ferraro}, S. and {Ereza}, J. and {Findlay}, N. and {Flaugher}, B. and {Font-Ribera}, A. and {Forero-S{\'a}nchez}, D. and {Forero-Romero}, J.~E. and {Garcia-Quintero}, C. and {Gazta{\~n}aga}, E. and {Gil-Mar{\'\i}n}, H. and {Gontcho a Gontcho}, S. and {Gonzalez-Morales}, A.~X. and {Gonzalez-Perez}, V. and {Gordon}, C. and {Green}, D. and {Gruen}, D. and {Gsponer}, R. and {Gutierrez}, G. and {Guy}, J. and {Hadzhiyska}, B. and {Hahn}, C. and {Hanif}, M.~M.~S. and {Herrera-Alcantar}, H.~K. and {Honscheid}, K. and {Howlett}, C. and {Huterer}, D. and {Ir{\v{s}}i{\v{c}}}, V. and {Ishak}, M. and {Juneau}, S. and {Kara{\c{c}}ayl{\i}}, N.~G. and {Kehoe}, R. and {Kent}, S. and {Kirkby}, D. and {Kong}, H. and {Kremin}, A. and {Krolewski}, A. and {Lai}, Y. and {Lan}, T.-W. and {Landriau}, M. and {Lang}, D. and {Lasker}, J. and {Le Goff}, J.~M. and {Le Guillou}, L. and {Leauthaud}, A. and {Levi}, M.~E. and {Li}, T.~S. and {Linder}, E. and {Lodha}, K. and {Magneville}, C. and {Manera}, M. and {Margala}, D. and {Martini}, P. and {Maus}, M. and {McDonald}, P. and {Medina-Varela}, L. and {Meisner}, A. and {Mena-Fern{\'a}ndez}, J. and {Miquel}, R. and {Moon}, J. and {Moore}, S. and {Moustakas}, J. and {Mueller}, E. and {Mu{\~n}oz-Guti{\'e}rrez}, A. and {Myers}, A.~D. and {Nadathur}, S. and {Napolitano}, L. and {Neveux}, R. and {Newman}, J.~A. and {Nguyen}, N.~M. and {Nie}, J. and {Niz}, G. and {Noriega}, H.~E. and {Padmanabhan}, N. and {Paillas}, E. and {Palanque-Delabrouille}, N. and {Pan}, J. and {Penmetsa}, S. and {Percival}, W.~J. and {Pieri}, M.~M. and {Pinon}, M. and {Poppett}, C. and {Porredon}, A. and {Prada}, F. and {P{\'e}rez-Fern{\'a}ndez}, A. and {P{\'e}rez-R{\`a}fols}, I. and {Rabinowitz}, D. and {Raichoor}, A. and {Ram{\'\i}rez-P{\'e}rez}, C. and {Ramirez-Solano}, S. and {Rashkovetskyi}, M. and {Ravoux}, C. and {Rezaie}, M. and {Rich}, J. and {Rocher}, A. and {Rockosi}, C. and {Roe}, N.~A. and {Rosado-Marin}, A. and {Ross}, A.~J. and {Rossi}, G. and {Ruggeri}, R. and {Ruhlmann-Kleider}, V. and {Samushia}, L. and {Sanchez}, E. and {Saulder}, C. and {Schlafly}, E.~F. and {Schlegel}, D. and {Schubnell}, M. and {Seo}, H. and {Sharples}, R. and {Silber}, J. and {Slosar}, A. and {Smith}, A. and {Sprayberry}, D. and {Swanson}, J. and {Tan}, T. and {Tarl{\'e}}, G. and {Trusov}, S. and {Vaisakh}, R. and {Valcin}, D. and {Valdes}, F. and {Vargas-Maga{\~n}a}, M. and {Verde}, L. and {Walther}, M. and {Wang}, B. and {Wang}, M.~S. and {Weaver}, B.~A. and {Weaverdyck}, N. and {Wechsler}, R.~H. and {Weinberg}, D.~H. and {White}, M. and {Wilson}, M.~J. and {Yu}, J. and {Yu}, Y. and {Yuan}, S. and {Y{\`e}che}, C. and {Zaborowski}, E.~A. and {Zarrouk}, P. and {Zhang}, H. and {Zhao}, C. and {Zhao}, R. and {Zhou}, R. and {Zou}, H. and {DESI Collaboration}},
        title = "{DESI 2024 III: baryon acoustic oscillations from galaxies and quasars}",
      journal = {\jcap},
         year = 2025,
        month = apr,
       volume = {2025},
       number = {4},
          eid = {012},
        pages = {012},
          doi = {10.1088/1475-7516/2025/04/012},
archivePrefix = {arXiv},
       eprint = {2404.03000},
 primaryClass = {astro-ph.CO},
       adsurl = {https://ui.adsabs.harvard.edu/abs/2025JCAP...04..012A}
}

@ARTICLE{DESI_ResultsII_BAO,
       author = {{Abdul Karim}, M. and {Aguilar}, J. and {Ahlen}, S. and {Alam}, S. and {Allen}, L. and {Allende Prieto}, C. and {Alves}, O. and {Anand}, A. and {Andrade}, U. and {Armengaud}, E. and {Aviles}, A. and {Bailey}, S. and {Baltay}, C. and {Bansal}, P. and {Bault}, A. and {Behera}, J. and {BenZvi}, S. and {Bianchi}, D. and {Blake}, C. and {Brieden}, S. and {Brodzeller}, A. and {Brooks}, D. and {Buckley-Geer}, E. and {Burtin}, E. and {Calderon}, R. and {Canning}, R. and {Rosell}, A. Carnero and {Carrilho}, P. and {Casas}, L. and {Castander}, F.~J. and {Charles}, M. and {Chaussidon}, E. and {Chaves-Montero}, J. and {Chebat}, D. and {Chen}, X. and {Claybaugh}, T. and {Cole}, S. and {Cooper}, A.~P. and {Cuceu}, A. and {Dawson}, K.~S. and {de la Macorra}, A. and {de Mattia}, A. and {Deiosso}, N. and {Della Costa}, J. and {Demina}, R. and {Dey}, A. and {Dey}, B. and {Ding}, Z. and {Doel}, P. and {Edelstein}, J. and {Eisenstein}, D.~J. and {Elbers}, W. and {Fagrelius}, P. and {Fanning}, K. and {Fern{\'a}ndez-Garc{\'\i}a}, E. and {Ferraro}, S. and {Font-Ribera}, A. and {Forero-Romero}, J.~E. and {Frenk}, C.~S. and {Garcia-Quintero}, C. and {Garrison}, L.~H. and {Gazta{\~n}aga}, E. and {Gil-Mar{\'\i}n}, H. and {Gontcho A Gontcho}, S. and {Gonzalez}, D. and {Gonzalez-Morales}, A.~X. and {Gordon}, C. and {Green}, D. and {Gutierrez}, G. and {Guy}, J. and {Hadzhiyska}, B. and {Hahn}, C. and {He}, S. and {Herbold}, M. and {Herrera-Alcantar}, H.~K. and {Ho}, M.-F. and {Honscheid}, K. and {Howlett}, C. and {Huterer}, D. and {Ishak}, M. and {Juneau}, S. and {Kamble}, N.~V. and {Kara{\c{c}}ayl{\i}}, N.~G. and {Kehoe}, R. and {Kent}, S. and {Kim}, A.~G. and {Kirkby}, D. and {Kisner}, T. and {Koposov}, S.~E. and {Kremin}, A. and {Krolewski}, A. and {Lahav}, O. and {Lamman}, C. and {Landriau}, M. and {Lang}, D. and {Lasker}, J. and {Le Goff}, J.~M. and {Le Guillou}, L. and {Leauthaud}, A. and {Levi}, M.~E. and {Li}, Q. and {Li}, T.~S. and {Lodha}, K. and {Lokken}, M. and {Lozano-Rodr{\'\i}guez}, F. and {Magneville}, C. and {Manera}, M. and {Martini}, P. and {Matthewson}, W.~L. and {Meisner}, A. and {Mena-Fern{\'a}ndez}, J. and {Menegas}, A. and {Mergulh{\~a}o}, T. and {Miquel}, R. and {Moustakas}, J. and {Mu{\~n}oz-Guti{\'e}rrez}, A. and {Mu{\~n}oz-Santos}, D. and {Myers}, A.~D. and {Nadathur}, S. and {Naidoo}, K. and {Napolitano}, L. and {Newman}, J.~A. and {Niz}, G. and {Noriega}, H.~E. and {Paillas}, E. and {Palanque-Delabrouille}, N. and {Pan}, J. and {Peacock}, J.~A. and {Pellejero Ibanez}, M. and {Percival}, W.~J. and {P{\'e}rez-Fern{\'a}ndez}, A. and {P{\'e}rez-R{\`a}fols}, I. and {Pieri}, M.~M. and {Poppett}, C. and {Prada}, F. and {Rabinowitz}, D. and {Raichoor}, A. and {Ram{\'\i}rez-P{\'e}rez}, C. and {Rashkovetskyi}, M. and {Ravoux}, C. and {Rich}, J. and {Rocher}, A. and {Rockosi}, C. and {Rohlf}, J. and {Rom{\'a}n-Herrera}, J.~O. and {Ross}, A.~J. and {Rossi}, G. and {Ruggeri}, R. and {Ruhlmann-Kleider}, V. and {Samushia}, L. and {Sanchez}, E. and {Sanders}, N. and {Schlegel}, D. and {Schubnell}, M. and {Seo}, H. and {Shafieloo}, A. and {Sharples}, R. and {Silber}, J. and {Sinigaglia}, F. and {Sprayberry}, D. and {Tan}, T. and {Tarl{\'e}}, G. and {Taylor}, P. and {Turner}, W. and {Ure{\~n}a-L{\'o}pez}, L.~A. and {Vaisakh}, R. and {Valdes}, F. and {Valogiannis}, G. and {Vargas-Maga{\~n}a}, M. and {Verde}, L. and {Walther}, M. and {Weaver}, B.~A. and {Weinberg}, D.~H. and {White}, M. and {Wolfson}, M. and {Y{\`e}che}, C. and {Yu}, J. and {Zaborowski}, E.~A. and {Zarrouk}, P. and {Zhai}, Z. and {Zhang}, H. and {Zhao}, C. and {Zhao}, G.~B. and {Zhou}, R. and {Zou}, H. and {DESI Collaboration}},
        title = "{DESI DR2 results. II. Measurements of baryon acoustic oscillations and cosmological constraints}",
      journal = {\prd},
         year = 2025,
        month = oct,
       volume = {112},
       number = {8},
          eid = {083515},
        pages = {083515},
          doi = {10.1103/tr6y-kpc6},
archivePrefix = {arXiv},
       eprint = {2503.14738},
 primaryClass = {astro-ph.CO},
       adsurl = {https://ui.adsabs.harvard.edu/abs/2025PhRvD.112h3515A}
}

@ARTICLE{DESI_DR1_paper,
       author = {{DESI Collaboration} and {Abdul Karim}, M. and {Adame}, A.~G. and {Aguado}, D. and {Aguilar}, J. and {Ahlen}, S. and {Alam}, S. and {Aldering}, G. and {Alexander}, D.~M. and {Alfarsy}, R. and {Allen}, L. and {Allende Prieto}, C. and {Alves}, O. and {Anand}, A. and {Andrade}, U. and {Armengaud}, E. and {Avila}, S. and {Aviles}, A. and {Awan}, H. and {Bailey}, S. and {Baleato Lizancos}, A. and {Ballester}, O. and {Bault}, A. and {Bautista}, J. and {Bean}, R. and {Behera}, J. and {BenZvi}, S. and {Beraldo e Silva}, L. and {Bermejo-Climent}, J.~R. and {Beutler}, F. and {Bianchi}, D. and {Blake}, C. and {Blum}, R. and {Bolton}, A.~S. and {Bonici}, M. and {Brieden}, S. and {Brodzeller}, A. and {Brooks}, D. and {Buckley-Geer}, E. and {Burtin}, E. and {Bystr{\"o}m}, A. and {Canning}, R. and {Carnero Rosell}, A. and {Carr}, A. and {Carrilho}, P. and {Casas}, L. and {Castander}, F.~J. and {Cereskaite}, R. and {Cervantes-Cota}, J.~L. and {Chaussidon}, E. and {Chaves-Montero}, J. and {Chen}, S. and {Chen}, X. and {Circosta}, C. and {Claybaugh}, T. and {Cole}, S. and {Cooper}, A.~P. and {Cousinou}, M.-C. and {Cuceu}, A. and {Davis}, T.~M. and {Dawson}, K.~S. and {de Belsunce}, R. and {de la Cruz}, R. and {de la Macorra}, A. and {de Mattia}, A. and {Deiosso}, N. and {Della Costa}, J. and {Demina}, R. and {Demirbozan}, U. and {DeRose}, J. and {Dey}, A. and {Dey}, B. and {Ding}, J. and {Ding}, Z. and {Doel}, P. and {Douglass}, K. and {Dowicz}, M. and {Ebina}, H. and {Edelstein}, J. and {Eisenstein}, D.~J. and {Elbers}, W. and {Emas}, N. and {Escoffier}, S. and {Fagrelius}, P. and {Fan}, X. and {Fanning}, K. and {Favole}, G. and {Fawcett}, V.~A. and {Fern{\'a}ndez-Garc{\'\i}a}, E. and {Ferraro}, S. and {Findlay}, N. and {Font-Ribera}, A. and {Forero-Romero}, J.~E. and {Forero-S{\'a}nchez}, D. and {Frenk}, C.~S. and {G{\"a}nsicke}, B.~T. and {Galbany}, L. and {Garc{\'\i}a-Bellido}, J. and {Garcia-Quintero}, C. and {Garrison}, L.~H. and {Gazta{\~n}aga}, E. and {Gil-Mar{\'\i}n}, H. and {Gloudemans}, A. and {Gnedin}, O.~Y. and {Gontcho A Gontcho}, S. and {Gonzalez}, D. and {Gonzalez-Morales}, A.~X. and {Gonzalez-Perez}, V. and {Gordon}, C. and {Graur}, O. and {Green}, D. and {Gruen}, D. and {Gsponer}, R. and {Guandalin}, C. and {Gutierrez}, G. and {Guy}, J. and {Hahn}, C. and {Han}, J.~J. and {Han}, J. and {He}, S. and {Herrera-Alcantar}, H.~K. and {Heydenreich}, S. and {Honscheid}, K. and {Hou}, J. and {Howlett}, C. and {Huterer}, D. and {Ir{\v{s}}i{\v{c}}}, V. and {Ishak}, M. and {Jacques}, A. and {Jiang}, L. and {Jimenez}, J. and {Jing}, Y.~P. and {Joachimi}, B. and {Joudaki}, S. and {Joyce}, R. and {Jullo}, E. and {Juneau}, S. and {Kara{\c{c}}ayl{\i}}, N.~G. and {Karim}, T. and {Kehoe}, R. and {Kent}, S. and {Khederlarian}, A. and {Kirkby}, D. and {Kisner}, T. and {Kitaura}, F.-S. and {Kizhuprakkat}, N. and {Kong}, H. and {Koposov}, S.~E. and {Kremin}, A. and {Krolewski}, A. and {Lahav}, O. and {Lai}, Y. and {Lamman}, C. and {Lan}, T.-W. and {Landriau}, M. and {Lang}, D. and {Lange}, J.~U. and {Lasker}, J. and {Le Goff}, J.~M. and {Le Guillou}, L. and {Leauthaud}, A. and {Levi}, M.~E. and {Li}, S. and {Li}, T.~S. and {Liu}, W. and {Lodha}, K. and {Lokken}, M. and {Luo}, Y. and {Magneville}, C. and {Manera}, M. and {Manser}, C.~J. and {Margala}, D. and {Martini}, P. and {Maus}, M. and {McCullough}, J. and {McDonald}, P. and {Medina}, G.~E. and {Medina-Varela}, L. and {Meisner}, A. and {Mena-Fern{\'a}ndez}, J. and {Menegas}, A. and {Meneses-Rizo}, J. and {Mezcua}, M. and {Miquel}, R. and {Montero-Camacho}, P. and {Moon}, J. and {Moustakas}, J. and {Mu{\~n}oz-Guti{\'e}rrez}, A. and {Mu noz-Santos}, D. and {Myers}, A.~D. and {Myles}, J. and {Nadathur}, S. and {Najita}, J. and {Napolitano}, L. and {Newman}, J.~A. and {Nikakhtar}, F. and {Nikutta}, R. and {Niz}, G. and {Noriega}, H.~E. and {Nugent}, P.},
        title = "{Data Release 1 of the Dark Energy Spectroscopic Instrument}",
      journal = {\aj},
         year = 2026,
        month = may,
       volume = {171},
       number = {5},
          eid = {285},
        pages = {285},
          doi = {10.3847/1538-3881/ae4c43},
archivePrefix = {arXiv},
       eprint = {2503.14745},
 primaryClass = {astro-ph.CO},
       adsurl = {https://ui.adsabs.harvard.edu/abs/2026AJ....171..285D}
}

@ARTICLE{Hou_Even_4PCF,
       author = {{Hou}, J. and {Cahn}, R.~N. and {Aguilar}, J. and {Ahlen}, S. and {Bianchi}, D. and {Brooks}, D. and {Claybaugh}, T. and {Doel}, P. and {Ferraro}, S. and {Forero-Romero}, J.~E. and {Gazta{\~n}aga}, E. and {Le Guillou}, L. and {Gutierrez}, G. and {Honscheid}, K. and {Huterer}, D. and {Ishak}, M. and {Joyce}, R. and {Juneau}, S. and {Kehoe}, R. and {Kirkby}, D. and {Kisner}, T. and {Kremin}, A. and {Lamman}, C. and {Landriau}, M. and {de la Macorra}, A. and {Manera}, M. and {de Mattia}, A. and {Miquel}, R. and {Mueller}, E. and {Nadathur}, S. and {Niz}, G. and {Percival}, W.~J. and {Prada}, F. and {P{\'e}rez-R{\`a}fols}, I. and {Ross}, A.~J. and {Rossi}, G. and {Sanchez}, E. and {Schlegel}, D. and {Schubnell}, M. and {Seo}, H. and {Silber}, J. and {Slepian}, Z. and {Sprayberry}, D. and {Tarl{\'e}}, G. and {Weaver}, B.~A. and {Zou}, H.},
        title = "{Study of the connected four-point correlation function of galaxies from the DESI Data Release 1 luminous red galaxy sample}",
      journal = {\prd},
         year = 2025,
        month = dec,
       volume = {112},
       number = {12},
          eid = {122005},
        pages = {122005},
          doi = {10.1103/1wyy-758f},
archivePrefix = {arXiv},
       eprint = {2508.09070},
 primaryClass = {astro-ph.CO},
       adsurl = {https://ui.adsabs.harvard.edu/abs/2025PhRvD.112l2005H}
}

@ARTICLE{Encyclopedia_Slepian,
       author = {{Slepian}, Zachary and {Kamalinejad}, Farshad and {Greco}, Alessandro},
        title = "{Power Spectrum, Bispectrum, 2- and 3-Point Correlation Function, and Beyond}",
      journal = {arXiv e-prints},
         year = 2025,
        month = aug,
          eid = {arXiv:2508.06762},
        pages = {arXiv:2508.06762},
          doi = {10.48550/arXiv.2508.06762},
archivePrefix = {arXiv},
       eprint = {2508.06762},
 primaryClass = {astro-ph.CO},
       adsurl = {https://ui.adsabs.harvard.edu/abs/2025arXiv250806762S}
}

@ARTICLE{Abacus_Mocks,
       author = {{Maksimova}, Nina A. and {Garrison}, Lehman H. and {Eisenstein}, Daniel J. and {Hadzhiyska}, Boryana and {Bose}, Sownak and {Satterthwaite}, Thomas P.},
        title = "{ABACUSSUMMIT: a massive set of high-accuracy, high-resolution N-body simulations}",
      journal = {\mnras},
         year = 2021,
        month = dec,
       volume = {508},
       number = {3},
        pages = {4017-4037},
          doi = {10.1093/mnras/stab2484},
archivePrefix = {arXiv},
       eprint = {2110.11398},
 primaryClass = {astro-ph.CO},
       adsurl = {https://ui.adsabs.harvard.edu/abs/2021MNRAS.508.4017M}
}

@ARTICLE{EZMock_Catalogue,
       author = {{Chuang}, Chia-Hsun and {Kitaura}, Francisco-Shu and {Prada}, Francisco and {Zhao}, Cheng and {Yepes}, Gustavo},
        title = "{EZmocks: extending the Zel'dovich approximation to generate mock galaxy catalogues with accurate clustering statistics}",
      journal = {\mnras},
         year = 2015,
        month = jan,
       volume = {446},
       number = {3},
        pages = {2621-2628},
          doi = {10.1093/mnras/stu2301},
archivePrefix = {arXiv},
       eprint = {1409.1124},
 primaryClass = {astro-ph.CO},
       adsurl = {https://ui.adsabs.harvard.edu/abs/2015MNRAS.446.2621C}
}

@ARTICLE{T2_statistic_Paper,
       author = {{Sellentin}, Elena and {Heavens}, Alan F.},
        title = "{On the insufficiency of arbitrarily precise covariance matrices: non-Gaussian weak-lensing likelihoods}",
      journal = {\mnras},
         year = 2018,
        month = jan,
       volume = {473},
       number = {2},
        pages = {2355-2363},
          doi = {10.1093/mnras/stx2491},
archivePrefix = {arXiv},
       eprint = {1707.04488},
 primaryClass = {astro-ph.CO},
       adsurl = {https://ui.adsabs.harvard.edu/abs/2018MNRAS.473.2355S}
}

@article{Fisher_Method_Brown,
  author  = {Brown, Morton B.},
  title   = {A Method for Combining Non-Independent, One-Sided Tests of Significance},
  journal = {Biometrics},
  year    = {1975},
  volume  = {31},
  number  = {4},
  pages   = {987--992},
  doi     = {10.2307/2529826},
  jstor   = {2529826}
}

\end{document}